\documentclass[twocolumn, trackchanges, preprint2]{aastex631}
\accepted{June 22, 2023}

\submitjournal{\apj}

\graphicspath{{./}{figures/}}

\usepackage{natbib}
\usepackage{gensymb}
\usepackage{amsmath, amssymb}
\usepackage{bigints}
\usepackage{indentfirst}
\usepackage{threeparttable,booktabs}
\usepackage{afterpage}
\usepackage{graphicx}
\usepackage{txfonts}
\usepackage[T1]{fontenc}
\usepackage{multirow}
\usepackage{footnote}
\usepackage{epstopdf}
\usepackage{longtable}

\usepackage{breakurl}
\usepackage{xcolor}
\usepackage{ulem}
\usepackage{soul}
\usepackage{float}
\usepackage{mathrsfs}
\usepackage{breqn} 
\usepackage{bm}

\def\HI{H~\textsc{i} }

\defcitealias{sternberg2014}{S14}
\defcitealias{bialy2016}{B16}

\begin{document}


\title{Probing the Conditions for the H~\textsc{i}-to-H$_{2}$ Transition in the Interstellar Medium}

\correspondingauthor{Min-Young Lee}
\email{mlee@kasi.re.kr}

\author{Gyueun Park}
\affiliation{Korea Astronomy and Space Science Institute, 776 Daedeok-daero, Daejeon 34055, Republic of Korea}
\affiliation{Department of Astronomy and Space Science, University of Science and Technology, 217 Gajeong-ro, Daejeon 34113, Republic of Korea}

\author{Min-Young Lee}
\affiliation{Korea Astronomy and Space Science Institute, 776 Daedeok-daero, Daejeon 34055, Republic of Korea}
\affiliation{Department of Astronomy and Space Science, University of Science and Technology, 217 Gajeong-ro, Daejeon 34113, Republic of Korea}

\author{Shmuel Bialy}
\affiliation{Department of Physics, Technion - Israel Institute of Technology, Haifa, 3200003, Israel}

\author{Blakesley Burkhart} 
\affiliation{Department of Physics and Astronomy, Rutgers University, Piscataway, NJ 08854, USA}
\affiliation{Center for Computational Astrophysics, Flatiron Institute, 162 Fifth Avenue, New York, NY 10010, USA} 

\author{J. R. Dawson}
\affiliation{School of Mathematical and Physical Sciences and Macquarie University Astrophysics and Space Technologies Research Centre, Macquarie University, NSW 2109, Australia}
\affiliation{Australia Telescope National Facility, CSIRO Space \& Astronomy, PO Box 76, Epping, NSW 1710, Australia}

\author{Carl Heiles}
\affiliation{Department of Astronomy, University of California, Berkeley, CA 94720, USA}

\author{Di Li}
\affiliation{National Astronomical Observatories, Chinese Academy of Sciences, Beijing 100101, China}
\affiliation{NAOC-UKZN Computational Astrophysics Centre, University of KwaZulu-Natal, Durban 4000, South Africa}
\affiliation{Research Center for Intelligent Computing, Zhejiang Laboratory, Hangzhou 311100, China}

\author{Claire Murray}
\affiliation{Space Telescope Science Institute, 3700 San Martin Drive, Baltimore, MD 21218, USA}
\affiliation{Department of Physics \& Astronomy, Johns Hopkins University, MD 21218, USA}

\author{Hiep Nguyen}
\affiliation{Research School of Astronomy and Astrophysics, The Australian National University, Canberra, ACT 2611, Australia}

\author{Anita Hafner}
\affiliation{Australia Telescope National Facility, CSIRO Space \& Astronomy, PO Box 76, Epping, NSW 1710, Australia}

\author{Daniel R. Rybarczyk}
\affiliation{Department of Astronomy, University of Wisconsin, Madison, WI 53706-15821, USA}

\author{Sne{\v{z}}ana Stanimirovi{\'c}}
\affiliation{Department of Astronomy, University of Wisconsin, Madison, WI 53706-15821, USA}



\begin{abstract}

\noindent 
In this paper, we investigate the conditions for the H~\textsc{i}-to-H$_{2}$ transition in the solar neighborhood by analyzing \HI emission and absorption measurements toward 58 Galactic lines of sight (LOSs) along with $^{12}$CO(1--0) (CO) and dust data. Based on the accurate column densities of the cold and warm neutral medium (CNM and WNM), we first perform a decomposition of gas into atomic and molecular phases and show that the observed LOSs are mostly H~\textsc{i}-dominated. In addition, we find that the CO-dark H$_{2}$, not the optically thick H~\textsc{i}, is a major ingredient of the dark gas in the solar neighborhood. To examine the conditions for the formation of CO-bright molecular gas, we analyze the kinematic association between \HI and CO and find that the CNM is kinematically more closely associated with CO than the WNM. When CNM components within CO line widths are isolated, we find the following characteristics: spin temperature $<$ 200~K, peak optical depth $>$ 0.1, CNM fraction of $\sim$0.6, and $V$-band dust extinction $>$ 0.5~mag. These results suggest that CO-bright molecular gas preferentially forms in environments with high column densities where the CNM becomes colder and more abundant. Finally, we confront the observed CNM properties with the steady-state H$_{2}$ formation model of Sternberg et al. and infer that the CNM must be clumpy with a small volume filling factor. Another possibility would be that missing processes in the model, such as cosmic-rays and gas dynamics, play an important role in the H~\textsc{i}-to-H$_{2}$ transition.

\end{abstract}




\section{Introduction} 
\label{s:intro}



As the most abundant molecule in the universe, molecular hydrogen (H$_{2}$) plays a key role in the heating and cooling of the interstellar medium (ISM), as well as in the formation of other heavier molecules \citep[e.g.,][]{sternberg1995, hollenbach1997}. In addition, H$_{2}$ is an essential ingredient for star formation, as extensively shown by Galactic and extragalactic observations \citep[e.g.,][]{kennicutt2012}. Considering this significance of H$_{2}$ in astrophysics, it is of critical importance to understand how H$_{2}$ forms out of the surrounding diffuse atomic (H~\textsc{i}) gas.

Observationally, the H~\textsc{i}-to-H$_{2}$ transition has been directly examined through ultraviolet (UV) absorption measurements toward early-type stars or active galactic nuclei \citep[e.g.,][]{savage1977, rachford2002, shull2021}. These measurements in the Lyman $\alpha$ (1216 $\AA$) and Lyman--Werner (LW; 912--1108 $\AA$) bands probe diffuse to translucent gas with the color excess $E(B-V)$ of $\sim$0.01--1.0 mag and were analyzed to derive \HI and H$_{2}$ column densities ($N$(H~\textsc{i}) and $N$(H$_{2}$)). The molecular fraction, $f{\rm (H_{2})} = 2N{\rm (H_{2})} / {\rm [}N{\rm (H~\textsc{i})} + 2N{\rm (H_{2})}{\rm ]}$, was then found to increase from very low ($\lesssim$ 0.01) to high values ($\gg$0.1) at the total hydrogen column density $N{\rm (H)} = N{\rm (H~\textsc{i})} + 2N{\rm (H_{2})}$ of $\sim$10$^{21}$~cm$^{-2}$ or $E(B-V)$ of $\sim$0.1~mag, indicating a sharp conversion from \HI to H$_{2}$. 

In addition, the H~\textsc{i}-to-H$_{2}$ transition has been indirectly inferred from the flattening of the \HI column density with respect to other dense gas tracers. For example, \citet{barriault2010} compared \HI and OH emission in infrared (IR) cirrus clouds and showed that the OH column density increases with the H~\textsc{i} column density up to $N$(OH) $\sim$~0.3~$\times$~10$^{14}$~cm$^{-2}$. At higher OH column densities, the \HI column density saturates to $\sim$5~$\times$~10$^{20}$~cm$^{-2}$, implying the presence of molecular gas not traced by \HI emission. Similarly, IR studies of diffuse clouds found a positive deviation from the linear relation between the \HI column density and IR emission \citep[e.g.,][]{reach1994, douglas2007}. The observed excess in IR emission indicates that a substantial amount of H$_{2}$ exists beyond the threshold \HI column density of $\sim$5~$\times$~10$^{20}$~cm$^{-2}$. 


Theoretically, the H~\textsc{i}-to-H$_{2}$ transition has been explored as one of the key processes in photodissociation regions (PDRs; e.g., \citealt{vandishoeck1986}; \citealt{draine1996}; \citealt{browning2003}; \citealt{goldsmith2007}; \citealt{liszt2007}). In interstellar space, molecular-dominated regions are found in dense regions where gas and dust grains provide sufficient shielding against dissociating UV radiation. These molecular regions are bound by PDRs, where the gas is primarily neutral. The structure of PDRs has been solved numerically and analytically, and recent analytical models (\citealt{krumholz2009}; \citealt{sternberg2014}; \citealt{bialy2016}) predict that the minimum \HI column density to shield H$_{2}$ from photodissociation depends on ISM conditions (e.g., $N$(H~\textsc{i}) $\sim$~10$^{21}$~cm$^{-2}$ for solar metallicity). Once this minimum \HI column density is accumulated, all excess \HI is converted into H$_{2}$, resulting in the uniform \HI distribution. 

While the observed threshold \HI column density of $\sim$(0.5--1)~$\times$~10$^{21}$~cm$^{-2}$ is consistent with what the analytical H$_{2}$ formation models predict for \HI shielding layers, the previous observational studies could not provide insights into what \HI conditions aside from the minimum column density are required for H$_{2}$ formation as they did not distinguish between different \HI phases. The distinct velocity structures between \HI emission and absorption spectral pairs have been interpreted as the presence of \HI gas with a range of temperatures and densities \citep[e.g.,][]{radhakrishnan1972}, and theoretical models of neutral atomic gas indeed have suggested that two \HI phases can coexist over the range of thermal pressure $P/k_{\rm B}$ $\sim$ 10$^{3}$--10$^{4}$~cm$^{-3}$~K ($k_{\rm B}$ = Boltzmann constant): cold neutral medium (CNM) and warm neutral medium (WNM) with densities and temperatures of ($n$, $T$) $\sim$ (5--120~cm$^{-3}$, 40--180~K) and (0.04--1~cm$^{-3}$, 7000--8000~K) 
\citep[e.g.,][]{wolfire1995,wolfire2003,bialy2019}. In addition to these stable phases, the thermally unstable medium (UNM) with intermediate densities and temperatures has been commonly observed \citep[e.g.,][]{murray2015,murray2018b}. As for the formation of molecular gas, the denser and colder CNM is expected to be crucial (e.g., H$_{2}$ formation $\propto$ \HI density), but the impact of the different \HI phases on the H~\textsc{i}-to-H$_{2}$ transition has been largely unexplored mainly because of a lack of observational constraints.

In this paper, we examine how the different \HI phases are related to the H~\textsc{i}-to-H$_{2}$ transition by analyzing \HI emission and absorption spectra along with $^{12}$CO($J = 1 \rightarrow 0$) (CO(1--0) hereafter) data toward 58 lines of sight (LOSs) at Galactic latitudes $b$~$<$~$-5{\degree}$. These data have been obtained as part of the Galactic Neutral Opacity and Molecular Excitation Survey (GNOMES) collaboration, whose primary science goal is to understand the properties of atomic and molecular gas in and around molecular clouds. So far the \HI and OH data were presented in \citet{stanimirovic2014}, \citet{nguyen2019}, and \citet{petzler2023}, and we make use of the derived \HI properties, such as the optical depth ($\tau_{\rm CNM}$) and spin temperature ($T_{\rm s}$) of the CNM and the column densities of the CNM and WNM ($N_{\rm CNM}$ and $N_{\rm WNM}$), to explore what conditions are required for the formation of CO-bright molecular gas. The observed \HI properties are also compared to the analytical model of \citet{sternberg2014} (\citetalias{sternberg2014} hereafter) to test if H$_{2}$ formation in steady state is indeed valid for solar neighborhood conditions. 



This paper is organized as follows. In Section \ref{s:background}, we summarize two of the most relevant studies, \citet{nguyen2019} and \citetalias{sternberg2014}, to provide background information. In Sections \ref{s:data} and \ref{s:env}, we present the H~\textsc{i}, CO, and dust data for our analyses and investigate the environmental conditions of the observed GNOMES LOSs. In Section \ref{s:results}, we describe the results from the CO observations and decompose the gas along each LOS into different atomic and molecular gas phases. The observed \HI and CO properties are compared to each other, as well as to the prediction from the \citetalias{sternberg2014} model, to provide observational and theoretical perspectives on the conditions for the formation of CO-bright molecular gas (Sections \ref{s:HI_CO_obs} and \ref{s:HI_CO_theory}). Finally, our results are discussed and summarized in Sections \ref{s:discussion} and \ref{s:summary}.

\section{Background}
\label{s:background}
In this section, we summarize recent observational and theoretical studies that are most relevant to our work. 

\subsection{CNM and WNM in and around molecular clouds}
\label{s:Nguyen19}

As part of GNOMES collaboration, \cite{nguyen2019} analyzed Arecibo \HI emission and absorption spectra toward 77 continuum sources located behind Perseus, Taurus, California, Rosette, NGC 2264, and Mon OB1. For their analyses, the authors divided the observed LOSs into the following three environments: (1) 22 LOSs at $b$~$>$~5${\degree}$ tracing the diffuse medium (``diffuse''); (2) 20 LOSs at |$b$|~$<$~5${\degree}$ penetrating the dense Galactic Plane with likely strong UV radiation field (``Plane''); (3) 35 LOSs at $b$~$<$~$-$5${\degree}$ probing the surroundings of local molecular clouds including Taurus and Perseus (``Perseus''). The \HI spectra along these LOSs were examined via the Gaussian decomposition method of \cite{heiles2003a} to estimate the physical properties of H~\textsc{i}, such as the optical depth, spin temperature, and column density of the CNM and WNM (see Section \ref{s:obs_HI_OH} for details on the observations and analysis methods). 

Strong \HI absorption was detected toward all the observed LOSs, and a total of 349 CNM and 327 WNM components were identified. For the identified CNM components, the peak optical depth ranges from $\sim$0.01 to $\sim$16.2 with a median of $\sim$0.4, and the spin temperature varies from $\sim$10~K to $\sim$480~K with the distribution peak at $\sim$50~K. Interestingly, these individual properties are comparable between the three environments and agree with results from previous measurements of random LOSs \citep[e.g.,][]{heiles2003b,murray2015, murray2018b}, implying that the CNM has universal properties throughout the Galaxy. On the other hand, the CNM fraction, which is defined as the ratio of the CNM to total \HI column density, is systematically higher in molecular cloud environments (median fractions of 0.43 and 0.37 for the Plane and Perseus LOSs versus 0.16 for the diffuse LOSs), suggesting a close association between the abundance of the CNM and the formation of molecular gas. 

\subsection{Theoretical Modeling of H$_{2}$ Formation in the Steady-State Medium}
\label{s:bialy}

\citetalias{sternberg2014} developed an analytical model of the H~\textsc{i}-to-H$_{2}$ transition in a one-dimensional plane-parallel slab of gas and dust and provided the following expression of the total \HI column density for two-sided isotropic UV radiation:

\begin{equation}
    \label{e:NHI_b16}
    N(\textrm{H~\textsc{i}})~(\textrm{cm}^{-2})= \frac{8.4 \times 10^{20}}{\tilde{\sigma_{\rm g}}}~\ln \left(\frac{\alpha G}{3.2} + 1\right) 
\end{equation}

\noindent where $\tilde{\sigma_{\rm g}}$ is the dust absorption cross-section per hydrogen nucleus in the LW band ($\sigma_{\rm g}$) normalized to the canonical solar metallicity value of $1.9 \times 10^{-21}$ cm$^{2}$.

The dimensionless parameter $\alpha$ in Equation (\ref{e:NHI_b16}) is the ratio of the unattenuated H$_{2}$ photodissociation rate to the H$_{2}$ formation rate, which can be expressed as

\begin{equation}
    \label{e:alpha}
    \begin{split}
        \alpha & = \frac{D_{0}}{Rn} \\ 
               & = 1.9 \times 10^{4}~\left(\frac{I_{\rm UV}}{\tilde{\sigma_{\rm g}}}
               \right) \left(\frac{100~\rm cm^{-3}}{n}\right), 
    \end{split}
\end{equation}

\noindent where $D_{\rm 0}$ is the free-space H$_{2}$ photodissociation rate, $R$ is the rate coefficient for H$_{2}$ formation on dust grains, $n = n_{1} + 2n_{2}$ is the total gas number density, $n_{1}$ is the \HI number density, $n_{2}$ is the H$_{2}$ number density, and $I_{\rm UV}$ is the strength of UV radiation relative to the Draine field \citep{draine1978, bialy2020}. On the other hand, the other dimensionless parameter $G$ can be interpreted as the average H$_{2}$ self-shielding factor. Here we employ the expression derived by \citet{bialy2016}, which uses a more accurate fitting function for the H$_{2}$ dissociation bandwidth and reads as

\begin{equation}
    \label{e:G}
    G = 3 \times 10^{-5}~\tilde{\sigma_{\rm g}} \left(\frac{9.9}{1+8.9\tilde{\sigma_{\rm g}}}\right)^{0.37}. 
\end{equation}


Combining Equations (\ref{e:alpha}) and (\ref{e:G}), $\alpha G$ can be written as\footnote{Equation (\ref{e:alphaG}) is taken from \citet{bialy2016}, who examined the \HI and H$_{2}$ density profiles of optically thick interstellar clouds based on the \citetalias{sternberg2014} model. While this expression was originally derived for beamed UV radiation, it is also applicable for isotropic UV radiation, considering that $\alpha$ is equal for beamed and isotropic UV radiation fields with the same strength and $G$ is independent of the UV field geometry (\citetalias{sternberg2014}).}

\begin{equation}
    \label{e:alphaG}
    \alpha G = 0.59~I_{\rm UV} \left(\frac{100~\textrm{cm}^{-3}}{n}\right) \left(\frac{9.9}{1 + 8.9 \tilde{\sigma_{\textrm{g}}}}\right)^{0.37} 
\end{equation}

\noindent and has the physical meaning of the ratio of the effective H$_{2}$ photodissociation rate (accounting for UV shielding) to the H$_{2}$ formation rate. For realistic ISM conditions, $\alpha G$ can range from large to small values. For example, when $\alpha G$ is small ($\ll$1; ``weak-field limit''), H$_{2}$ self-shielding primarily protects H$_{2}$ from dissociating UV photons, and the H~\textsc{i}-to-H$_{2}$ transition is gradual. In other words, most of the \HI column density is built up beyond the transition point where the gas is mainly molecular. On the contrary, when $\alpha G$ is large ($\gg$1; ``strong-field limit''), dust absorption becomes important, resulting in a sharp H~\textsc{i}-to-H$_{2}$ transition due to the exponential reduction of UV radiation with cloud column density. In this case, the \HI column density is built up in the outer layer of the gas slab prior to the transition point. We refer to \citetalias{sternberg2014} and \citet{bialy2016} for details on the model and the parameters.

\begin{figure*}[htbp]
    \centering
    \includegraphics[scale=0.39]{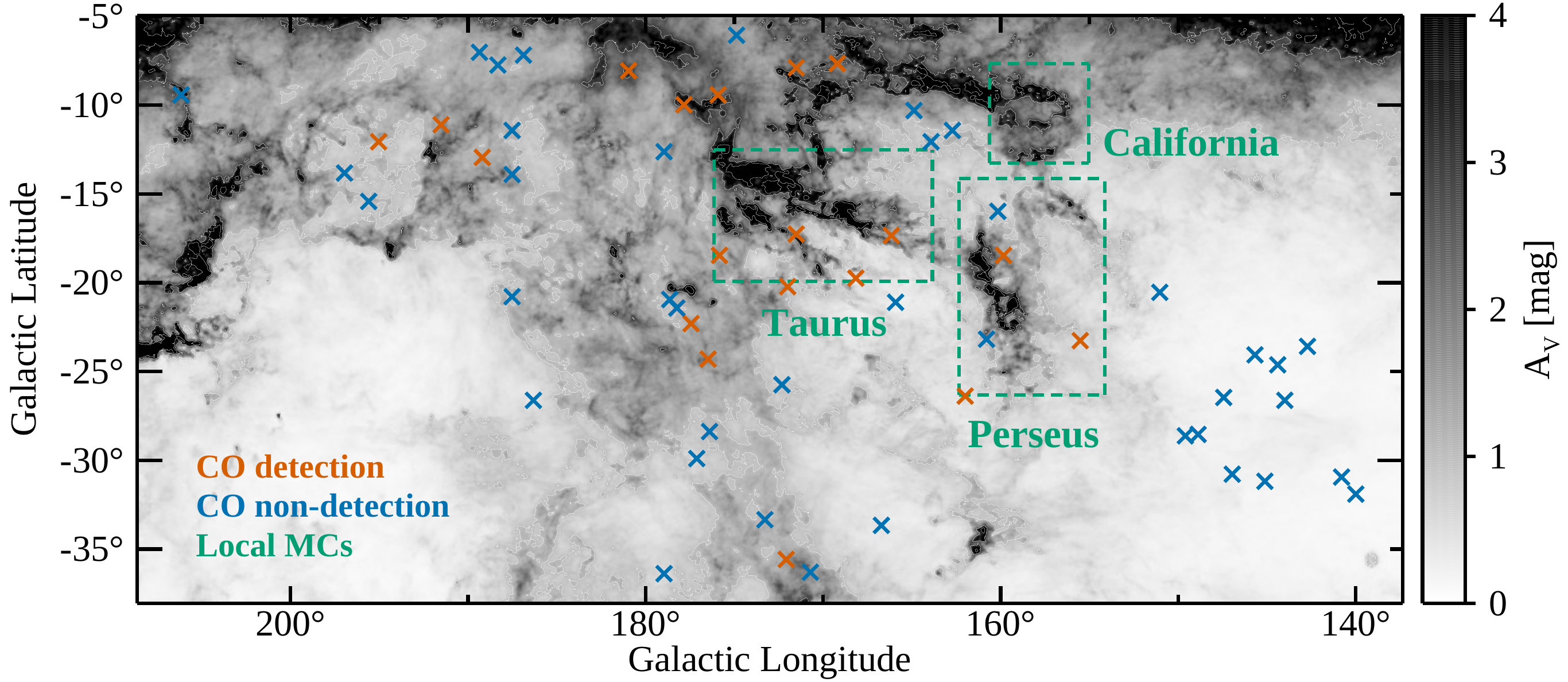}
    \vspace{0.2cm}
    \caption{\label{planckmap} GNOMES LOSs at $b < -5{\degree}$ overlaid on the \textit{Planck} $A_{V}$ image (Section \ref{s:total_col}; 1, 3, and 5 mag as the white contours). Among the total 58 LOSs, 19 LOSs where CO(1--0) emission is clearly detected are shown as the tan crosses. The remaining 39 LOSs without CO detection are indicated as the blue crosses. The green boxes represent the approximate extents of local molecular clouds (top to bottom: California, Taurus, and Perseus; \citealt{lee2018}).}
\end{figure*}

\section{Data} \label{s:data}

\subsection{GNOMES: {H~\textsc{i}} and OH}
\label{s:obs_HI_OH}

In this study, we make use of the \HI (1.4204~GHz) and OH (1.6654~and 1.6673~GHz) emission$/$absorption spectra from \citet{stanimirovic2014} and \citet{nguyen2019}. These spectra were obtained with the 305~m Arecibo telescope (providing angular and velocity resolutions of 3.5$'$ and 0.16~km~s$^{-1}$) toward 100 extragalactic continuum sources that were selected from the NRAO VLA Sky Survey (NVSS; \citealt{condon1998}) with 1.4~GHz flux densities $S_{1.4}$ $\gtrsim$ 0.6~Jy. Among the observed sources, 58 at $b$~$<$~$-$5{$\degree$} probing the surroundings of the Perseus, Taurus, and California molecular clouds were considered for our study (Figure \ref{planckmap} and Table \ref{t:58LOSs}). 

The methodology of the observations and data reduction in \citet{stanimirovic2014} and \citet{nguyen2019} is essentially based on \citet{heiles2003a}, and we provide here a summary for the \HI data. For each source, 1 on-source and 16 off-source measurements were made to obtain optical depth ($\tau_{\rm CNM}$) and ``expected'' emission ($T_{\rm exp}$) spectra. The expected emission spectrum is the one that would be observed at the source position if the source were turned off, and was derived by approximating the off-source spectra as a second-order Taylor expansion of the expected emission spectrum. This approximation was to consider spatial variations in \HI emission, and the derivatives were used to estimate the uncertainty spectrum of expected emission. The median 1$\sigma$ uncertainties in the measured optical depth ($\sigma_{e^{-\tau}}$) and expected emission ($\sigma_{T_{\rm exp}}$) at a velocity resolution of 0.16~km~s$^{-1}$ are 0.02 and 0.36~K, respectively. 

The obtained \HI absorption and emission spectra were analyzed through the Gaussian decomposition method of \citet{heiles2003a}. This method simultaneously fits the absorption and emission spectra with individual Gaussian components under the assumption that the CNM is detected in both absorption and emission, while the WNM contributes to the emission spectrum only. In the fitting process, all possible permutations of the CNM components are considered to find the best-fit model with a minimum chi-square value. In addition, the fitting takes into account the possibility that a certain fraction of the WNM ($F$) could be located in front of the CNM by assuming three cases $F$ = 0, 0.5, and 1 (e.g., $F = 1$ means that the WNM is not absorbed by the CNM at all). The final parameters from the fitting process include the velocities, widths, spin temperatures, peak optical depths, and \HI column densities of individual Gaussian components\footnote{For WNM components, lower and upper limits are provided on the spin temperature and peak optical depth, respectively.}, and we refer to Section 3 of \citet{stanimirovic2014} for details on the fitting procedure. For our analyses, we mostly used the derived \HI properties and utilized the OH spectra only to separate LOSs with molecular gas (Section \ref{s:total_col}).  

Finally, we note that 3C092, 3C131, and 4C$+$27.14 were observed both in \citet{stanimirovic2014} and \citet{nguyen2019}. These observations are essentially consistent within uncertainties, and we used the spectra from \citet{nguyen2019} for our analyses since they have better sensitivities. 

\startlongtable
\begin{deluxetable*}{l c c c c c c c}
    \centering
    \tablecaption{\label{t:58LOSs}58 LOSs in our study}
    \tablewidth{0pt}
    \setlength{\tabcolsep}{12pt}
    \tabletypesize{\small}
    \tablehead{
    \colhead{Source} & \colhead{R.A. (J2000)} & \colhead{Decl. (J2000)} & \colhead{$l$} &
    \colhead{$b$} & \colhead{$S_{\rm 1.4}$} & \colhead{$T_{\rm sky}$} & \colhead{CO(1--0)} \\
    \colhead{} & \colhead{(hh:mm:ss)} & \colhead{(dd:mm:ss)} & \colhead{($\degree$)} & \colhead{($\degree$)} & \colhead{(Jy)} & \colhead{(K)} & \colhead{} \\ 
    \colhead{(1)} & \colhead{(2)} & \colhead{(3)} & \colhead{(4)} & \colhead{(5)} & \colhead{(6)} & \colhead{(7)} & \colhead{(8)} 
    }
    \startdata
    J034053+073525 (4C+07.13)      & 03:40:53.73 & 07:35:25.40 & 178.87 & $-$36.27 & 1.0 & 4.07 &\\
    J032153+122114 (PKS0319+12)    & 03:21:53.11 & 12:21:14.00 & 170.59 & $-$36.24 & 1.9 & 4.51 &\\
    J032723+120835 (4C+11.15)      & 03:27:23.11 & 12:08:35.80 & 171.98 & $-$35.48 & 1.2 & 4.17 & \checkmark\\
    J031857+162833 (4C+16.09)      & 03:18:57.77 & 16:28:33.10 & 166.64 & $-$33.60 & 8.0 & 6.93 &\\
    J033626+130233 (3C090)         & 03:36:26.56 & 13:02:33.20 & 173.15 & $-$33.29 & 2.0 & 4.67 &\\
    J015712+285138 (NV0157+28)     & 01:57:12.85 & 28:51:38.49 & 139.90 & $-$31.83 & 1.4  & 2.78 &\\ 
    J021701+280458 (4C+27.07)	   & 02:17:01.89 & 28:04:59.12 & 145.01 & $-$31.09 & 1.0  & 2.79 &\\
    J020136+293340 (4C+29.05)	   & 02:01:35.91 & 29:33:44.18 & 140.72 & $-$30.88 & 1.2  & 2.79 &\\
    J022412+275011 (3C067)	       & 02:24:12.31 & 27:50:11.69 & 146.82 & $-$30.70 & 3.0  & 2.79 &\\
    J035613+130535                 & 03:56:13.81 & 13:05:35.80 & 177.02 & $-$29.78 & 0.9 & 4.14 &\\
    J023752+284809 (4C+28.07)	   & 02:37:52.42 & 28:48:09.16 & 149.47 & $-$28.53 & 2.2  & 2.79 &\\
    J023535+290857 (4C+28.06)	   & 02:35:35.41 & 29:08:57.73 & 148.78 & $-$28.44 & 1.3	& 2.79 &\\
    J035900+143622 (3C096)         & 03:59:00.91 & 14:36:22.50 & 176.27 & $-$28.26 & 1.2  & 4.37 &\\
    J022048+324106 (5C06.237)	   & 02:20:48.06 & 32:41:06.64 & 143.88 & $-$26.53 & 0.9  & 2.79 &\\
    J042725+085330 (4C+08.15)      & 04:27:25.05 & 08:53:30.30 & 186.21 & $-$26.51 & 0.9 & 4.08 &\\
    J023423+313418 (3C068.2)  	   & 02:34:23.87 & 31:34:17.62 & 147.33 & $-$26.38 & 1.0  & 2.79 &\\
    J032504+244445 (4C+24.06)      & 03:25:04.35 & 24:44:45.60 & 161.92 & $-$26.26 & 0.8 & 4.13 & \checkmark\\
    J035633+190034 (4C+18.11)      & 03:56:33.46 & 19:00:34.60 & 172.23 & $-$25.66 & 1.1 & 4.15 &\\
    J022610+342130 (4C+34.07)	   & 02:26:10.34 & 34:21:30.45 & 144.31 & $-$24.55 & 2.9	& 2.79 &\\
    J041140+171405 (4C+17.23)      & 04:11:40.77 & 17:14:05.10 & 176.36 & $-$24.24 & 1.0 & 4.26 & \checkmark\\
    J023228+342405 (NV0232+34)    & 02:32:28.72 & 34:24:06.08 & 145.60 & $-$23.98 & 2.6  & 2.79 &\\
    J022105+355613 (B20218+35)	   & 02:21:05.48 & 35:56:13.91 & 142.60 & $-$23.49 & 1.7	& 2.79 &\\
    J031135+304320 (4C+30.04)	   & 03:11:35.19 & 30:43:20.62 & 155.40 & $-$23.17 & 1.0  & 2.79 & \checkmark\\
    J032957+275615 (B20326+27)	   & 03:29:57.69 & 27:56:15.64 & 160.70 & $-$23.07 & 1.3  & 2.79 &\\
    J042022+175355 (3C114)         & 04:20:22.17 & 17:53:55.20 & 177.30 & $-$22.24 & 1.1 & 4.23 & \checkmark\\
    J042524+175525 (4C+17.25)      & 04:25:24.43 & 17:55:25.30 & 178.11 & $-$21.31 & 0.9 & 4.16 &\\
    J035204+262418 (4C+26.12)	   & 03:52:04.36 & 26:24:18.11 & 165.82 & $-$21.06 & 1.4  & 2.78 &\\
    J042756+175242 (4C+17.26)      & 04:27:56.98 & 17:52:42.80 & 178.56 & $-$20.88 & 1.0 & 4.22 &\\
    J044907+112128 (PKS0446+11)    & 04:49:07.65 & 11:21:28.20 & 187.43 & $-$20.74 & 0.9 & 4.16 &\\
    J030142+351219 (4C+34.09)      & 03:01:42.38 & 35:12:20.84 & 150.94 & $-$20.49 & 1.9  & 2.79 &\\
    J041243+230506 (3C108)	       & 04:12:43.69 & 23:05:05.53 & 171.87 & $-$20.12 & 1.5  & 2.79 & \checkmark\\
    J040305+260001 (B20400+25)	   & 04:03:05.61 & 26:00:01.61 & 168.03 & $-$19.65 & 0.9  & 2.79 & \checkmark\\
    J034008+320901 (3C092)         & 03:40:08.54 & 32:09:01.30 & 159.74 & $-$18.41 & 1.6 & 3.95 & \checkmark\\
    J042846+213331 (4C+21.17)      & 04:28:46.64 & 21:33:31.40 & 175.70 & $-$18.36 & 1.3 & 4.35 & \checkmark\\
    J040442+290215 (4C+28.11)      & 04:04:42.82 & 29:02:15.90 & 166.06 & $-$17.22 & 1.0 & 3.69 & \checkmark\\
    J042049+252627 (4C+25.14)	   & 04:20:49.30 & 25:26:27.63 & 171.37 & $-$17.16 & 1.0	& 2.79 & \checkmark\\
    J034846+335315 (3C093.1)	   & 03:48:46.93 & 33:53:15.41 & 160.04 & $-$15.91 & 2.4	& 2.80 &\\
    J052424+074957 (4C+07.16)      & 05:24:24.04 & 07:49:57.10 & 195.51 & $-$15.35 & 0.8 & 4.25 &\\
    J051240+151723 (PKS0509+152)   & 05:12:40.99 & 15:17:23.80 & 187.41 & $-$13.79 & 1.0 & 4.11 &\\
    J053239+073243                 & 05:32:39.01 & 07:32:43.50 & 196.84 & $-$13.74 & 2.7 & 4.96 &\\
    J051930+142829 (4C+14.14)      & 05:19:30.95 & 14:28:29.00 & 189.04 & $-$12.85 & 0.9 & 4.15 & \checkmark\\
    J045643+224922 (3C132)	       & 04:56:43.08 & 22:49:22.27 & 178.86 & $-$12.52 & 3.4	& 2.80 &\\
    J053450+100430 (4C+09.21)      & 05:34:50.82 & 10:04:30.30 & 194.89 & $-$11.98 & 1.1 & 4.62 & \checkmark\\
    J041437+341851 (B20411+34)	   & 04:14:37.28 & 34:18:51.31 & 163.80 & $-$11.98 & 1.9  & 2.79 &\\
    J041236+353543 (4C+35.07)      & 04:12:36.28 & 35:35:43.20 & 162.58 & $-$11.36 & 0.9 & 3.93 &\\
    J052109+163822 (3C138)         & 05:21:09.93 & 16:38:22.20 & 187.41 & $-$11.34 & 8.6  & 7.59 &\\
    J053056+133155 (PKS0528+134)   & 05:30:56.44 & 13:31:55.30 & 191.37 & $-$11.01 & 1.6 & 4.64 & \checkmark\\
    J042353+345144 (3C115)         & 04:23:53.25 & 34:51:44.80 & 164.76 & $-$10.24 & 1.3  & 3.88 &\\
    J050258+251624 (3C133)	       & 05:02:58.51 & 25:16:25.16 & 177.73 & $-$9.91  & 5.8  & 2.80 & \checkmark\\
    J060536+014512 (4C+01.17)      & 06:05:36.56 & 01:45:12.70 & 206.08 & $-$9.37  & 0.6 & 4.07 &\\
    J045956+270602 (4C+27.14)       & 04:59:56.09 & 27:06:02.90 & 175.83 & $-$9.36  & 0.9 & 3.90 & \checkmark\\
    J051740+235110 (4C+23.14)      & 05:17:40.81 & 23:51:10.20 & 180.86 & $-$8.01  & 1.0 & 4.32 & \checkmark\\
    J045323+312924 (3C131)         & 04:53:23.34 & 31:29:24.20   & 171.44 & $-$7.80  & 2.9 & 4.04 & \checkmark\\
    J053557+175600 (4C+17.33)      & 05:35:57.42 & 17:56:00.70 & 188.22 & $-$7.67  & 0.8 & 4.23 &\\
    J044708+332747 (4C+33.10)	   & 04:47:08.90 & 33:27:46.85 & 169.05 & $-$7.57	 & 1.2  & 2.80 & \checkmark\\
    J053444+192721 (PKS0531+19)    & 05:34:44.51 & 19:27:21.70 & 186.76 & $-$7.11  & 7.0 & 6.48 &\\
    J054046+172839 (4C+17.34)      & 05:40:46.05 & 17:28:39.20 & 189.21 & $-$6.93  & 1.5 & 4.50 &\\
    J050929+295755 (4C+29.16)      & 05:09:29.51 & 29:57:55.80 & 174.77 & $-$5.97  & 1.1 & 4.03 &\\
    \enddata
    \begin{tablenotes}
        \item \tablecomments{\small (1) Source name; (2, 3) Right ascension (R.A.) and declination (Decl.) coordinates; (4, 5) Galactic coordinates; (6) Flux density at 1.4~GHz; (7) Diffuse background radio continuum emission; (8) Detection of CO(1--0) emission. Here the columns (1)--(7) are from \citet{stanimirovic2014} and \citet{nguyen2019}.}
    \end{tablenotes}
\end{deluxetable*}

\subsection{CO} 
\label{s:CO} 

Single-pointing observations of the CO(1--0) transition at 115.2712~GHz were carried out toward the 58 GNOMES LOSs at $b$~$<$~$-5{\degree}$ using the 13.7~m telescopes at the Taeduk Radio Astronomy Observatory (TRAO) and Purple Mountain Observatory (PMO). The TRAO observations were made in March and December 2020 and in February 2021, while the PMO observations were performed from March to May 2020. During these observations, the system temperature was 400--900~K and 200--300~K for the TRAO and PMO telescopes, respectively. 

The obtained CO(1--0) spectra were processed using the GILDAS CLASS software\footnote{\url{https://www.iram.fr/IRAMFR/GILDAS/}}. For the TRAO data, a beam efficiency of $\eta_{\rm MB}$ = 0.40 was adopted to convert the corrected antenna temperature into the main-beam brightness temperature ($T_{\rm MB}$ = $T_{\rm A}^{\ast}$ / $\eta_{\rm MB}$). On the other hand, no conversion was made for the PMO data since they were delivered in units of main-beam brightness temperature. The final spectra on 48$''$ scales were smoothed to a velocity resolution of 0.32~km~s$^{-1}$ and have a median root-mean-square (rms) noise level of 0.1~K. A comparison between the CO(1--0) emission and \HI absorption spectra is presented in Appendix \ref{s:appendix1}. 

To determine the presence of CO emission, we adopted a 3$\sigma$ threshold and considered components whose peak-to-rms ratios are equal to or higher than three as detections. Once the presence of CO emission was confirmed, we fitted Gaussians to the spectra to derive line parameters such as the central velocity ($\varv_{\rm CO}$), full width at half maximum (FWHM; $\Delta \varv_{\rm CO}$), and peak main-beam brightness temperature ($T_{\rm peak,CO})$. The derived line parameters, as well as the CO integrated intensity ($I({\rm CO})$; calculated by integrating CO(1--0) emission over a velocity range where the emission is clearly visible) and rms noise, are presented in Table \ref{t:CO_properties}.

Finally, we note that two of our target sources (3C092 and 3C108) were observed using both telescopes to check the calibration levels of the TRAO and PMO observations. The difference between the TRAO and PMO observations was 10--20\%, which is within the calibration uncertainty of $\sim$20$\%$ for the TRAO telescope at 115~GHz. This suggests that the obtained CO(1--0) spectra are well calibrated and can be used for further analyses. 

\startlongtable
\begin{deluxetable*}{l c c c c c c}
    \centering
    \tablecaption{\label{t:CO_properties} Derived CO(1--0) properties}
    \tablewidth{0pt}
    \setlength{\tabcolsep}{15pt}
    \tabletypesize{\small}
    \tablehead{
     \colhead{Source} & \colhead{$\varv_{\rm CO}$} & \colhead{$\Delta \varv_{\rm CO}$} & \colhead{$T_{\rm peak,CO}$} & \colhead{$I({\rm CO})$} & \colhead{$\sigma_{\rm rms}$} & \colhead{Telescope} \\
    \colhead{ } & \colhead{(km~s$^{-1}$)} & \colhead{(km~s$^{-1}$)} & \colhead{(K)} & \colhead{(K~km~s$^{-1}$)} & \colhead{(K)} & \colhead{} \\ 
    \colhead{(1)} & \colhead{(2)} & \colhead{(3)} & \colhead{(4)} & \colhead{(5)} & \colhead{(6)} & \colhead{(7)} 
    }
    \startdata
    4C+07.13 & $\cdots$ & $\cdots$ & $\cdots$ & $\cdots$ & 0.10 & PMO\\
    PKS0319+12 & $\cdots$ & $\cdots$ & $\cdots$ & $\cdots$ & 0.11 & PMO\\
    4C+11.15 & 6.96 $\pm$ 0.02 & 0.41 $\pm$ 0.06 & 1.14 $\pm$ 0.13 & 0.49 $\pm$ 0.04 & 0.07 & PMO\\
    4C+16.09 & $\cdots$ & $\cdots$ & $\cdots$ & $\cdots$ & 0.17 & TRAO\\
    3C090 & $\cdots$ & $\cdots$ & $\cdots$ & $\cdots$ & 0.10 & PMO\\
    NV0157+28 & $\cdots$ & $\cdots$ & $\cdots$ & $\cdots$ & 0.10 & PMO\\
    4C+27.07 & $\cdots$ & $\cdots$ & $\cdots$ & $\cdots$ & 0.10 & PMO\\
    4C+29.05 & $\cdots$ & $\cdots$ & $\cdots$ & $\cdots$ & 0.10 & PMO\\
    3C067 & $\cdots$ & $\cdots$ & $\cdots$ & $\cdots$ & 0.10 & PMO\\
    J035613+130535 & $\cdots$ & $\cdots$ & $\cdots$ & $\cdots$ & 0.09 & PMO\\
    4C+28.07 & $\cdots$ & $\cdots$ & $\cdots$ & $\cdots$ & 0.08 & PMO\\
    4C+28.06 & $\cdots$ & $\cdots$ & $\cdots$ & $\cdots$ & 0.10 & PMO\\
    3C096 & $\cdots$ & $\cdots$ & $\cdots$ & $\cdots$ & 0.08 & PMO\\
    5C06.237 & $\cdots$ & $\cdots$ & $\cdots$ & $\cdots$ & 0.10 & PMO\\
    4C+08.15 & $\cdots$ & $\cdots$ & $\cdots$ & $\cdots$ & 0.20 & TRAO\\
    3C068.2 & $\cdots$ & $\cdots$ & $\cdots$ & $\cdots$ & 0.07 & PMO\\
    4C+24.06 & 6.95 $\pm$ 0.09 & 1.02 $\pm$ 0.21 & 0.44 $\pm$ 0.08 & 0.50 $\pm$ 0.09 & 0.10 & PMO\\
    4C+18.11 & $\cdots$ & $\cdots$ & $\cdots$ & $\cdots$ & 0.10 & PMO\\
    4C+34.07 & $\cdots$ & $\cdots$ & $\cdots$ & $\cdots$ & 0.08 & PMO\\
    4C+17.23$^{\rm a}$ & 9.10 $\pm$ 0.01 & 0.64 $\pm$ 0.02 & 4.96 $\pm$ 0.11 & 3.43 $\pm$ 0.07 & 0.11 & PMO\\
    4C+17.23$^{\rm a}$ & 11.17 $\pm$ 0.02 & 0.92 $\pm$ 0.05 & 2.07 $\pm$ 0.09 & 2.04 $\pm$ 0.07 & 0.11 & PMO\\
    NV0232+34 & $\cdots$ & $\cdots$ & $\cdots$ & $\cdots$ & 0.08 & PMO\\
    B20218+35 & $\cdots$ & $\cdots$ & $\cdots$ & $\cdots$ & 0.07 & PMO\\
    4C+30.04$^{\rm a}$ & 0.69 $\pm$ 0.01 & 0.62 $\pm$ 0.03 & 2.64 $\pm$ 0.12 & 1.77 $\pm$ 0.06 & 0.10 & PMO\\
    4C+30.04$^{\rm a}$ & 0.89 $\pm$ 0.19 & 2.79 $\pm$ 0.53 & 0.37 $\pm$ 0.09 & 1.10 $\pm$ 0.13 & 0.10 & PMO\\
    B20326+27 & $\cdots$ & $\cdots$ & $\cdots$ & $\cdots$ & 0.10 & PMO\\
    3C114$^{\rm a}$ & 8.55 $\pm$ 0.11 & 0.70 $\pm$ 0.27 & 0.33 $\pm$ 0.10 & 0.25 $\pm$ 0.06 & 0.10 & PMO\\
    3C114$^{\rm a}$ & 9.48 $\pm$ 0.05 & 0.46 $\pm$ 0.15 & 0.66 $\pm$ 0.16 & 0.33 $\pm$ 0.05 & 0.10 & PMO\\
    4C+17.25 & $\cdots$ & $\cdots$ & $\cdots$ & $\cdots$ & 0.10 & PMO\\
    4C+26.12 & $\cdots$ & $\cdots$ & $\cdots$ & $\cdots$ & 0.11 & TRAO\\
    4C+17.26 & $\cdots$ & $\cdots$ & $\cdots$ & $\cdots$ & 0.11 & PMO\\
    PKS0446+11 & $\cdots$ & $\cdots$ & $\cdots$ & $\cdots$ & 0.12 & PMO\\
    4C+34.09 & $\cdots$ & $\cdots$ & $\cdots$ & $\cdots$ & 0.10 & PMO\\
    3C108$^{\rm b}$ & 6.14 $\pm$ 0.11 & 0.40 $\pm$ 0.21 & 0.69 $\pm$ 0.29 & 0.30 $\pm$ 0.07 & 0.13 & TRAO\\
    3C108$^{\rm b}$ & 9.42 $\pm$ 0.01 & 1.13 $\pm$ 0.02 & 11.10 $\pm$ 0.15 & 13.49 $\pm$ 0.07 & 0.13 & TRAO\\
    B20400+25 & 7.07 $\pm$ 0.06 & 1.01 $\pm$ 0.13 & 0.59 $\pm$ 0.07 & 0.69 $\pm$ 0.08 & 0.09 & PMO\\
    3C092 & 8.80 $\pm$ 0.01 & 1.66 $\pm$ 0.02 & 9.68 $\pm$ 0.10 & 17.22 $\pm$ 0.12 & 0.10 & PMO\\
    4C+21.17 & 10.17 $\pm$ 0.06 & 0.65 $\pm$ 0.13 & 0.66 $\pm$ 0.12 & 0.58 $\pm$ 0.09 & 0.12 & PMO\\
    4C+28.11 & 6.63 $\pm$ 0.01 & 1.10 $\pm$ 0.02 & 9.83 $\pm$ 0.12 & 11.78 $\pm$ 0.12 & 0.10 & PMO\\
    4C+25.14$^{\rm a}$ & 3.59 $\pm$ 0.09 & 1.33 $\pm$ 0.21 & 0.47 $\pm$ 0.06 & 0.67 $\pm$ 0.09 & 0.09 & PMO\\
    4C+25.14$^{\rm a}$ & 6.78 $\pm$ 0.01 & 0.74 $\pm$ 0.02 & 4.78 $\pm$ 0.10 & 3.80 $\pm$ 0.09 & 0.09 & PMO\\
    4C+25.14$^{\rm a}$ & 7.92 $\pm$ 0.02 & 1.14 $\pm$ 0.06 & 2.64 $\pm$ 0.07 & 3.24 $\pm$ 0.08 & 0.09 & PMO\\
    3C093.1 & $\cdots$ & $\cdots$ & $\cdots$ & $\cdots$ & 0.09 & PMO\\
    4C+07.16 & $\cdots$ & $\cdots$ & $\cdots$ & $\cdots$ & 0.09 & PMO\\
    PKS0509+152 & $\cdots$ & $\cdots$ & $\cdots$ & $\cdots$ & 0.12 & PMO\\
    J053239+073243 & $\cdots$ & $\cdots$ & $\cdots$ & $\cdots$ & 0.09 & PMO\\
    4C+14.14 & 2.17 $\pm$ 0.13 & 0.34 $\pm$ 0.42 & 0.60 $\pm$ 0.57 & 0.28 $\pm$ 0.08 & 0.15 & PMO\\
    3C132 & $\cdots$ & $\cdots$ & $\cdots$ & $\cdots$ & 0.10 & PMO\\
    4C+09.21 & 1.99 $\pm$ 0.26 & 2.59 $\pm$ 0.61 & 0.23 $\pm$ 0.05 & 0.64 $\pm$ 0.11 & 0.10 & PMO\\
    B20411+34 & $\cdots$ & $\cdots$ & $\cdots$ & $\cdots$ & 0.09 & PMO\\
    4C+35.07 & $\cdots$ & $\cdots$ & $\cdots$ & $\cdots$ & 0.11 & PMO\\
    3C138 & $\cdots$ & $\cdots$ & $\cdots$ & $\cdots$ & 0.15 & PMO\\
    PKS0528+134 & 9.63 $\pm$ 0.02 & 0.89 $\pm$ 0.06 & 2.64 $\pm$ 0.15 & 2.77 $\pm$ 0.15 & 0.17 & PMO\\
    3C115 & $\cdots$ & $\cdots$ & $\cdots$ & $\cdots$ & 0.11 & TRAO\\
    3C133 & 7.45 $\pm$ 0.02 & 0.84 $\pm$ 0.04 & 3.57 $\pm$ 0.15 & 3.18 $\pm$ 0.14 & 0.18 & PMO\\
    4C+01.17 & $\cdots$ & $\cdots$ & $\cdots$ & $\cdots$ & 0.11 & TRAO\\
    4C+27.14$^{\rm a}$ & 6.03 $\pm$ 0.04 & 1.00 $\pm$ 0.10 & 1.11 $\pm$ 0.08 & 1.19 $\pm$ 0.07 & 0.09 & PMO\\
    4C+27.14$^{\rm a}$ & 7.78 $\pm$ 0.01 & 1.27 $\pm$ 0.02 & 7.16 $\pm$ 0.07 & 9.77 $\pm$ 0.08 & 0.09 & PMO\\
    4C+23.14$^{\rm b}$ & -3.70 $\pm$ 0.13 & 1.32 $\pm$ 0.30 & 0.36 $\pm$ 0.07 & 0.51 $\pm$ 0.09 & 0.10 & TRAO\\
    4C+23.14$^{\rm b}$ & 1.15 $\pm$ 0.13 & 0.95 $\pm$ 0.31 & 0.41 $\pm$ 0.08 & 0.42 $\pm$ 0.09 & 0.10 & TRAO\\
    4C+23.14$^{\rm b}$ & 2.37 $\pm$ 0.13 & 0.84 $\pm$ 0.30 & 0.39 $\pm$ 0.09 & 0.35 $\pm$ 0.07 & 0.10 & TRAO\\
    3C131$^{\rm a}$ & 4.79 $\pm$ 0.12 & 1.60 $\pm$ 0.31 & 0.51 $\pm$ 0.06 & 0.87 $\pm$ 0.10 & 0.10 & PMO\\
    3C131$^{\rm a}$ & 6.86 $\pm$ 0.02 & 1.31 $\pm$ 0.04 & 3.67 $\pm$ 0.07 & 5.16 $\pm$ 0.09 & 0.10 & PMO\\
    4C+17.33 & $\cdots$ & $\cdots$ & $\cdots$ & $\cdots$ & 0.09 & TRAO\\
    4C+33.10$^{\rm b}$ & -2.29 $\pm$ 0.01 & 1.22 $\pm$ 0.02 & 5.53 $\pm$ 0.09 & 7.24 $\pm$ 0.09 & 0.10 & PMO\\
    4C+33.10$^{\rm b}$ & 5.91 $\pm$ 0.01 & 0.42 $\pm$ 0.06 & 3.09 $\pm$ 0.35 & 1.38 $\pm$ 0.09 & 0.10 & PMO\\
    4C+33.10$^{\rm b}$ & 6.61 $\pm$ 0.01 & 0.68 $\pm$ 0.02 & 6.24 $\pm$ 0.13 & 4.55 $\pm$ 0.07 & 0.10 & PMO\\
    PKS0531+19 & $\cdots$ & $\cdots$ & $\cdots$ & $\cdots$ & 0.09 & TRAO\\
    4C+17.34 & $\cdots$ & $\cdots$ & $\cdots$ & $\cdots$ & 0.11 & TRAO\\
    4C+29.16 & $\cdots$ & $\cdots$ & $\cdots$ & $\cdots$ & 0.18 & TRAO\\
    \enddata
    \begin{tablenotes}
        \item \tablecomments{(1) Source name; (2) Central velocity; (3) FWHM; (4) Peak brightness temperature; (5) Integrated intensity; (6) rms noise level; (7) Telescope that was used to obtained the spectrum. $^{\rm a}$ Multiple peaks are close enough in velocity to be considered as one component based on our threshold (Section \ref{s:obs_CO}). $^{\rm b}$ These sources are considered to have two distinct peaks.}
    \end{tablenotes}
\end{deluxetable*}

\subsection{Planck Data}
\label{s:planck}

To estimate the environmental conditions of the GNOMES LOSs such as the strength of UV radiation ($I_{\rm UV}$), $V$-band dust extinction ($A_{V}$), and dust-to-gas ratio (DGR), we used \textit{Planck} data. Specifically, we employed the images of dust temperature ($T_{\rm dust}$), spectral index ($\beta$), and dust opacity at 353~GHz ($\tau_{353}$) from  \citet{planck_xlviii} and extracted the values toward the 58 LOSs by using the Python package ``dustmaps'' of \citet{green2018}. These extracted values are on 5$'$ scales. 

\section{Environmental Conditions}
\label{s:env} 

\subsection{Dust and Gas Properties}
\label{s:total_col}

Before comparing the observed \HI and CO properties, we probed the environmental conditions of the GNOMES LOSs based on the \textit{Planck} data. As the first step, we calculated dust abundances by converting the 353~GHz dust opacity into the $V$-band dust extinction: 

\begin{equation}
\begin{split}
\label{e:Av} 
A_{V}~(\textrm{mag}) & = R_{V}~E(B-V) \\ 
                    & = 3.1 \times (1.5 \times 10^{4}~\tau_{353}).
\end{split} 
\end{equation} 

\noindent For this calculation, the total-to-selective extinction ratio $R_{V} = 3.1$ for the diffuse ISM is assumed \citep{mathis1990}. 
In addition, the conversion factor of 1.5~$\times$~10$^{4}$~mag is adopted to translate the 353~GHz dust opacity $\tau_{353}$ into the reddening $E(B-V)$ based on \citet{planck_xi}. The derived $A_{V}$ ranges from 0.2~mag to 4~mag with a median of 1~mag, suggesting that our LOSs probe diffuse to dense interstellar gas.  

Next we estimated the DGR by deriving $A_{V}/N$(H~\textsc{i}) for diffuse LOSs where gas is dominated by atomic gas. To identify such LOSs, the following criteria were applied: (1) no CO or OH detection; (2) $A_{V}$ $<$ 0.5~mag. The second threshold is motivated by observational and theoretical studies that found H$_{2}$ formation at $A_{V}$ $\sim$ 0.5~mag in the solar neighborhood \citep[e.g.,][]{lee2012,sternberg2014}. With the two criteria, we found 13 atomic-dominated LOSs and calculated $N({\rm H~\textsc{i}})$ by considering both CNM and WNM column densities: 

\begin{gather}
\begin{split}
\label{e:N_HI_true}
N(\rm \HI)~(\rm cm^{-2}) & = N_{\rm CNM} + N_{\rm WNM} \\
            & = 1.823 \times 10^{18} \int\Biggl( \sum_{0}^{N-1}T_{s,n}\tau_{0,n}e^{-\left[(\varv-\varv_{0,n})/\delta\varv_{n}\right]^{2}} \\
            &~~~+ \sum_{0}^{K-1}T_{0,k}e^{-\left[(\varv-\varv_{0,k})/\delta\varv_{k}\right]^{2}} \Biggr)~ d\varv~ \left({\rm K~km~s^{-1}}\right).
\end{split}
\end{gather}

\noindent Here the subscripts $n$ and $k$ refer to CNM and WNM components, $\tau_{0}$ is the peak optical depth, $\varv_{0}$ is the central velocity, $T_{0}$ is the peak brightness temperature, and $\delta\varv$ is the $1/e$ width of the component. 
The derived $A_{V}/N$(H~\textsc{i}) in units of mag~cm$^{2}$ has a range of (0.3--0.5)~$\times$~10$^{-21}$ with a median of 0.4~$\times$~10$^{-21}$ and is in good agreement with typical Galactic DGR values \citep[e.g.,][]{bohlin1978,liszt2014,lenz2017,nguyen2018}. 

Finally, we estimated total gas column densities toward the GNOMES LOSs by dividing the \textit{Planck}-based $A_{V}$ by the representative DGR of 0.4~$\times$~10$^{-21}$~mag~cm$^{2}$: 

\begin{equation}
    \label{eq:total_gas}
    \begin{split}
    N({\rm H})_{A_{V}{\rm-}{\rm based}}~({\rm cm^{-2}}) & = N({\rm H~\textsc{i}}) + 2N({\rm H_{2}}) \\ 
               & = \frac{A_{V}}{0.4 \times 10^{-21}~{\rm mag~cm}^{2}} .\\
    \end{split}
\end{equation}

\subsection{UV Radiation}
\label{s:UV}

We estimated the strength of UV radiation in units of the Draine field using the expression derived for dust grains at high Galactic latitudes \citep[e.g.,][]{boulanger1996,paradis11}:

\begin{equation}
\label{eq:guv}
I_{\rm UV} = \left(\frac{T_{\rm dust}}{\rm 17.5~K}\right)^{\beta + 4}. 
\end{equation}

\noindent Except for four LOSs with relatively high dust temperatures (20--24~K), $T_{\rm dust}$ is mostly 18~K, resulting in the typical solar neighborhood condition of $I_{\rm UV}$ $\sim$ 1. 

\subsection{Uncertainties}
\label{s:uncertainties}

Our estimation of the strength of UV radiation and total gas column density likely suffers from several systematic uncertainties. Firstly, the \textit{Planck} dust data ($T_{\rm dust}$, $\beta$, and $\tau_{\rm 353}$) were derived based on the model of modified blackbody emission, and the main assumptions for this derivation, such as a single dust temperature along a LOS, could be invalid under some circumstances. Secondly, the conversion of $\tau_{\rm 353}$ into $N$(H) involves a few steps, which are likely reasonable for diffuse gas, but could be less appropriate for H$_{2}$-dominated LOSs. For example, $R_{V}$ could be higher than 3.1 in dense regions due to grain growth \citep[e.g.,][]{chapman2009,steinacker2010}. Grain growth could also cause an underestimation of the DGR \citep[e.g.,][]{roman-duval2014}. Finally, the different angular resolutions of the \textit{Planck} and \HI measurements could hinder a derivation of accurate gas and dust properties.  

This discussion on possible uncertainties demonstrates that various factors affect our derivation of the dust and gas properties. However, it is not straightforward to evaluate the impact of each factor based on the currently available data, and we hence proceeded bearing in mind the uncertainty sources. 

\section{Results}
\label{s:results}

\subsection{Observed CO Properties} 
\label{s:obs_CO} 

CO(1--0) is detected toward 19 sources (tan crosses in Figure \ref{planckmap}), suggesting a detection rate of 33$\%$ at a rms level of 0.1~K. Among these sources, ten show simple spectra with single Gaussians, while nine have multiple peaks. For these peaks, we examined the difference between their central velocities and considered them as one component if the velocity difference is smaller than the sum of their 2~$\times$~FWHMs. Based on this threshold, only the peaks toward 3C108, 4C$+$23.14, and 4C$+$33.10 are regarded as sufficiently distinct components. 


The derived central velocities are mostly between 7~km~s$^{-1}$ and 11~km~s$^{-1}$ with a few components at lower velocities ($-$4~km~s$^{-1}$ to 4~km~s$^{-1}$), suggesting that the observed CO(1--0) emission is associated with Perseus, Taurus, California, and their surrounding regions \citep[e.g.,][]{ridge2006,narayanan2008,lada2009}. The FWHM line widths are generally small, with a median of 1~km~s$^{-1}$. Finally, the peak main-beam brightness temperature ranges from 0.2~K to 11.1~K, indicating that we are tracing diffuse ($\lesssim$ 1~K) to dense ($\gtrsim$ 5~K) molecular gas.

\subsection{Gas Phases \label{s:gas_composition}}


With the available multi-wavelength data, we can examine gas phases toward the GNOMES LOSs where 

\begin{equation}
\label{e:gas_composition}
\begin{split}
N({\rm H}) & = N({\rm H~\textsc{i}}) + 2N({\rm H_{2}}) \\ 
         & = N({\rm H~\textsc{i}})_{\rm thin} + N({\rm H~\textsc{i}})_{\rm thick} + 2N({\rm H_{2}})_{\rm dark} + 2N({\rm H_{2}})_{\rm bright}. 
\end{split}
\end{equation}

\begin{figure*}
    \centering
    \includegraphics[scale=0.4]{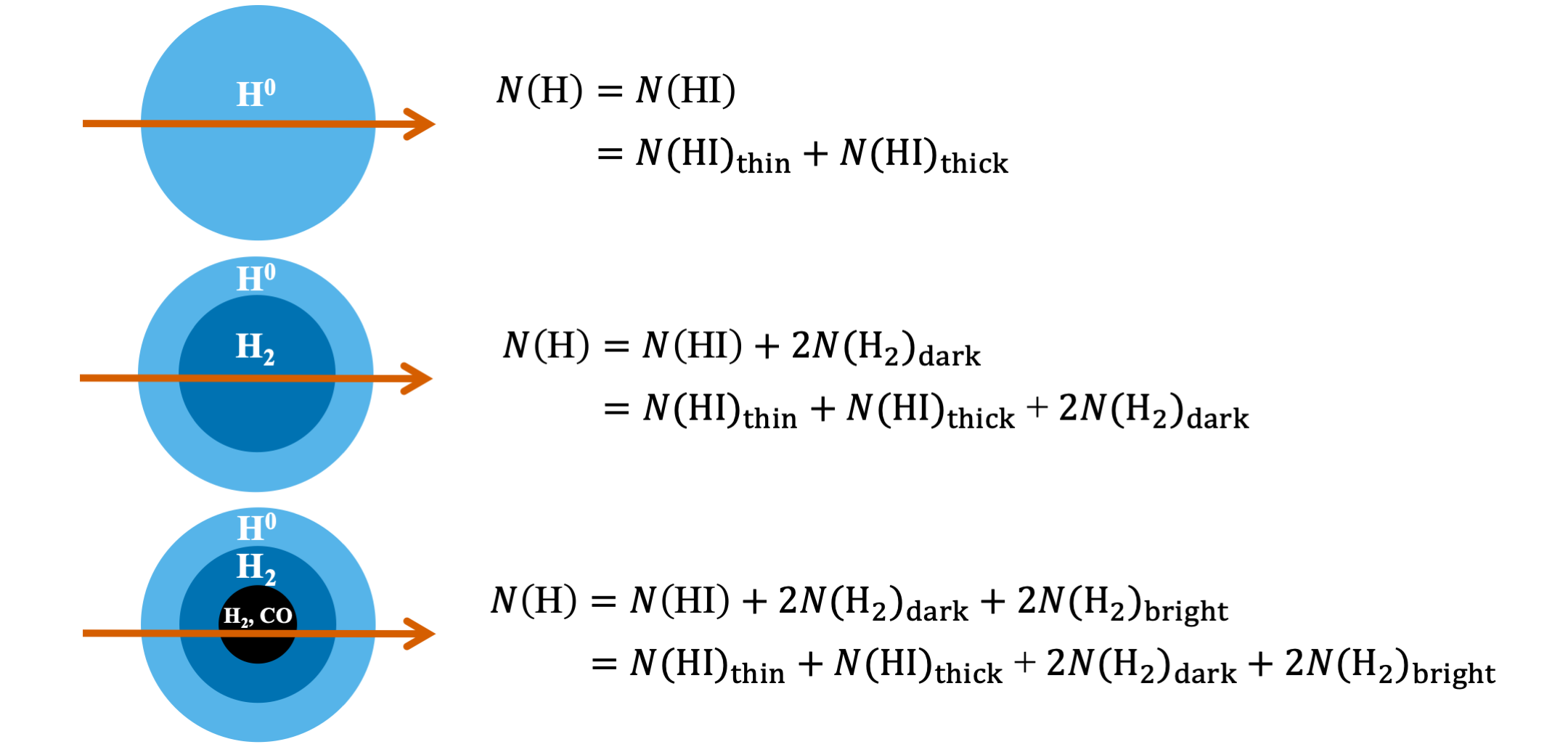}
    \caption{\label{f:gas_phases} Illustration of LOSs with different gas phases. (Top) H~\textsc{i}-only LOSs. (Middle) LOSs with \HI and CO-dark H$_{2}$ gas. (Bottom) LOSs with H~\textsc{i}, CO-dark H$_{2}$, and CO-bright H$_{2}$ gas.}
\end{figure*}

\noindent For our examination, we used $A_{V}$ as a tracer of total gas column density and estimated $N$(H) by dividing $A_{V}$ by the representative DGR of 0.4~$\times$~10$^{-21}$~mag~cm$^{2}$ (Section \ref{s:total_col}). In the case of atomic gas, its ``true'' column density is a sum of the following two: (1) $N$(H~\textsc{i})$_{\rm thin}$ that is calculated by assuming optically thin emission; (2) $N$(H~\textsc{i})$_{\rm thick}$ that is missing in the optically thin approximation due to a high opacity. Similarly, molecular gas consists of two types of H$_{2}$: (1) $N$(H$_{2}$)$_{\rm dark}$ that is invisible in CO(1--0) emission; (2) $N$(H$_{2}$)$_{\rm bright}$ that is traced by CO(1--0) emission. 
H$_{2}$ with faint or no CO(1--0) emission (``CO-dark'' H$_{2}$) is expected due to the different locations of H$_{2}$ and CO formation in interstellar clouds ($A_{V}$ $\sim$ 0.5~mag and 2~mag) and exists along with C$^{+}$ and C$^{0}$ \citep[e.g.,][]{tielens1985,grenier2005,wolfire2010,bolatto2013}. Among the four components of gas in Equation (\ref{e:gas_composition}), the optically thick \HI and CO-dark H$_{2}$ together are called ``dark gas'', since they are not probed by traditional gas tracers such as \HI and CO(1--0) emission. Our examination of the different gas phases is illustrated in Figure \ref{f:gas_phases}. 

As the first step of our examination, we derived $N{\rm(H~\textsc{i})} = N{\rm(H~\textsc{i})_{thin}} + N\rm{(H~\textsc{i})_{thick}}$ based on Equation (\ref{e:N_HI_true}) and separated $N$(H~\textsc{i})$_{\rm thin}$ and $N$(H~\textsc{i})$_{\rm thick}$ for the observed 58 LOSs by: 

\begin{equation}
    \label{eq:N_HI_thin}
    N({\rm \HI})_{\rm thin}~({\rm cm^{-2}}) = 1.823 \times 10^{18} \int T_{\rm exp}~\left({\rm K~km~s^{-1}}\right)
\end{equation}

\noindent and 

\begin{equation}
    \label{eq:N_HI_thick}
    N({\rm \HI})_{\rm thick} = N({\rm \HI}) - N({\rm \HI})_{\rm thin}.
\end{equation}

\begin{figure}
    \centering
    \includegraphics[scale=0.45]{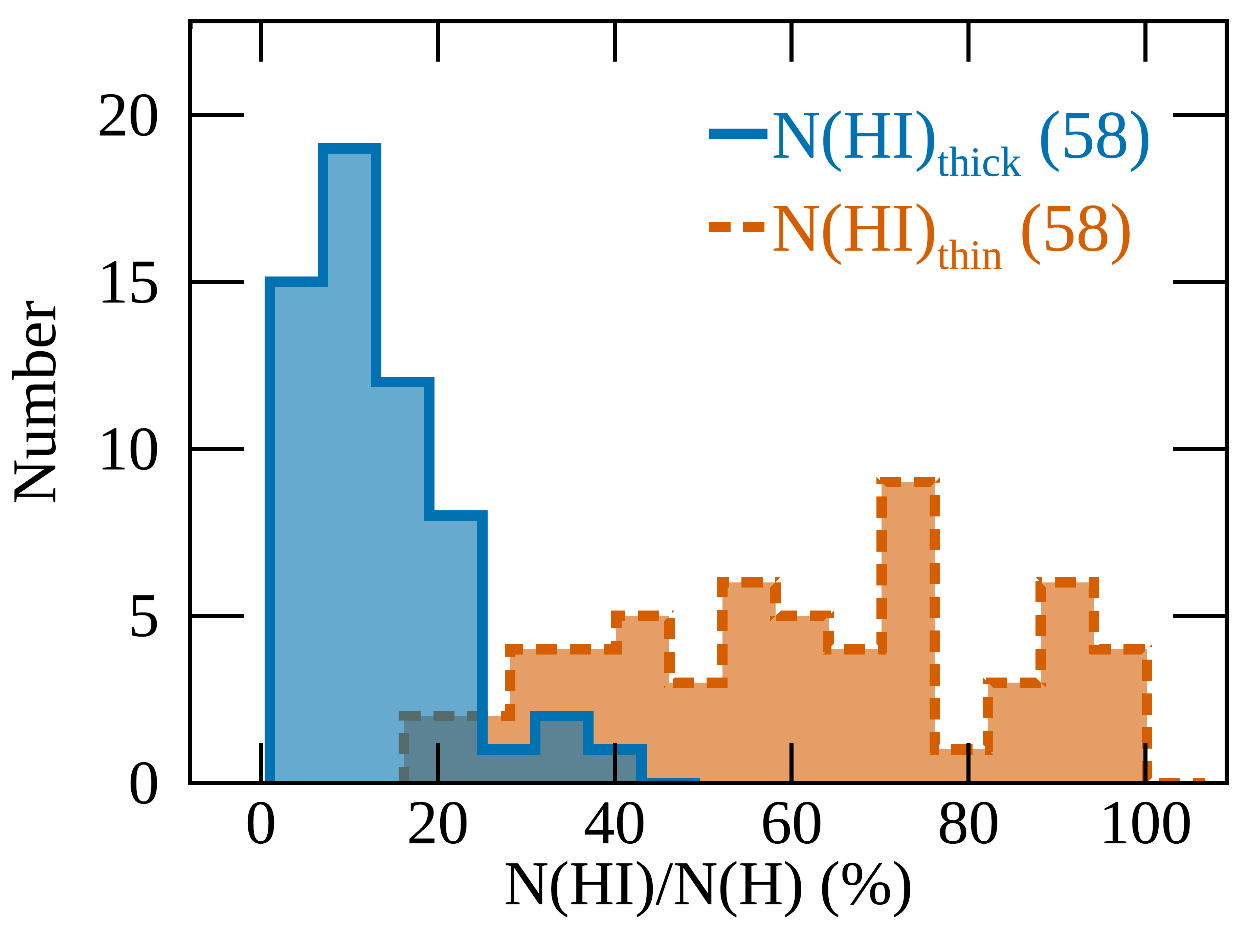}
    \includegraphics[scale=0.45]{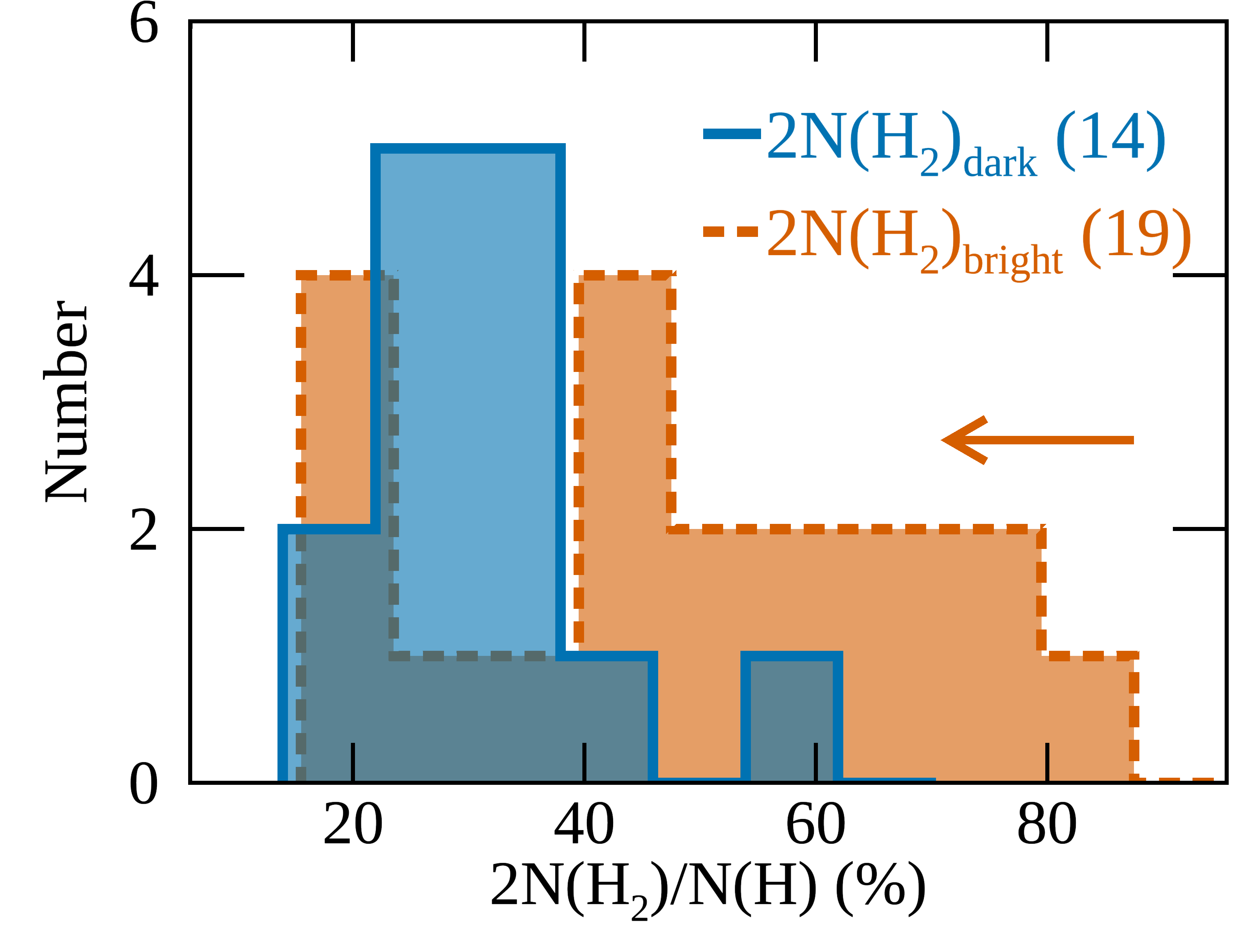}
    \caption{\label{f:gas_fractions} Fraction of each gas phase with respect to the total hydrogen. (Top) Optically thick and thin \HI in solid blue and dashed tan. (Bottom) CO-dark and CO-bright H$_{2}$ in solid blue and dashed tan. The leftward arrow indicates that the derived fractions for CO-bright H$_{2}$ are upper limits.}
    \label{f:gas_phase_histogram}
\end{figure}

\noindent 
The derived optically thin H~\textsc{i} column densities make up 16--99\% (median of 62\%) of the total $N$(H), while the optically thick H~\textsc{i} column densities constitute only 1--38\% (median of 12\%). These results suggest that observed LOSs are mostly H~\textsc{i}-dominated and the contribution from the optically thick \HI to the total $N$(H) is small (Figure \ref{f:gas_fractions} and Table \ref{t:gas_fractions}). 

\begin{deluxetable}{l c c c}[t]
\centering
\tablecaption{\label{t:gas_fractions}Derived properties of the four gas phases}
    \tablewidth{0pt}
    \setlength{\tabcolsep}{8pt}
    \tabletypesize{\small}
    \tablehead{
    \colhead{Gas Phase} & \colhead{Minimum} & \colhead{Maximum} & \colhead{Median} \\
    \colhead{} & \colhead{(${\rm cm^{-2}}$)} & \colhead{(${\rm cm^{-2}}$)} & \colhead{(${\rm cm^{-2}}$)} \\ 
     \colhead{(1)} & \colhead{(2)} & \colhead{(3)} & \colhead{(4)} 
    }
    \startdata
    $N({\rm H~\textsc{i}})_{\rm thin}$ (58) & 4.61 $\times$ 10$^{20}$ & 2.97 $\times$ 10$^{21}$ & 1.49 $\times$ 10$^{21}$\\
    & 16\% & 99\% & 62\%\\
    $N({\rm H~\textsc{i}})_{\rm thick}$ (58) & 5.94 $\times$ 10$^{18}$ & 1.67 $\times$ 10$^{21}$ & 3.52 $\times$ 10$^{20}$\\
    & 1\% & 38\% & 12\%\\
    $2N({\rm H_{2}})_{\rm dark}$ (14) & 5.34 $\times$ 10$^{20}$ & 2.03 $\times$ 10$^{21}$ & 9.60 $\times$ 10$^{20}$\\
    & 14\% & 54\% & 31\%\\
    $2N({\rm H_{2}})_{\rm bright}^{\rm a}$ (19) & 1.84 $\times$ 10$^{20}$ & 6.35 $\times$ 10$^{21}$ & 2.85 $\times$ 10$^{21}$\\
     & 16\% & 81\% & 44\%\\
    \enddata
    \begin{tablenotes}
        \item \tablecomments{(1) Gas phase. The number of LOSs that were used for the derivation is indicated in parenthesis; (2) Minimum values of the derived properties. The column density and the fraction with respect to the total hydrogen are presented in row; (3) Same as the second column, but for maximum values; (4) Same as the second column, but for median values. $^{\rm a}$ The derived values are upper limits, as CO-dark H$_{2}$ was not considered for the derivation.}
    \end{tablenotes} 
\end{deluxetable}

\begin{figure}
\centering
\includegraphics[scale=0.35]{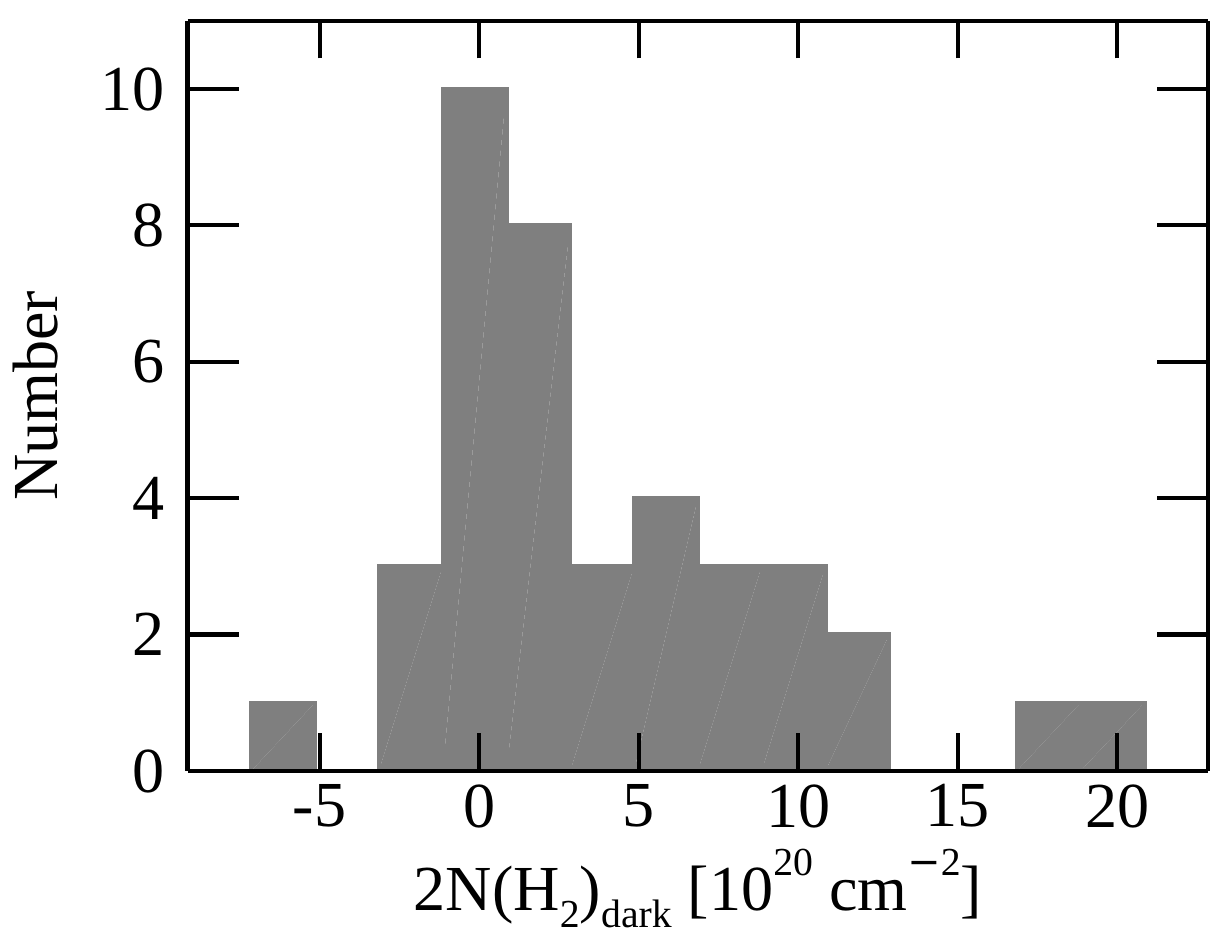}
\caption{\label{f:av_nh} Histogram of the CO-dark H$_{2}$ column density for the 39 CO non-detected LOSs.}
\end{figure}

In addition, we calculated 2$N$(H$_{2}$)$_{\rm dark}$ for the 39 CO non-detected LOSs, which include H~\textsc{i}-only and H~\textsc{i} $+$ CO-dark H$_{2}$ LOSs, by 

\begin{equation}
    \label{eq:N_H2_dark}
    \begin{split} 
    2N({\rm H_{2}})_{\rm dark} & = N({\rm H})_{A_{V}{\rm-}{\rm based}} - N({\rm \HI}) \\ 
                               & = N({\rm H})_{A_{V}{\rm-}{\rm based}} - (N({\rm \HI})_{\rm thin} + N({\rm \HI})_{\rm thick}) 
    \end{split}
\end{equation}

\noindent and present its distribution in Figure \ref{f:av_nh}. 

Figure \ref{f:av_nh} shows that the CO-dark H$_{2}$ column density distribution is approximately Gaussian from $-$5~$\times$~10$^{20}$~cm$^{-2}$ to 5~$\times$~10$^{20}$~cm$^{-2}$ with a peak of $\sim$0~cm$^{-2}$, suggesting that the 39 LOSs are dominated by atomic-only LOSs and the dispersion of the Gaussian distribution likely results from a slight variation in the DGR. In other words, our adopted DGR of 0.4~$\times$~10$^{-21}$~mag~cm$^{2}$ is indeed representative, and 2$N$(H$_{2}$)$_{\rm dark}$ values larger than 5~$\times$~10$^{20}$~cm$^{-2}$ are likely reliable.
For 14 CO non-detected LOSs with 2$N$(H$_{2}$)$_{\rm dark}$ $>$ 5~$\times$~10$^{20}$~cm$^{-2}$, we then found that the ratio of 2$N$(H$_{2}$)$_{\rm dark}$ to $N$(H) changes from 14\% to 54\% with a median of 31\% (Figure \ref{f:gas_fractions} and Table \ref{t:gas_fractions}). As compared to the CO-dark H$_{2}$, the contribution from the optically thick H~\textsc{i} to the total $N$(H) is minor (7--34\% with a median of 15\%). This finding of the CO-dark H$_{2}$ as a major constituent of the dark gas in the solar neighborhood is in agreement with previous studies such as \citet{lee2015}, \citet{liszt2018}, and \citet{murray2018a}. In addition, our median CO-dark H$_{2}$ fraction of 31\% is consistent with the Galactic average value of $\sim$30\% derived from the \textit{Herschel} GOT C$^{+}$ survey \citep{langer2014}. 

Finally, we estimated upper limits on 2$N$(H$_{2}$)$_{\rm bright}$ for the 19 CO-detected LOSs by

\begin{equation}
    \label{eq:N_H2_bright}
    \begin{split}
    2N({\rm H_{2}})_{\rm bright} & = N({\rm H})_{A_{V}{\rm-}{\rm based}} - N({\rm \HI}) - 2N({\rm H_{2}})_{\rm dark} \\
                               & < N({\rm H})_{A_{V}{\rm-}{\rm based}} - N({\rm \HI})
    \end{split}
\end{equation}

\noindent and summarize the results in Figure \ref{f:gas_fractions} and Table \ref{t:gas_fractions}. As shown in Figure \ref{f:gas_phases}, the CO-detected LOSs probe CO-free H$_{2}$ shells, as well as CO-bright H$_{2}$ cores. Separating these two components is not straightforward though, unless a CO-to-H$_{2}$ conversion factor $X_{\rm CO}$ is applied to the measured CO integrated intensity to calculate the CO-bright H$_{2}$ column density. Considering that $X_{\rm CO}$ could change by more than a factor of 100 over the measured $A_{V}$ $\sim$ 0.5--4~mag for the CO-detected LOSs \citep[e.g.,][]{lee2014}, we do not take the $X_{\rm CO}$ approach and provide upper limits on 2$N$(H$_{2}$)$_{\rm bright}$ by assigning all the measured H$_{2}$ to CO-bright H$_{2}$ (inequality sign in Equation \ref{eq:N_H2_bright}). In this case, the ratio of the upper limit on the CO-bright H$_{2}$ column density to the total hydrogen column density ranges from 16\% to 81\% with a median of 44\%.

\section{Conditions for the Formation of Molecular Gas: Observational Perspective}
\label{s:HI_CO_obs} 

\subsection{Kinematic Association between \HI and CO}
\label{s:HI_CO_vel_compare} 

\begin{figure}
\centering
\includegraphics[scale=0.35]{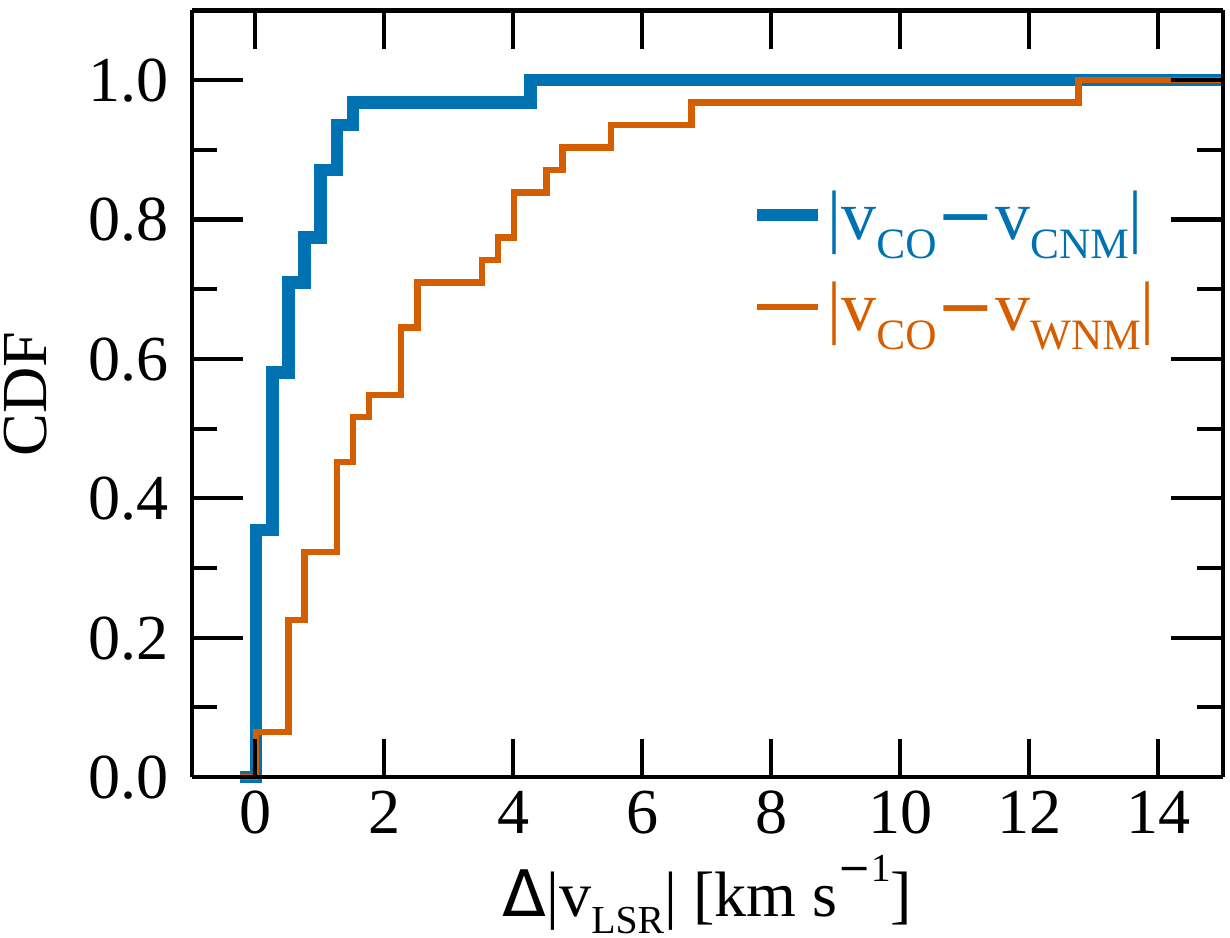}
\caption{\label{hi_co_vel} CDF of the absolute velocity difference between \HI and CO (CNM and CO in thick blue; WNM and CO in thin tan). For these CDFs, \HI components that are closest to the observed CO emission in velocity were considered.}
\end{figure}

To investigate the conditions for the formation of molecular gas, we first compared \HI and CO central velocities. For our analysis, we selected CNM and WNM components that are closest to the detected CO(1--0) emission in velocity and calculated absolute velocity differences between \HI and CO. The cumulative distribution function (CDF) of these velocity differences is shown in Figure \ref{hi_co_vel}. 

Figure \ref{hi_co_vel} shows that the velocity difference between the CNM and CO is systematically smaller than that between the WNM and CO. Specifically, the CNM--CO velocity difference ranges from 0.01~km~s$^{-1}$ to 4.3~km~s$^{-1}$ (median of 0.4~km~s$^{-1}$), while the WNM is offset from CO by 0.04--12.8~km~s$^{-1}$ (median of 1.7~km~s$^{-1}$). This difference between the CNM and WNM becomes more significant when additional components are considered (e.g., including the first and second closest components to CO results in median velocity differences of 1.3~km~s$^{-1}$ and 4.7~km~s$^{-1}$ for the CNM and WNM), demonstrating that the CNM is kinematically more closely associated with CO emission. If we take velocity as a proxy for position (e.g., CNM components at different velocities would be located in different places), our result implies that CO-bright molecular gas likely forms in CNM environments. 

\begin{deluxetable}{l c c}[t]
\centering
\tablecaption{\label{t:HI_number}Number of \HI Components}
    \tablewidth{0pt}
    \setlength{\tabcolsep}{12pt}
    \tabletypesize{\small}
    \tablehead{
    \colhead{Classification} & \colhead{CNM} & \colhead{WNM}
    }
    \startdata
    \multicolumn{3}{c}{Individual Properties}\\
    \hline
    CO non-detection & 146 & 128\\
    CO detection & 100 & 69\\
    Case A & 33 & 13\\
    Case B & 25 & 8\\
    \hline
    \multicolumn{3}{c}{Integrated Properties}\\
    \hline
    CO non-detection & 39 & 39 \\
    CO detection & 19 & 19 \\
    Case A & 22 & 22 \\
    Case B & 22 & 22 \\
    \enddata
\end{deluxetable}

\subsection{Individual \HI Properties}
\label{s:individual_HI} 

Next we examined the properties of individual \HI components ($T_{\rm s}$, $\tau_{\rm CNM}$, $N_{\rm CNM}$, and $N_{\rm WNM}$) in the presence of CO(1--0) emission. For our examination, we classified the observed Gaussian components into four groups: (1) CO non-detection (all components toward the 39 CO non-detected LOSs); (2) CO detection (all components toward the 19 CO-detected LOSs); (3) Case A (components whose central velocities fall between $\varv_{\rm CO}-2{\Delta \varv_{\rm CO}}$ and $\varv_{\rm CO}+2{\Delta \varv_{\rm CO}}$; subset of CO detection); (4) Case B (similar to Case A, but \HI central velocities are within $\pm \Delta \varv_{\rm CO}$ from $\varv_{\rm CO}$; subset of CO detection and Case A). This classification is motivated to probe the individual \HI properties required for the formation of CO-bright molecular gas, and in particular, Cases A and B are designed to select \HI components that are kinematically closely associated with CO emission with small velocity differences (e.g., Figure \ref{f:gas_morphology}).
The number of CNM and WNM components for each group is summarized in Table \ref{t:HI_number}. 

\begin{figure*}[t]
    \centering
    \includegraphics[scale=0.35]{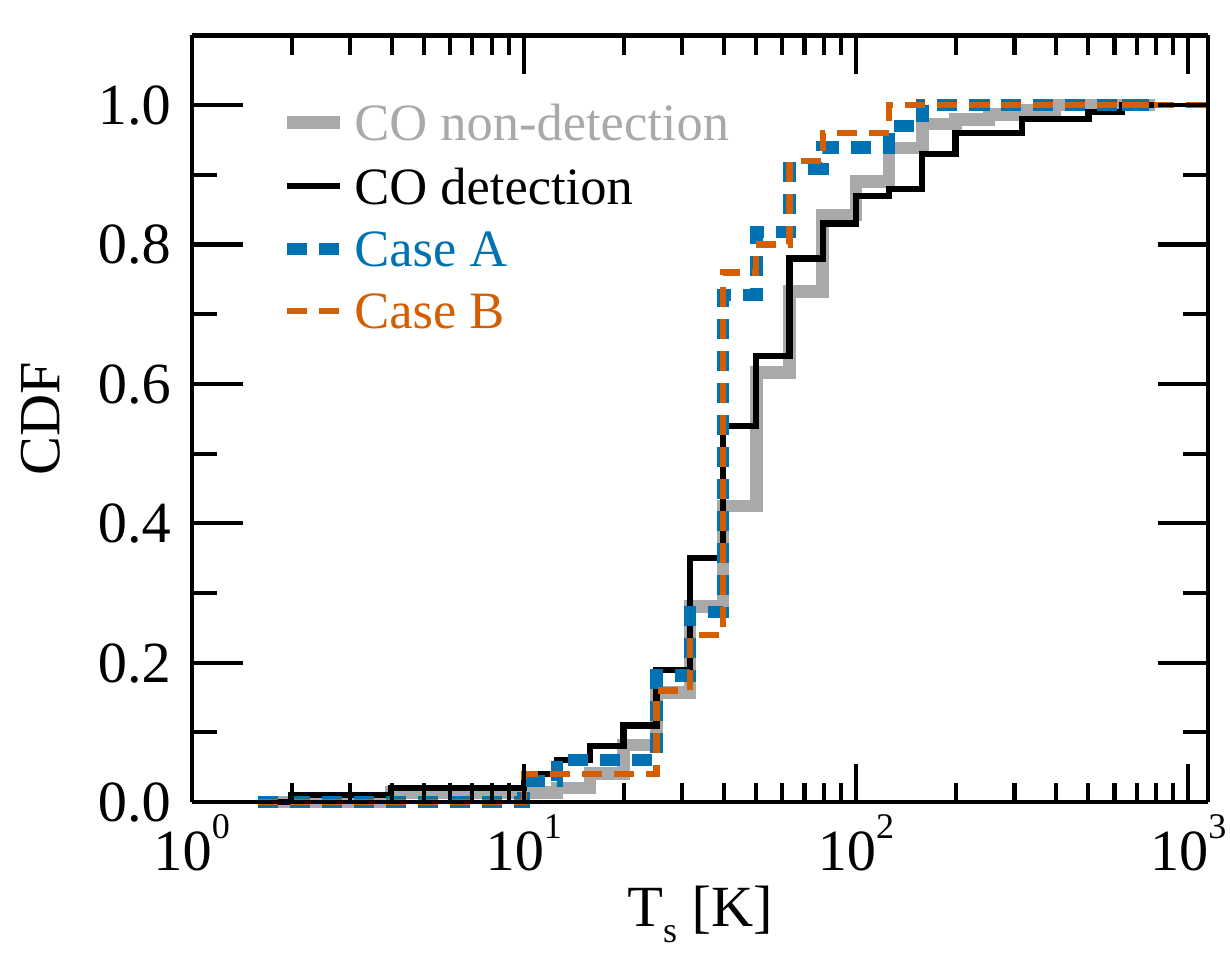}\hspace{0.3cm}
    \includegraphics[scale=0.35]{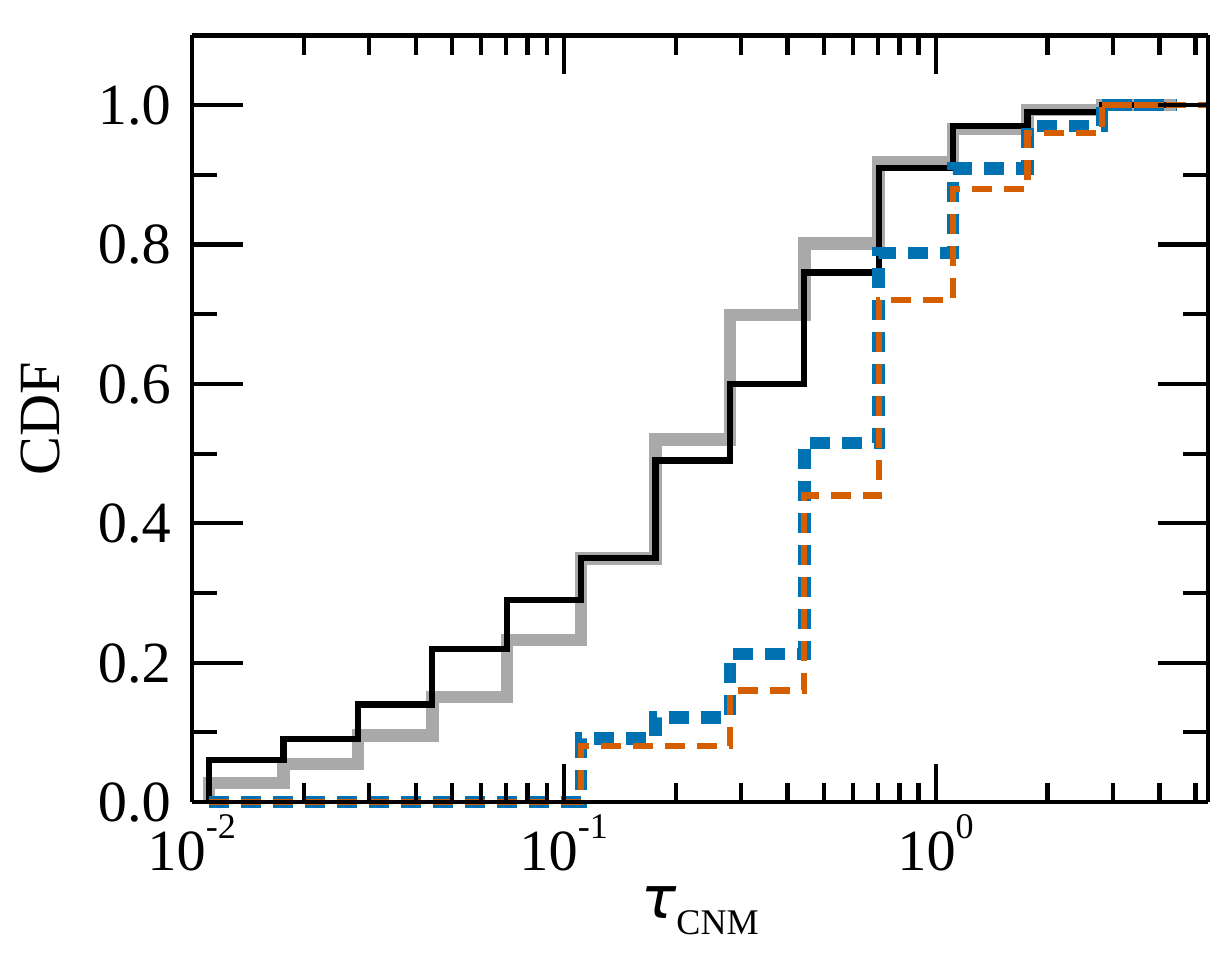}\vspace{0.2cm}
    \includegraphics[scale=0.35]{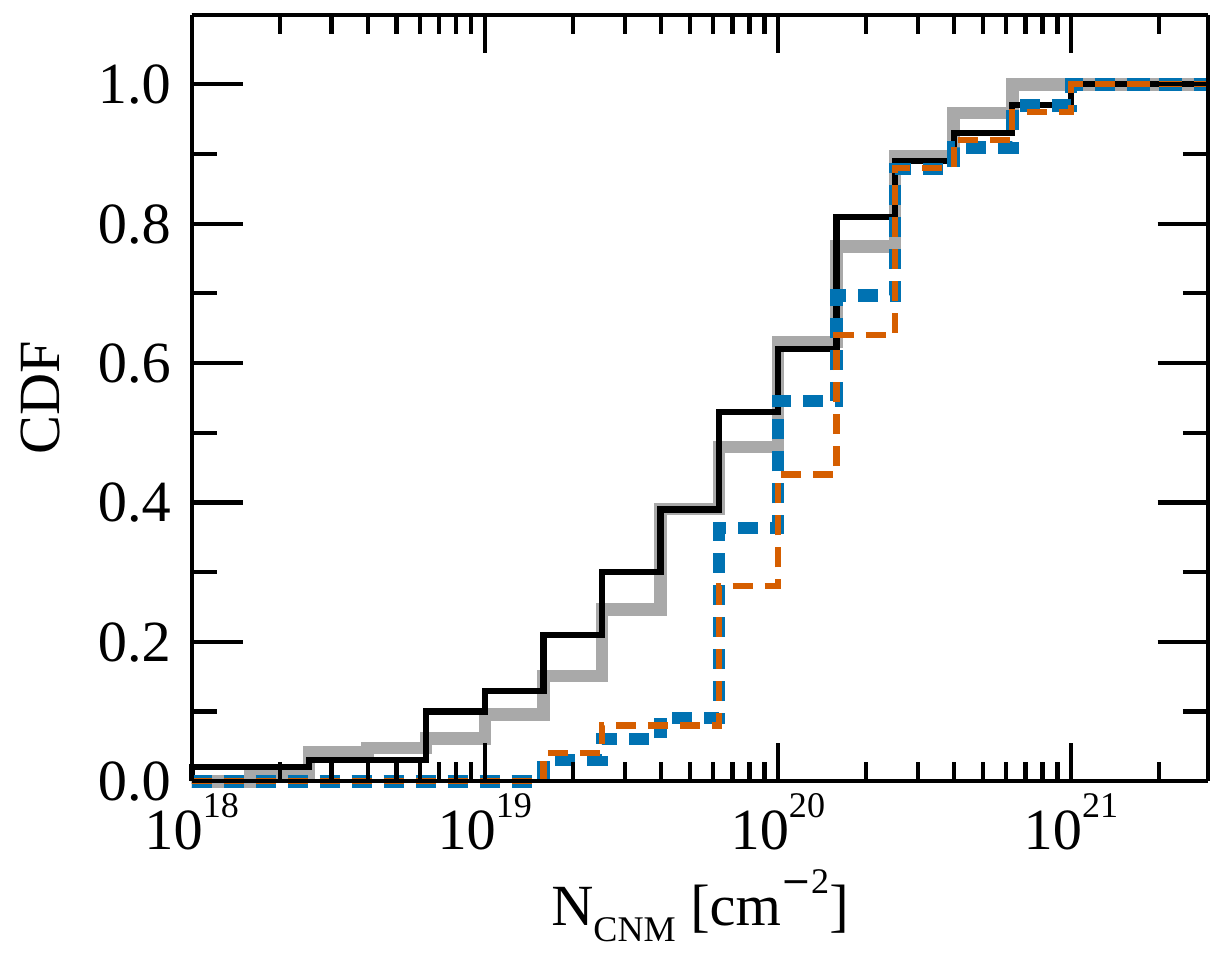}\hspace{0.3cm}
    \includegraphics[scale=0.35]{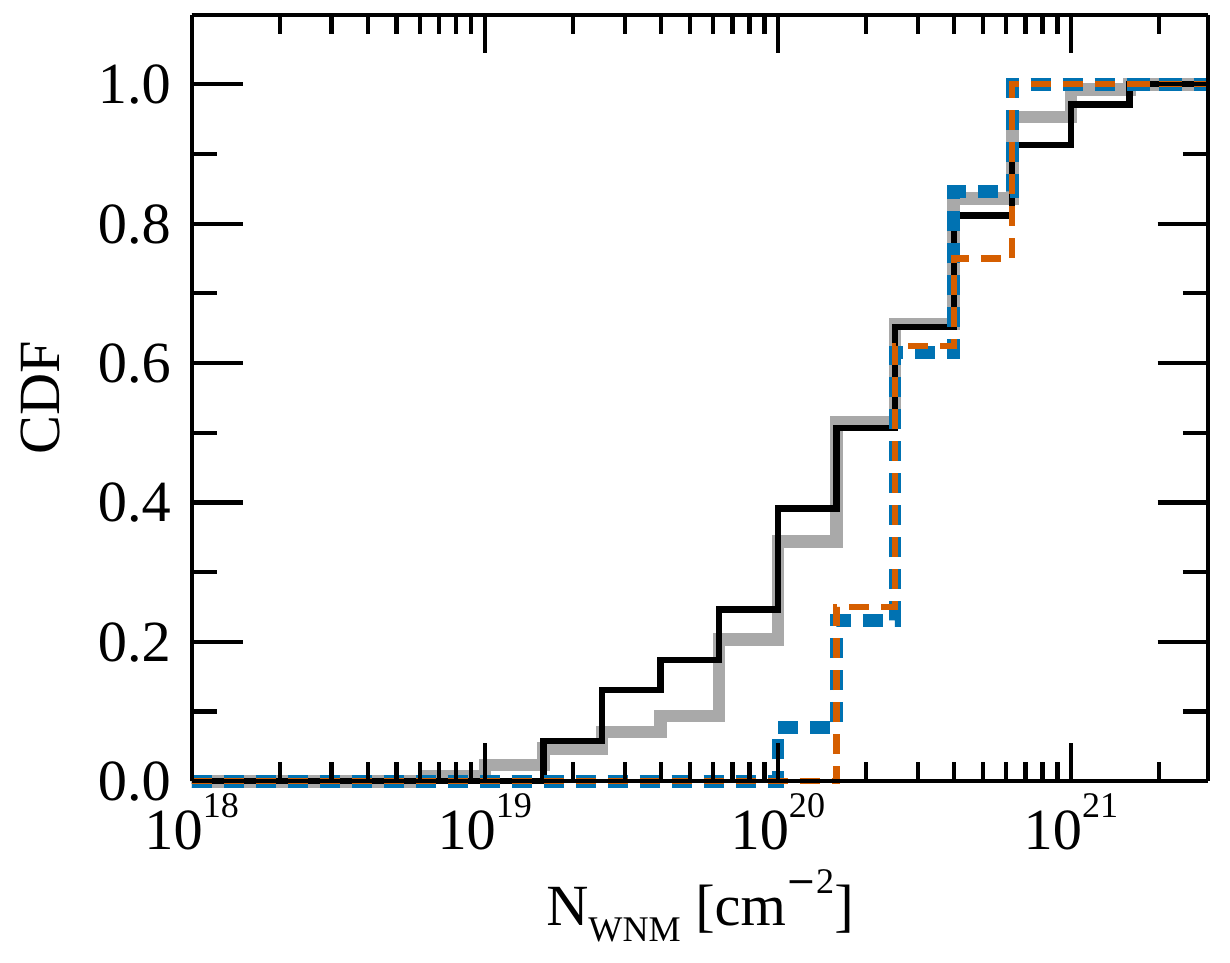}
    \caption{\label{individual_cdfs} CDFs of the spin temperature, optical depth, CNM and WNM column density. In each panel, the four groups are shown in different colors and line styles: CO non-detection (thick solid gray), CO detection (thin solid black), Case A (thick dashed blue), and Case B (thin dashed tan). 
    }
\end{figure*}

For each group, we examined the distributions of spin temperature, optical depth, CNM and WNM column density and presented them in Figure \ref{individual_cdfs} and Table \ref{t:HI_properties}. In general, we found that the CO non-detection and detection groups are almost indistinguishable in terms of their \HI properties. 
On the other hand, Cases A and B have several distinctive features compared to the CO non-detection and detection groups. For example, they do not have CNM components with $T_{\rm s}$ $>$ 200~K and show a factor of 2--5 smaller dispersion in $T_{\rm s}$ compared to the CO non-detection and detection groups. In addition, their minimum $\tau_{\rm CNM}$ = 0.1, $N_{\rm CNM}$ = 2~$\times$~10$^{19}$~cm$^{-2}$, and $N_{\rm WNM}$ = 2~$\times$~$^{20}$~cm$^{-2}$ are an order of magnitude higher than those for the CO non-detection and detection groups. These distinctive features of Cases A and B are not pronounced in the comparison between the CO non-detection and detection groups, mainly because Cases A and B are only a small fraction of the individual \HI components (e.g., the Case B CNM and WNM are 25\% and 12\% of the CO detection CNM and WNM). 


All in all, our result implies that CO-bright molecular gas forms in regions where individual CNM components evolve toward colder temperature and higher column density. However, only $\sim$20\% of the CNM components with $T_{\rm s}$ $<$ 200~K, $\tau_{\rm CNM}$ $>$ 0.1, and $N_{\rm CNM}$ $>$ 2~$\times$~10$^{20}$~cm$^{-2}$ are associated with CO emission (Cases A and B), suggesting that individual CNM components with low temperature and high column density are necessary, but not sufficient for the formation of CO-bright molecular gas. This conclusion is consistent with what \citet{rybarczyk2022} found from \HI and HCO$^{+}$ observations of diffuse Galactic LOSs (see Appendix \ref{s:SPONGE_comparison} for details). 

\begin{deluxetable*}{c c c c c}
    \centering
    \tablecaption{\label{t:HI_properties}Physical Properties of \HI Components}
    \tablewidth{2pt}
    \setlength{\tabcolsep}{10pt}
    \tabletypesize{\small}
    \tablehead{
    \colhead{Properties} & \colhead{CO non-detection} & \colhead{CO detection} & \colhead{Case A} & \colhead{Case B} \\
    \colhead{(1)} & \colhead{(2)} & \colhead{(3)} & \colhead{(4)} & \colhead{(5)}
    }
    \startdata
    \multicolumn{5}{c}{Individual Properties}\\
    \hline
    $T_{\rm s}$~({$\rm K$}) & 4.11$-$479.28 & 1.99$-$725.42 & 10.82$-$188.27 & 10.82$-$130.63\\
    & 53.83 & 48.13 & 46.21 & 46.21\\
    $\tau_{\rm CNM}$ & 0.01$-$3.14 & 0.01$-$3.41 & 0.11$-$3.41 & 0.11$-$3.41\\
    & 0.27 & 0.29 & 0.68 & 0.78\\
    $N_{\rm CNM}$~($\rm 10^{20}~cm^{-2}$) & 0.02$-$8.59 & 0.01$-$13.70 & 0.17$-$10.02 & 0.17$-$10.02\\
    & 1.07 & 0.87 & 1.50 & 1.91\\
    $N_{\rm WNM}$~($\rm 10^{20}~cm^{-2}$) & 0.10$-$18.57 & 0.17$-$16.70 & 1.23$-$7.33 & 1.86$-$7.33\\
    & 2.22 & 2.49 & 3.52 & 3.79\\
    \hline
    \multicolumn{5}{c}{Integrated Properties}\\
    \hline
    $N_{\rm CNM}$~($\rm 10^{20}~cm^{-2}$) & 0.58$-$19.90 & 1.97$-$22.45 & 0.17$-$9.93 & 0.01$-$8.31\\
    & 5.48 & 7.99 & 2.82 & 1.69\\
    $N_{\rm WNM}$~($\rm 10^{20}~cm^{-2}$) & 1.67$-$27.48 & 1.85$-$25.54 & 0.03$-$5.96 & 0.01$-$4.44\\
    & 9.58 & 11.90 & 2.41 & 1.29\\
    $N(\rm \HI)$~($\rm 10^{20}~cm^{-2}$) & 4.70$-$38.57 & 7.33$-$38.08 & 0.81$-$14.81 & 0.38$-$11.41\\
    & 16.99 & 24.21 & 5.51 & 2.96\\
    $f_{\rm CNM}$ & 0.12$-$0.69 & 0.19$-$0.77 & 0.05$-$ 0.99 & 0.01$-$0.99\\
    & 0.36 & 0.37 & 0.56 & 0.57\\
    $f_{\rm \#CNM}$ & 0.20$-$0.80 & 0.38$-$0.82 & 0.00$-$1.00$^{\rm a}$ & 0.00$-$1.00$^{\rm a}$\\
    & 0.56 & 0.57 & 0.75 & 1.00\\
    \enddata
    \begin{tablenotes}
        \item \tablecomments{(1) Physical properties. Each row displays the range of values for each physical property, with the median value indicated below the range; (2) \HI components toward the 39 CO non-detected LOSs; (3) \HI components toward the 19 CO-detected LOSs; (4) \HI components whose central velocities fall between $\varv_{\rm CO}-2{\Delta \varv_{\rm CO}}$ and $\varv_{\rm CO}+2{\Delta \varv_{\rm CO}}$; (5) \HI components whose central velocities are in the range of $\varv_{\rm CO}\pm{\Delta \varv_{\rm CO}}$. $^{\rm a}$There are two (Case A) and three (Case B) CO peaks where there is no associated CNM component. $f_{\rm \#CNM}$ is set to zero accordingly, and these cases are still considered for the calculation of the median values. 
        }
    \end{tablenotes}    
\end{deluxetable*}

\subsection{Integrated \HI Properties} 
\label{s:integrated_HI} 

\begin{figure*}[t]
    \centering
    \includegraphics[scale=0.35]{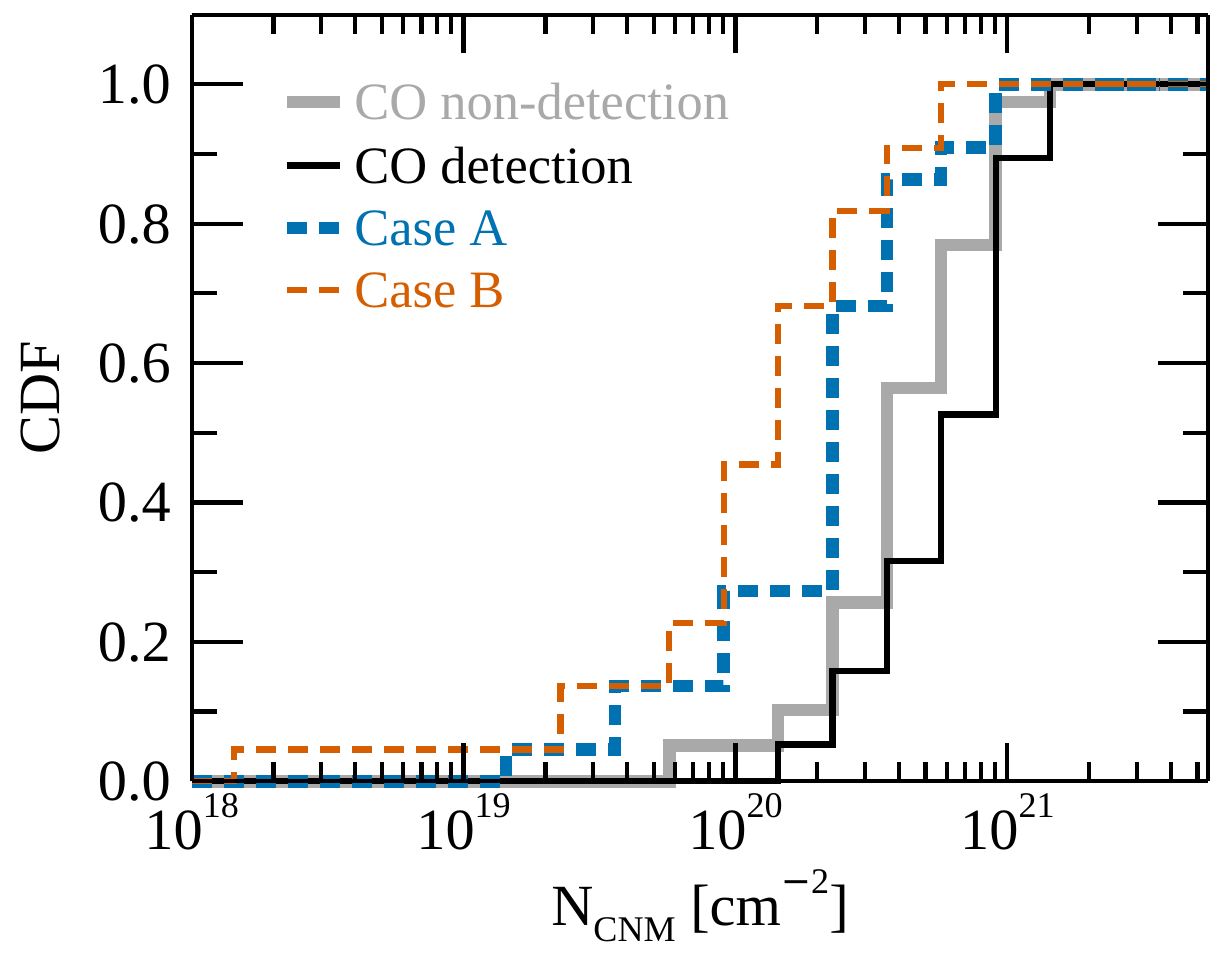}\hspace{0.3Cm}
    \includegraphics[scale=0.35]{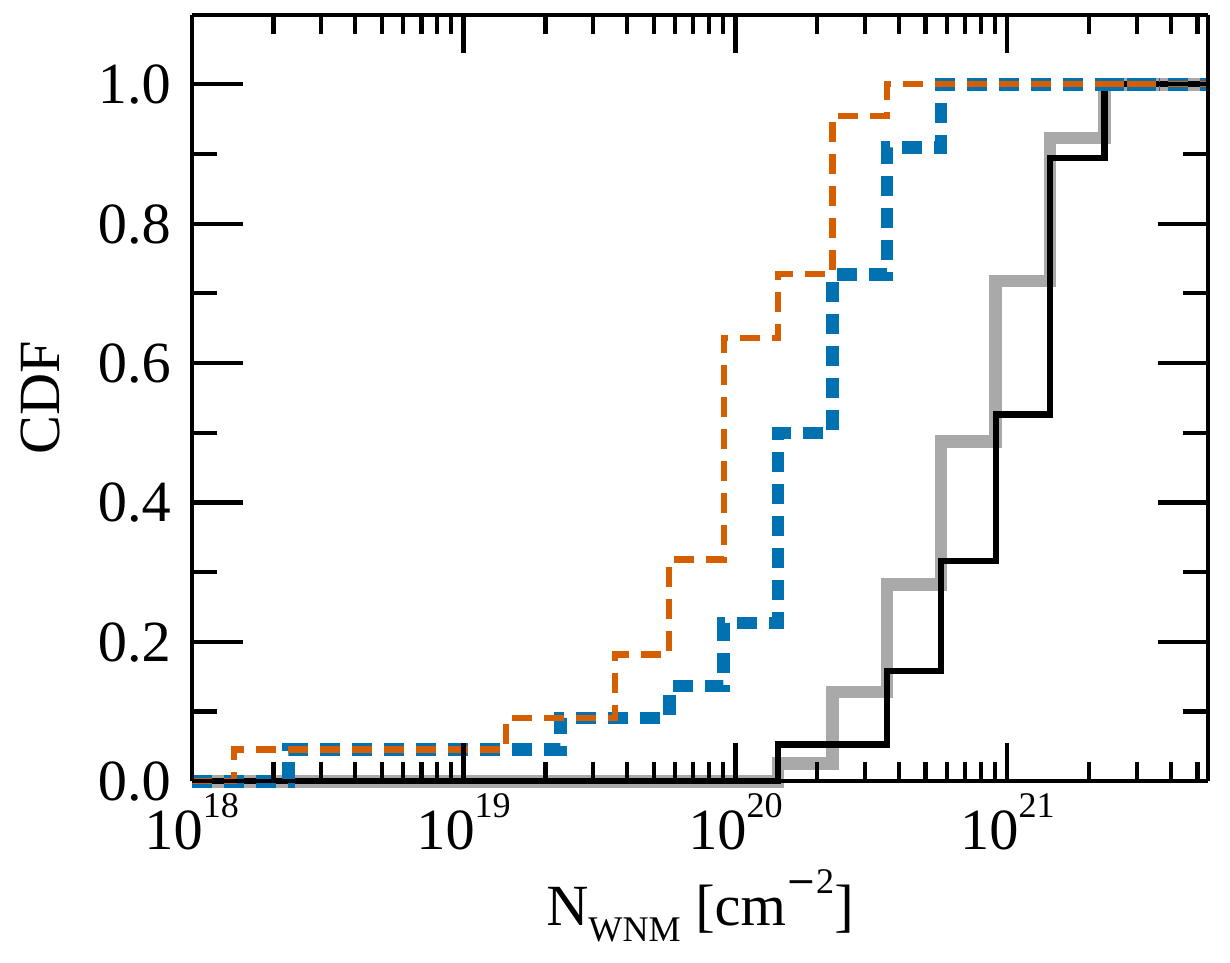}\vspace{0.2Cm}
    \includegraphics[scale=0.35]{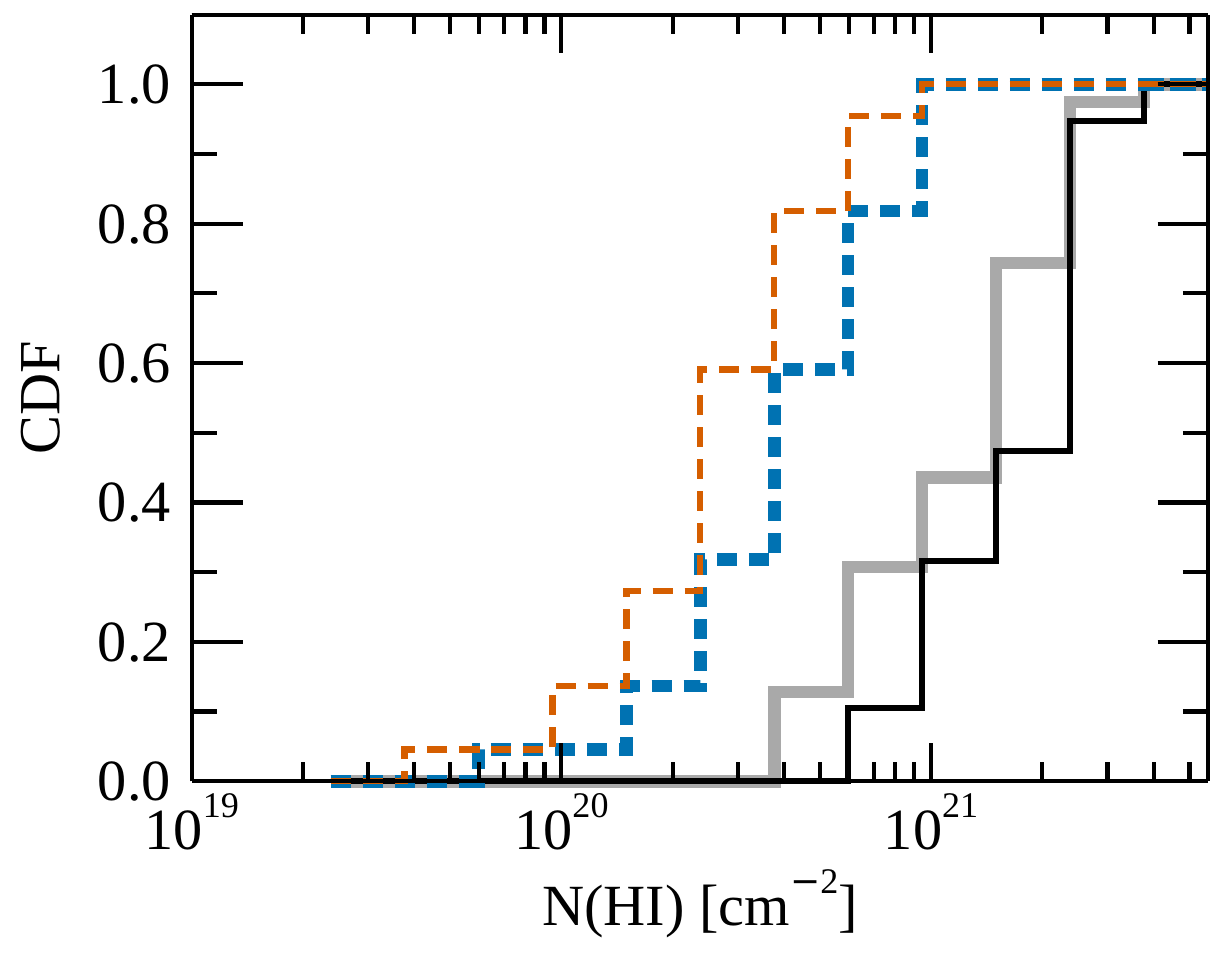}\hspace{0.3Cm}
    \includegraphics[scale=0.35]{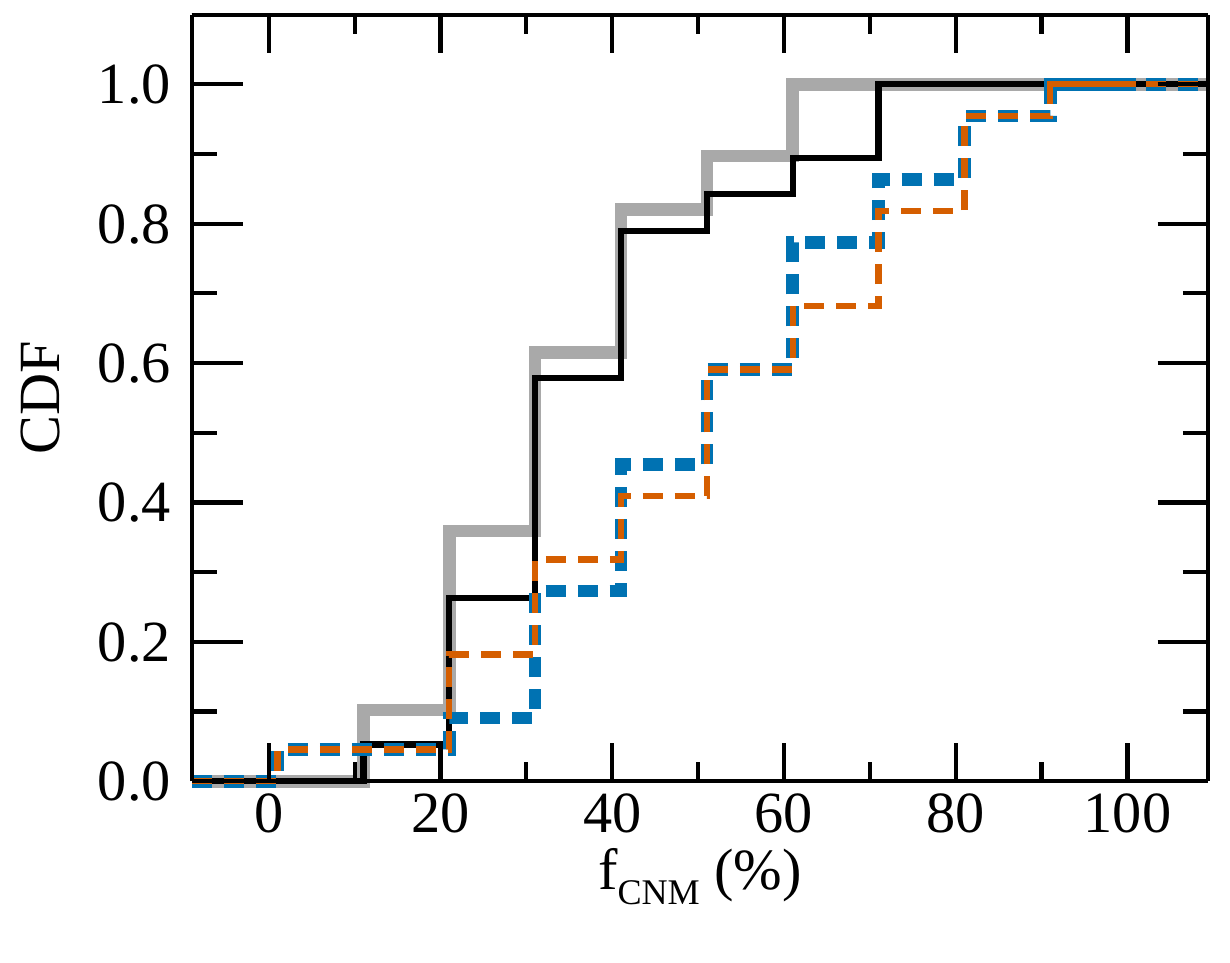}\vspace{0.2Cm}
    \includegraphics[scale=0.35]{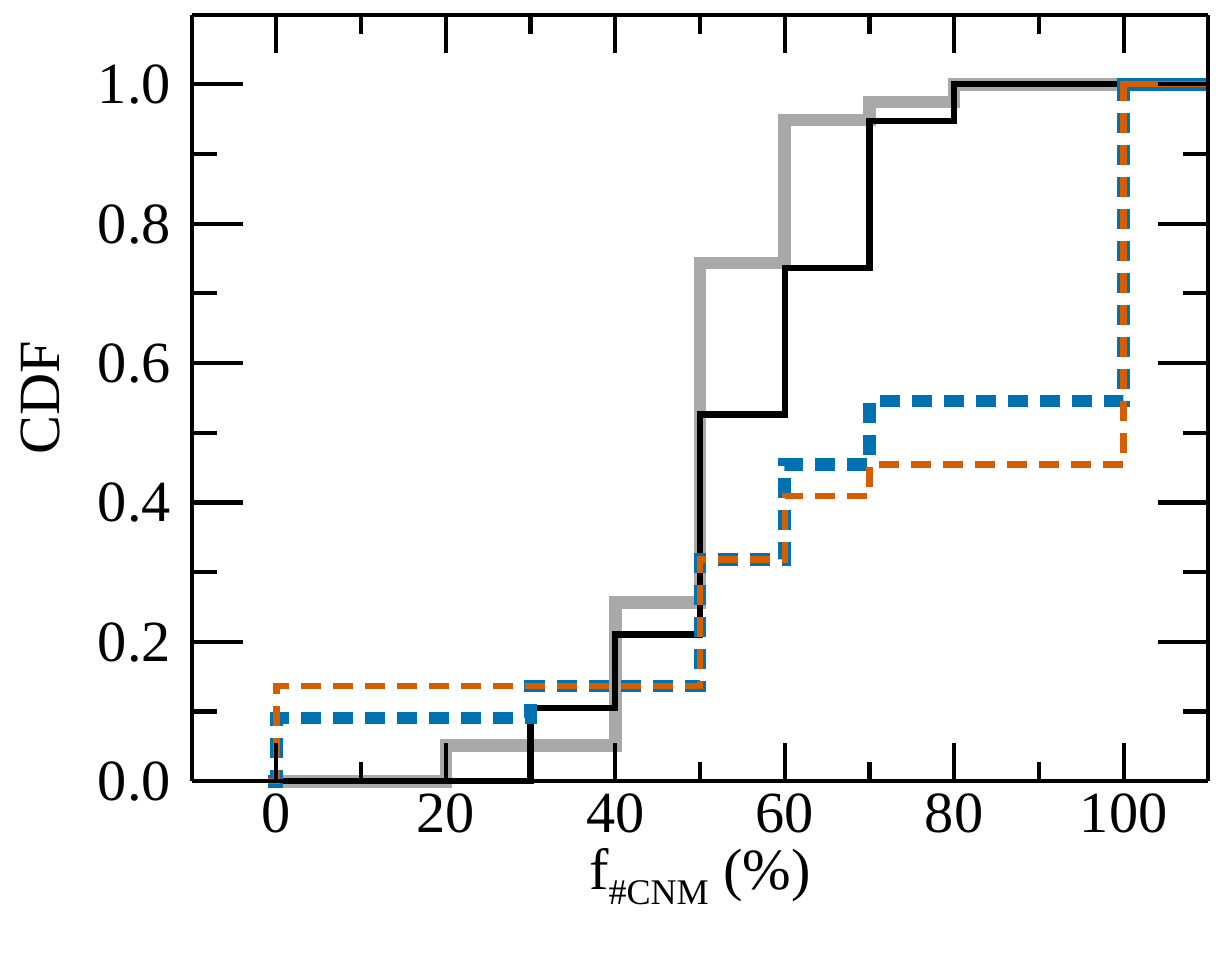}
    \caption{\label{integrated_cdfs} CDFs of the CNM, WNM, CNM$+$WNM column density, CNM fraction, and CNM component density. As in Figure \ref{individual_cdfs}, the four groups are shown in different colors and line styles: CO non-detection (thick solid gray), CO detection (thin solid black), Case A (thick dashed blue), and Case B (thin dashed tan). The column densities of Cases A and B are lower than those of the CO non-detection and detection by design (integrated over smaller velocity ranges) and are shown here only for the sake of completeness.
    }
\end{figure*}

Finally, we examined the integrated \HI properties required for the formation of CO-bright molecular gas by comparing the four groups in terms of total CNM, WNM, CNM$+$WNM column densities, and CNM fraction (Figure \ref{integrated_cdfs} and Table \ref{t:HI_properties}). In contrast to the analysis in Section \ref{s:individual_HI}, these integrated \HI properties were derived by considering all (CO detection and non-detection) or several (Cases A and B) Gaussian components. Specifically, we used Equation (\ref{e:N_HI_true}) and applied the relevant velocity limits for integration (whole LOS for the CO non-detection and detection groups and $\varv_{\rm CO} \pm 2\Delta\varv_{\rm CO}$ and $\varv_{\rm CO} \pm \Delta\varv_{\rm CO}$ for Cases A and B) to calculate the total CNM and WNM column densities. In addition, we defined the CNM fraction $f_{\rm CNM} = N_{\rm CNM} / N({\rm \HI})$, where $N$(H~\textsc{i}) is given by Equation (\ref{e:N_HI_true}).

\begin{figure}
    \centering
    \includegraphics[scale=0.35]{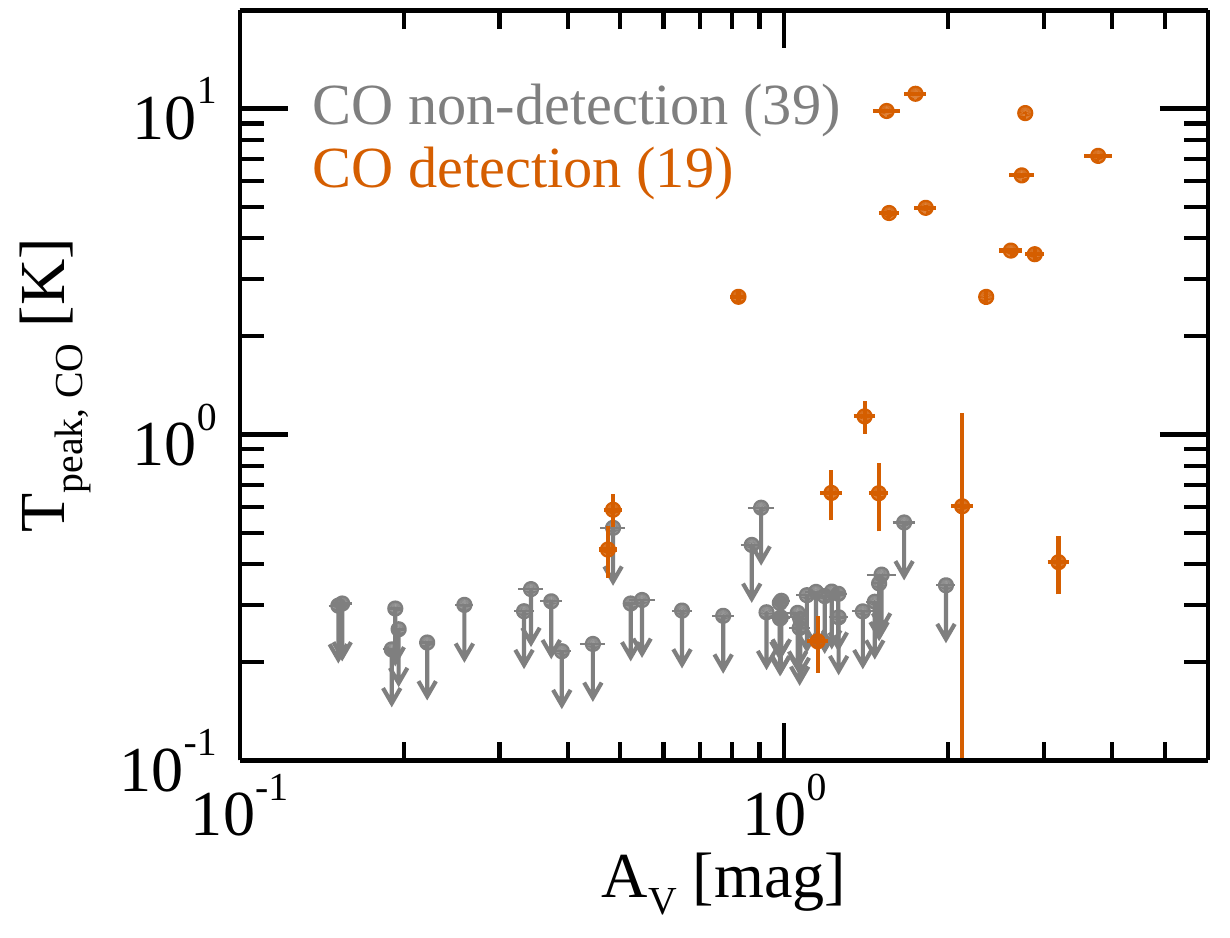} 
     \caption{\label{CO_peak_Av} CO peak brightness temperature as a function of \textit{Planck}-based $A_{V}$. The 19 CO-detected LOSs are shown in tan, while upper limits based on 3$\sigma$ values are indicated as the downward arrows for the 39 CO non-detected LOSs.
     }
\end{figure}

Figure \ref{integrated_cdfs} shows that the CO detection group generally has slightly higher \HI column densities than the CO non-detection group. For example, the median total CNM, WNM, CNM$+$WNM column densities of the CO detection group are a factor of 1.2--1.5 higher than those of the CO non-detection group. This difference in the integrated column densities is in contrast with the almost identical distributions of the individual \HI properties for the two groups (Section \ref{s:individual_HI}) and implies that the total amount of gas along a LOS (and consequently associated dust extinction) could be one of the important factors for the formation of CO-bright molecular gas. An examination of the CO peak brightness temperature as a function of $A_{V}$ (Figure \ref{CO_peak_Av}) indeed reveals that CO emission is detected primarily toward LOSs with $A_{V}$ $\gtrsim$ 0.5--1~mag, which is comparable to the threshold dust extinction for CO formation in the solar neighborhood \citep[e.g.,][]{pineda2008,lee2014,lee2018}.  

Another interesting finding is that Cases A and B have systematically higher CNM fractions than the other groups (e.g., median CNM fraction of 0.4 for the CO non-detection and detection groups and 0.6 for Cases A and B). These higher CNM fractions could result from two cases: (1) an increase in the column density of individual CNM components; (2) an increase in the relative number of CNM components. As for the first case, Figure \ref{individual_cdfs} and Table \ref{t:HI_properties} confirm that the column density of individual \HI components increases toward CO more in the CNM than in the WNM. For example, from the CO detection to Case B, the median CNM and WNM column densities increase by a factor of 2.2 and 1.5, respectively. 
To evaluate the second case, we then estimated the CNM component density ($f_{\rm \#CNM}$) by dividing the number of CNM components by the number of total \HI components and compared its distribution between the four groups (Figure \ref{integrated_cdfs} and Table \ref{t:HI_properties}). Our analysis shows that the CNM component density is indeed systematically higher for Cases A and B than for the CO non-detection and detection groups, which is in line with our previous finding of the CNM being kinematically more closely associated with CO emission (Section \ref{s:HI_CO_vel_compare}). Based on these results, we conclude that an increase in both the individual CNM column density and the relative number of CNM components could contribute to the higher CNM fraction toward CO. 

In summary, our comparison between \HI and CO suggests that the formation of CO-bright molecular gas is favored in high column density environments that are able to provide significant shielding against dissociating UV radiation. In these environments, the CNM becomes colder (lower temperature) and more abundant (higher density), facilitating H$_{2}$ and consequently CO formation.
We will discuss further on the conditions for the formation of molecular gas in Section \ref{s:conditions_molecular_gas}.

\section{Conditions for the Formation of Molecular Gas: Theoretical Perspective}
\label{s:HI_CO_theory}

In this section, we compare the observed CNM properties to the prediction from the \citetalias{sternberg2014} model with the aim of investigating the fundamental principles of the H~\textsc{i}-to-H$_{2}$ transition. Specifically, our approach is to estimate the density expected for H$_{2}$ formation from \citetalias{sternberg2014}($n^{\rm exp}$) and to confront it with the CNM density inferred from our observations ($n_{\rm CNM}^{\rm obs}$). As for the theoretically expected density, we recall that the total \HI column density of a plane-parallel slab of gas and dust in the \citetalias{sternberg2014} model is controlled by the dimensionless parameter $\alpha G$ (Equation \ref{e:NHI_b16}). As $\alpha G$ is a function of $I_{\rm UV}$ and $n$ (Equation \ref{e:alphaG}; $\tilde{\sigma_{\rm g}}$ $\sim$ 1 for our case of the solar neighborhood conditions), $n^{\rm exp}$ can be expressed as 

\begin{equation}
    \label{e:n_exp}
    \begin{split}
        n^{\textrm{exp}}~(\textrm{cm}^{-3}) & = \frac{18.4~I_{\textrm{UV}}}{\textrm{exp}\left(N_{\textrm{CNM}}/8.4\times10^{20} ~{\rm{cm^{-2}}}\right)-1}, \\ 
    \end{split}
\end{equation}

\noindent where $N$(H~\textsc{i}) is substituted with $N_{\rm CNM}$. This substitution is motivated by the fact that our \textit{Planck}-based $I_{\rm UV}$ estimates are mostly $\sim$1 (Section \ref{s:UV}). The nearly uniform $I_{\rm UV}$ values suggest isotropic UV radiation that is most likely attenuated by the widespread WNM. The impact of the WNM on the H~\textsc{i}-to-H$_{2}$ transition is already taken into account in this manner, and we therefore proceeded by replacing $N$(H~\textsc{i}) with $N_{\rm CNM}$. As for the observationally inferred density, we took the thermal pressure log$_{10}$($P/k_{\rm B}$~cm$^{-3}$~K) = 3.58 $\pm$ 0.18 from \citet{jenkins2011} (estimated for the CNM based on \textit{Hubble Space Telescope} observations of C~\textsc{i} multiplets at UV wavelengths) and calculated $n_{\rm CNM}^{\rm obs}$ by

\begin{equation}
    \label{e:n_cnm_obs} 
    n_{\textrm{CNM}}^{\textrm{obs}}~(\textrm{cm}^{-3}) = \left(\frac{10^{3.58\pm0.18}}{T_{\textrm{s}}}\right). 
\end{equation}

\noindent Since $n^{\rm exp} = n_{1} + 2n_{2}$ is the total number density, $n^{\rm exp}$ should be higher than $n_{\rm CNM}^{\rm obs}$ for CO-detected LOSs. 

\subsection{Density versus CNM column density}
\label{s:n_vs_CNM} 

\begin{figure}
    \centering
    \includegraphics[scale=0.16]{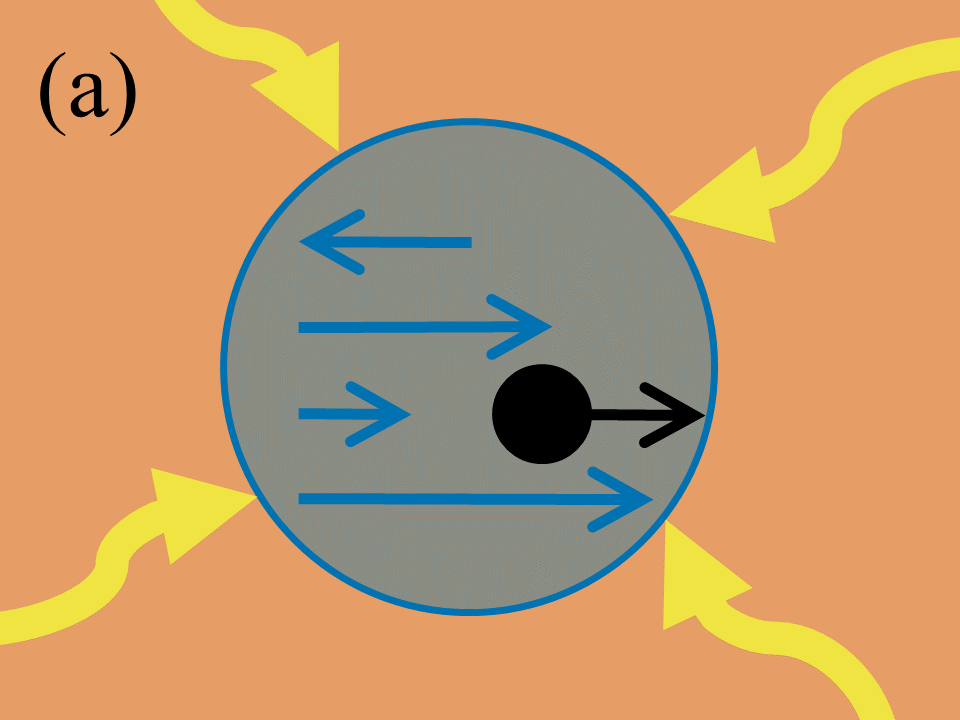}
    \includegraphics[scale=0.16]{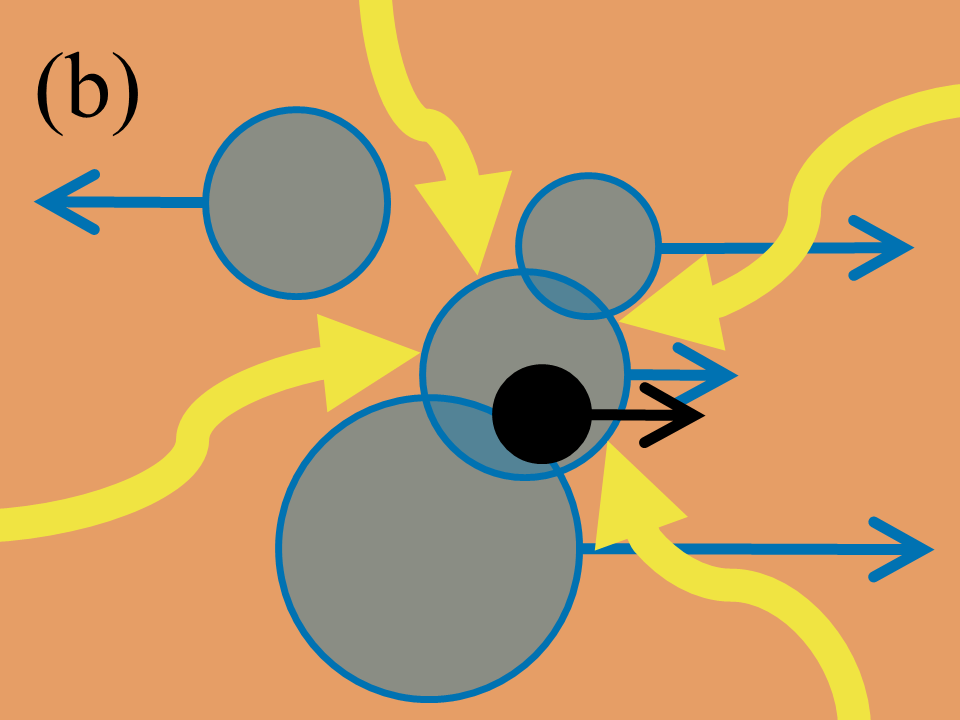}\vspace{0.2cm}
    \caption{\label{f:gas_morphology} Possible distributions of the WNM, CNM, and CO (tan, blue, and black, respectively). Based on our finding of the kinematic association between \HI and CO (Section \ref{s:HI_CO_vel_compare}), the WNM is represented as the diffuse background, while the CNM and CO are shown as the smaller embedded structures. The velocities of the CNM and CO are indicated as the blue and black arrows with arbitrary sizes, and UV radiation is presented as the yellow arrows. (Left) Scenario for the Entire LOS, where CNM components at different velocities belong to one large structure and contribute to the shielding of the CO core. (Right) Scenario for Cases A and B, where CNM components close to CO only provide shielding against UV radiation.
    }\vspace{0.3Cm}
\end{figure}

We estimated $n^{\rm exp}$ and $n_{\rm CNM}^{\rm obs}$ for the following three cases: (1) Entire LOS; (2) Case A; (3) Case B. For each LOS, all CNM components are considered for the Entire LOS, while CNM components within $\varv_{\rm CO}\pm2{\Delta \varv_{\rm CO}}$ and $\varv_{\rm CO}\pm{\Delta \varv_{\rm CO}}$ are examined for Cases A and B. We assessed these three cases mainly because of a lack of knowledge on the geometry of the CNM. For example, the Entire LOS would correspond to a case where all CNM components at different velocities belong to one large structure and absorb dissociating UV photons (Figure \ref{f:gas_morphology}(a)). Meanwhile, Cases A and B would be equivalent to a case where only CNM components near CO clumps provide shielding against UV photons (Figure \ref{f:gas_morphology}(b)). While being simple pictures, these two scenarios cover small (Cases A and B) to large (Entire LOS) volume filling factors for the CNM. Finally, we used the \textit{Planck}-based $I_{\rm UV}$ values for Equation (\ref{e:n_exp}) and the opacity-weighted mean spin temperature ($T_{\textrm{s}, \tau}$) for Equation (\ref{e:n_cnm_obs}): 

\begin{equation}
    T_{\textrm{s}, \tau}~(\textrm{K}) = \frac{\sum_{n=0}^{N-1} \tau_{0,n} T_{\textrm{s},n}}{\sum_{n=0}^{N-1} \tau_{0,n}}.
\end{equation}

\noindent The derived $n^{\rm exp}$ and $n_{\rm CNM}^{\rm obs}$ values for the three cases are presented as a function of $N_{\rm CNM}$ in Figure \ref{f:n_vs_nh1}. 

We found that the three cases show similar trends. For example, the total densities expected from \citetalias{sternberg2014} are higher than the inferred CNM densities at low column densities ($N_{\rm CNM}$~$\lesssim$~10$^{20}$~cm$^{-2}$). In other words, the model is in agreement with the lower limits on the total densities constrained by our observations. 
On the contrary, the \citetalias{sternberg2014}-based total densities are lower than the inferred CNM densities at high column densities ($N_{\rm CNM}$ $\gtrsim$ 10$^{20}$~cm$^{-2}$), resulting in a discrepancy between the model and our observations. This discrepancy becomes more significant from Case B to Case A to the Entire LOS case and reaches up to one or two orders of magnitude at the highest column density of $\sim$10$^{21}$~cm$^{-2}$. 

\begin{figure}[t]
    \centering
    \includegraphics[scale=0.32]{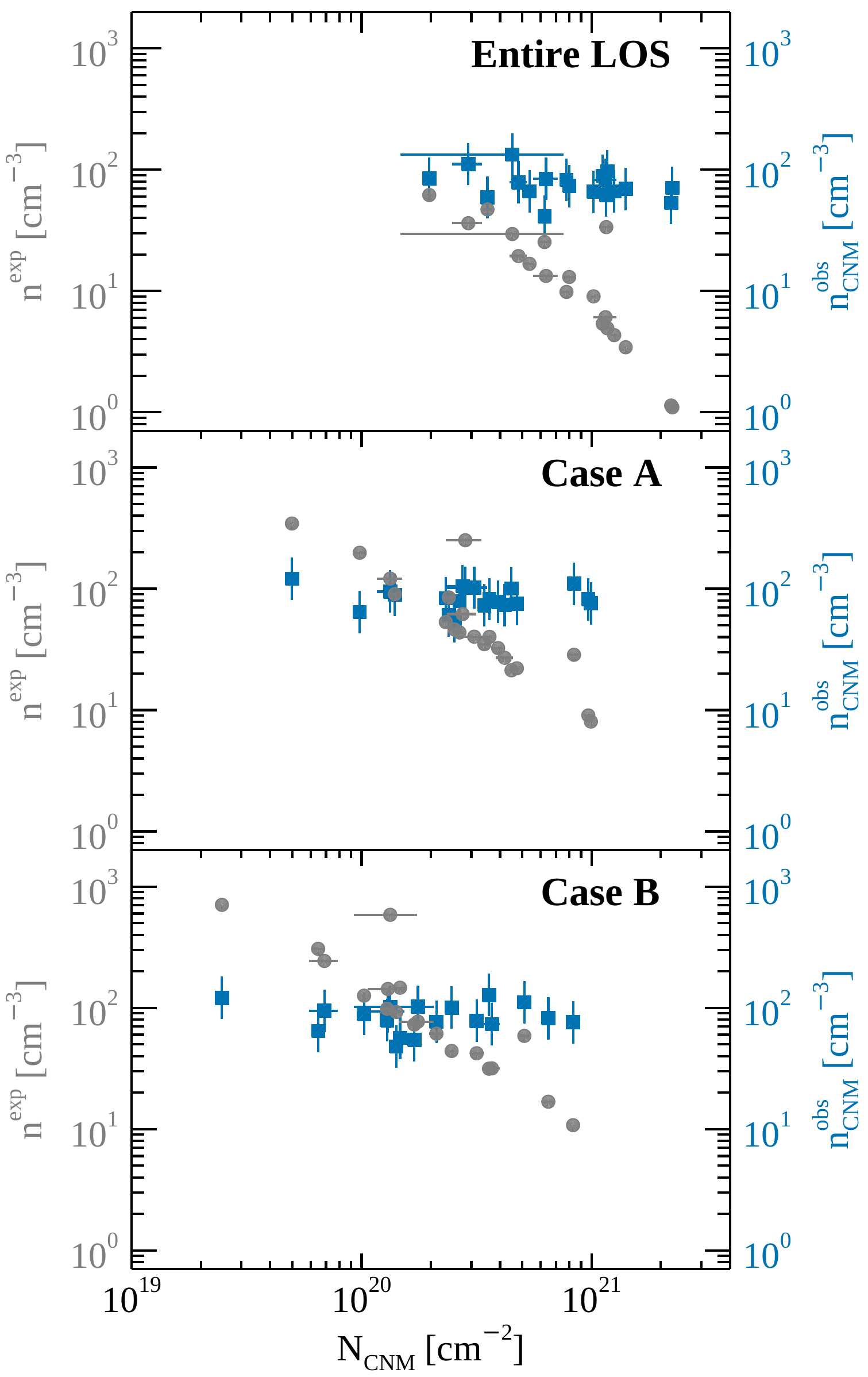}\hspace{0.6cm}
    \caption{\label{f:n_vs_nh1} Comparison between the \citetalias{sternberg2014}-based total number densities ($n^{\rm exp}$; gray circles) and the observationally inferred CNM densities ($n^{\rm obs}_{\rm CNM}$; blue squares) as a function of the CNM column density for the 19 CO-detected LOSs. Since the predicted $n^{\rm exp}$ corresponds to the total gas number density ($n_{1} + 2n_{2}$) at which H$_{2}$ formation is expected to occur, it should be higher than $n_{\rm CNM}^{\rm obs}$.
    Finally, the observed variation in the thermal pressure, log$_{10}$($P/k_{\rm B}$~cm$^{-3}$~K) = 3.58 $\pm$ 0.18, is indicated as the 1$\sigma$ error bars for $n_{\rm CNM}^{\rm obs}$. (Top) Entire LOS. (Middle) Case A. (Bottom) Case B.  
    }\vspace{0.3Cm}
\end{figure}

\subsection{Limitations and Implications} 
\label{s:B16_implications}

Taken at face value, the discrepancy at high CNM column densities implies that only a small fraction of the total CNM along a LOS participates in H$_{2}$ formation ($\lesssim$ 13\% on average; this fraction is estimated from the LOSs with large discrepancies at $N_{\rm CNM}$ $>$ 10$^{20}$~cm$^{-2}$ for the Entire LOS case). 
In other words, the CNM must be clumpy with a small volume filling factor. While this is a reasonable interpretation, however, our analysis is not without limitations. Below we discuss other sources of the discrepancy and their implications.  

One possible source of the observed discrepancy is the assumed thermal pressure of 2500--5800~cm$^{-3}$~K for the CNM. While we considered this factor of two variation in the thermal pressure, \citet{goldsmith2018} recently found a larger variation ($\sim$10$^{3}$--10$^{4}$~cm$^{-3}$~K) from SOFIA [C~\textsc{ii}] 158~$\mu$m observations of Galactic LOSs (3C131, one of our CO-detected LOSs, was found to have $P/k_{\rm B}$ $\sim$ 2500--3200~cm$^{-3}$~K). If the thermal pressure of the CNM varies by an order of magnitude as the SOFIA observations suggest, the discrepancy between the \citetalias{sternberg2014} prediction and our observations would certainly decrease, but it is likely that the discrepancy would still persist at the highest column density of $\sim$10$^{21}$~cm$^{-2}$.
 
Another possible source of the discrepancy is cosmic-rays. Cosmic-rays ionize atoms and molecules in collisions and can destruct H$_{2}$ as follows \citep[e.g.,][]{sternberg2021}: 

\begin{align}
    \textrm{H}_{2} + \textrm{CR} \longrightarrow \textrm{H}_{2}^{+} + e, \\ 
    \textrm{H}_{2}^{+} + \textrm{H}_{2} \longrightarrow \textrm{H}_{3}^{+} + \textrm{H}. 
\end{align}

\noindent In addition, cosmic-rays can directly dissociate H$_{2}$: 

\begin{align}
    \textrm{H}_{2} + \textrm{CR} \longrightarrow \textrm{H} + \textrm{H}.
\end{align}

\noindent A preliminary study of the impact of cosmic-rays on the H~\textsc{i}-to-H$_{2}$ transition suggests that a combination of UV photons and cosmic-rays could increase the total \HI column density by up to a factor of ten compared to the case with UV photons only (when examined over a reasonable parameter space with the density $n$ = 10$^{1}$--10$^{3}$~cm$^{-3}$, total column density $N$(H) = (1--4)~$\times$~10$^{21}$~cm$^{-2}$, UV radiation field $I_{\rm UV}$ = 1, and cosmic-ray ionization rate $\zeta$ = (0.5--2)~$\times$~10$^{-16}$~s$^{-1}$; Sternberg \& Bialy, in preparation). Interestingly, considering a realistic density increase by a factor of ten from the envelope to the core of a cloud reduces the impact of cosmic-rays significantly, making it almost negligible at the envelope density of $\sim$10$^{2}$~cm$^{-3}$ (comparable to the CNM densities inferred from our observations). These results suggest that detailed studies are needed to properly evaluate the impact of cosmic-rays on H$_{2}$ formation. 

Finally, the steady-state approximation in \citetalias{sternberg2014} could be invalid. A wide range of dynamical processes operate in the ISM, producing continuous flows of gas. For example, dense molecular clouds can undergo gravitational collapse on the free-fall timescale $t_{\rm ff}$ $\sim$ 1~Myr ($n$/10$^{4}$~cm$^{-3}$)$^{-1/2}$. Similarly, interstellar turbulence dissipates its kinetic energy on the eddy turnover timescale $t_{\rm turb}$ $\sim$ 1~Myr ($L$/pc)$^{1/2}$ where $L$ is the eddy size \citep[e.g.,][]{chevance22}. For the CNM with $n$~$\sim$~10$^{2}$~cm$^{-3}$ distributed within $\sim$100~pc scale molecular clouds, $t_{\rm ff}$ and $t_{\rm turb}$ are approximately 10~Myr, which are comparable to the H$_{2}$ formation timescale $t_{\rm H_{2}}$ $\sim$ (10$^{9}$/$n$)~yr \citep[e.g.,][]{hollenbach71}. These rough estimates illustrate that the CNM could be heavily perturbed over time, making the steady-state approximation for H$_{2}$ formation inappropriate \citep[e.g.,][]{valdivia2016,bialy2021}.


In summary, we conclude that the CNM must be clumpy with a small volume filling factor if H$_{2}$ formation in the solar neighborhood is determined by a combination of UV radiation, gas density, and metallicity ($\alpha G$), as the simple steady-state \citetalias{sternberg2014} model predicts. Otherwise, missing elements in the \citetalias{sternberg2014} model, such as cosmic-rays and dynamical processes, could play an important role and need to be considered more comprehensively for a better understanding of H$_{2}$ formation.

\section{Discussion} 
\label{s:discussion}

\subsection{Conditions for the Formation of Molecular Gas}
\label{s:conditions_molecular_gas}

In Section \ref{s:HI_CO_vel_compare}, we concluded that CO-bright molecular gas likely forms in CNM environments based on a close association between the CNM and CO in velocity. This conclusion is consistent with \citet{savage1977}, who measured an H$_{2}$ kinetic temperature of 45--128~K with a median of 77~K (comparable to our median spin temperature of $\sim$50~K) for the medium within 1~kpc of the Sun by analyzing \textit{Copernicus} UV absorption observations. Similarly, \citet{bellomi20} compared UV measurements of the H~\textsc{i}-to-H$_{2}$ transition to a suite of magnetohydrodynamic simulations and claimed that H$_{2}$ within 2~kpc of the Sun is built up in CNM structures with a size of $\sim$3--10~pc. 

In Sections \ref{s:individual_HI} and \ref{s:integrated_HI}, we then went further and showed that the formation of CO-bright molecular gas is favored in high column density environments where the CNM becomes colder and more abundant (which is as expected). As to the conditions for more abundant CNM, \citet{saury2014} examined a large set of hydrodynamic simulations and found that the fraction of the CNM increases with increasing initial density and decreasing turbulent velocity, implying that high densities ($\gtrsim$ 2~cm$^{-3}$) along with a moderate level of gas compression are required for the formation of the CNM and consequently molecular gas. 

Last but not least, our finding of the minimum dust extinction $A_{V}$ $\gtrsim$ 0.5--1~mag for CO detection (Section \ref{s:integrated_HI}) implies the importance of the total amount of gas available for the formation of CO-bright molecular gas. All things considered, we conclude that accumulating a large amount of atomic gas by dynamical processes (e.g., spiral arms, supernova explosions, and expanding shells that lead to gas compression) and building up cold and dense structures would be a key step in the formation of molecular gas. This conclusion is consistent with what previous observational and theoretical studies suggested \citep[e.g.,][]{mckee2007,chevance22}.  

\subsection{H~\textsc{i} Absorption as a Diagnostic Tool for Probing the Formation and Evolution of Molecular Clouds} 
\label{s:HI_abs}

While a range of dynamical processes (e.g., spiral arms on large scales and expanding shells and bubbles on small scales) certainly play a role in the formation and evolution of molecular clouds \citep[e.g.,][]{mckee2007}, it remains unclear exactly how they operate and which process dominates. As an accessible tracer of atomic gas, the raw ingredient of molecular clouds, \HI emission has been frequently employed to address this issue. For example, \citet{fukui2009} found that the \HI mass of molecular clouds in the Large Magellanic Cloud (LMC) increases with evolutionary stages of star formation and estimated an \HI accretion rate of 0.05~$M_{\sun}$~yr$^{-1}$ based on \HI line widths. In addition, \citet{tahani22} examined the difference in velocity between \HI and CO emission for the Perseus molecular cloud and interpreted a systematic positive offset of $\varv_{\rm CO} - \varv_{\rm H~\textsc{i}}$ as an indication of the formation of molecular gas behind compressed \HI bubbles. In comparison to \HI emission that traces all three phases of neutral atomic gas and often exhibits broad and featureless spectra, \HI absorption mostly arises from the CNM (which is more closely associated with molecular gas) and shows relatively narrow and structured spectra, making it an excellent probe for the formation and evolution of molecular clouds. As an example, we showed that there is an absolute velocity difference of 0.01--4.3~km~s$^{-1}$ with a median of 0.4~km~s$^{-1}$ between the CNM and CO (Section \ref{s:HI_CO_vel_compare}). Considering that the CNM has a comparable velocity difference of 0.06--2.64~km~s$^{-1}$ with a median of 0.5~km~s$^{-1}$ with OH absorption as well (estimated from 10 of our 58 GNOMES LOSs where OH absorption is clearly detected; \citealt{petzler2023}), this velocity difference between the CNM and CO is most likely real (not due to different beam sizes) and could suggest that the CNM and CO-bright molecular gas are in slightly different regions and/or dynamically decoupled \citep[e.g.,][]{soler2019,beuther20, wang2020}. Unfortunately, our GNOMES LOSs are scattered over a relatively large area of sky and cannot provide insights into how the CNM is distributed and moves about in individual molecular clouds. 

The power of \HI absorption as a diagnostic tool for probing the formation and evolution of molecular clouds could be harnessed by getting a large number of \HI absorption spectra over a fine grid of continuum sources located behind molecular clouds. These spectra could then be analyzed with synthetic \HI data from numerical simulations of multiphase gas \citep[e.g.,][]{kim2017,seifried2022}, enabling us to examine the signature of the formation and evolution process imprinted on the properties of the CNM (e.g., kinematics and distribution). Such observations as we propose will be routinely carried out by next generation radio telescopes with a wide field of view, including the Square Kilometre Array (SKA), as \citet{dickey2022} recently demonstrated with the Australian Square Kilometre Array Pathfinder (ASKAP).

\section{Summary} 
\label{s:summary}

This paper presents a detailed study on the formation of molecular gas in the solar neighborhood. To probe the conditions for the H~\textsc{i}-to-H$_{2}$ transition, \HI emission and absorption spectra toward 58 LOSs at $b$~<~$-5{\degree}$ (Arecibo) were analyzed along with CO(1--0) and dust data (TRAO, PMO, and \textit{Planck}). These multi-wavelength data were compared to the one-dimensional steady-state H$_{2}$ formation model of \citet{sternberg2014} as well to provide insights into the fundamental principles of the H~\textsc{i}-to-H$_{2}$ transition. Our key results are as follows. 

\begin{enumerate}
    \item Among the observed 58 sources, 19 sources show clear CO(1--0) emission, suggesting a detection rate of 33\% at a rms level of 0.1~K (angular and spectral resolutions of 48$''$ and 0.32~km~s$^{-1}$, respectively). 
    \item The decomposition of gas into atomic and molecular phases shows that the observed LOSs are mostly H~\textsc{i}-dominated. In addition, the CO-dark H$_{2}$, not the optically thick H~\textsc{i}, is found as a major constituent of the dark gas in the solar neighborhood. 
    \item The CNM shows a systematically smaller velocity difference from CO emission than the WNM. When CO-closest components are considered, a median value of the absolute velocity difference between the CNM and CO is 0.4~km~s$^{-1}$, as opposed to 1.7~km~s$^{-1}$ for the WNM and CO. This implies that the CNM is kinematically (and spatially if we take velocity as a proxy for position) more closely associated with CO. 
    \item When CO-associated components (ones within CO velocity ranges) are considered, the CNM and WNM exhibit distinctive properties. Namely, the CO-associated components have the spin temperature $T_{\rm s}$~$<$~200~K, optical depth $\tau_{\rm CNM}$~$>$~0.1, and column densities $N_{\rm CNM}$~$>$~2~$\times$~10$^{19}$~cm$^{-2}$ and $N_{\rm WNM}$~$>$~2~$\times$~10$^{20}$~cm$^{-2}$. This suggests that CO-bright molecular gas forms in environments where individual CNM components evolve toward colder temperature and higher column density.
    \item The CO-associated components have higher total column densities ($V$-band dust extinction $A_{V}$ $\gtrsim$ 0.5~mag) and CNM fractions (median of 0.6) than those outside CO emission, indicating that high column density environments where the CNM becomes more abundant facilitate the formation of CO-bright molecular gas. 
    \item A comparison with the prediction from \citet{sternberg2014} infers that the CNM must be clumpy with a small volume filling factor. An alternative possibility would be that missing ingredients in the model, such as cosmic-rays and dynamical processes, play an important role in the H~\textsc{i}-to-H$_{2}$ transition in the solar neighborhood. 
\end{enumerate}

\begin{acknowledgments}
We thank Chang-Goo Kim, Jeong-Gyu Kim, and Amiel Sternberg for insightful discussions and the referee for helpful comments that improved this work. In addition, we acknowledge Interstellar Institute's program ``With Two Eyes'' and the Paris-Saclay University's Institut Pascal for hosting discussions that nourished the development of the ideas behind this work. Part of the CO data were obtained with the 13.7 m telescope of the Qinghai Station of Purple Mountain Observatory, and we appreciate the help from Dr. Sun, Jixian and all the staff members of the PMO-13.7m telescope. S.B. thanks the Physics department at the Technion, Israel, and the Center for Theory and Computations (CTC) at the University of Maryland, College Park, for financial support. B.B. acknowledges support from NSF grant AST-2009679 and NASA grant No. 80NSSC20K0500 and is grateful for the generous support of the David and Lucile Packard Foundation and the Alfred P. Sloan Foundation. D.L. acknowledges support from the National Natural Science Foundation of China project NSFC11988101. D.R.R. acknowledges support by the NSF through award SOSPA6-023 from the NRAO. S.S. acknowledges the support by the National Aeronautics and Space Administration under Grant No. 4200766703 and the University of Wisconsin-Madison Office of the Vice Chancellor for Research and Graduate Education with funding from the Wisconsin Alumni Research Foundation.
\end{acknowledgments}

\appendix
\restartappendixnumbering

\section{Comparison between the GNOMES \HI and CO Spectra}
\label{s:appendix1}

\begin{figure*}
    \centering
    \figurenum{A1}
    \includegraphics[scale=0.25]{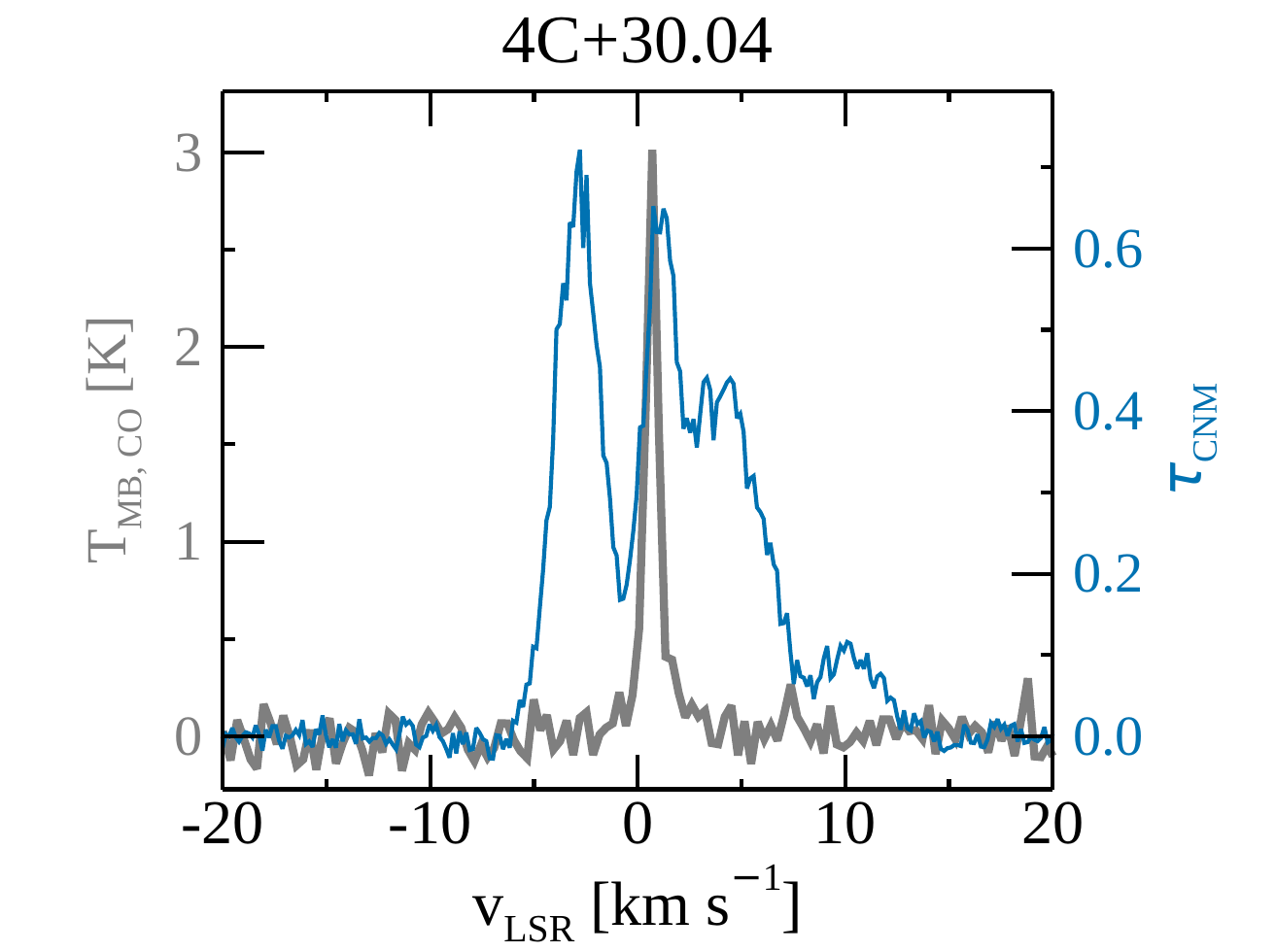}\vspace{0.1cm}\hspace{0.2cm} 
    \includegraphics[scale=0.25]{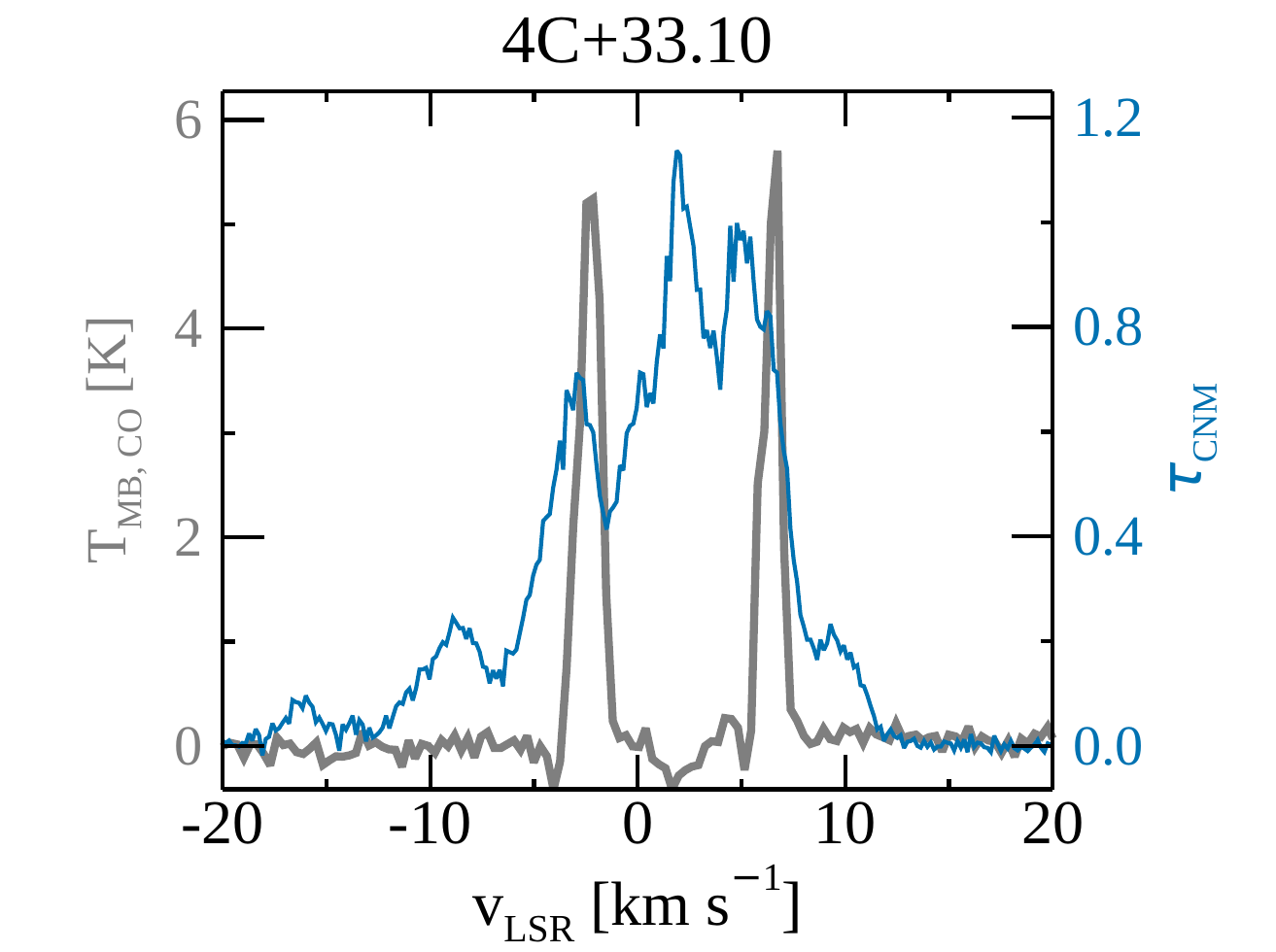}\vspace{0.1cm}\hspace{0.2cm}
    \includegraphics[scale=0.25]{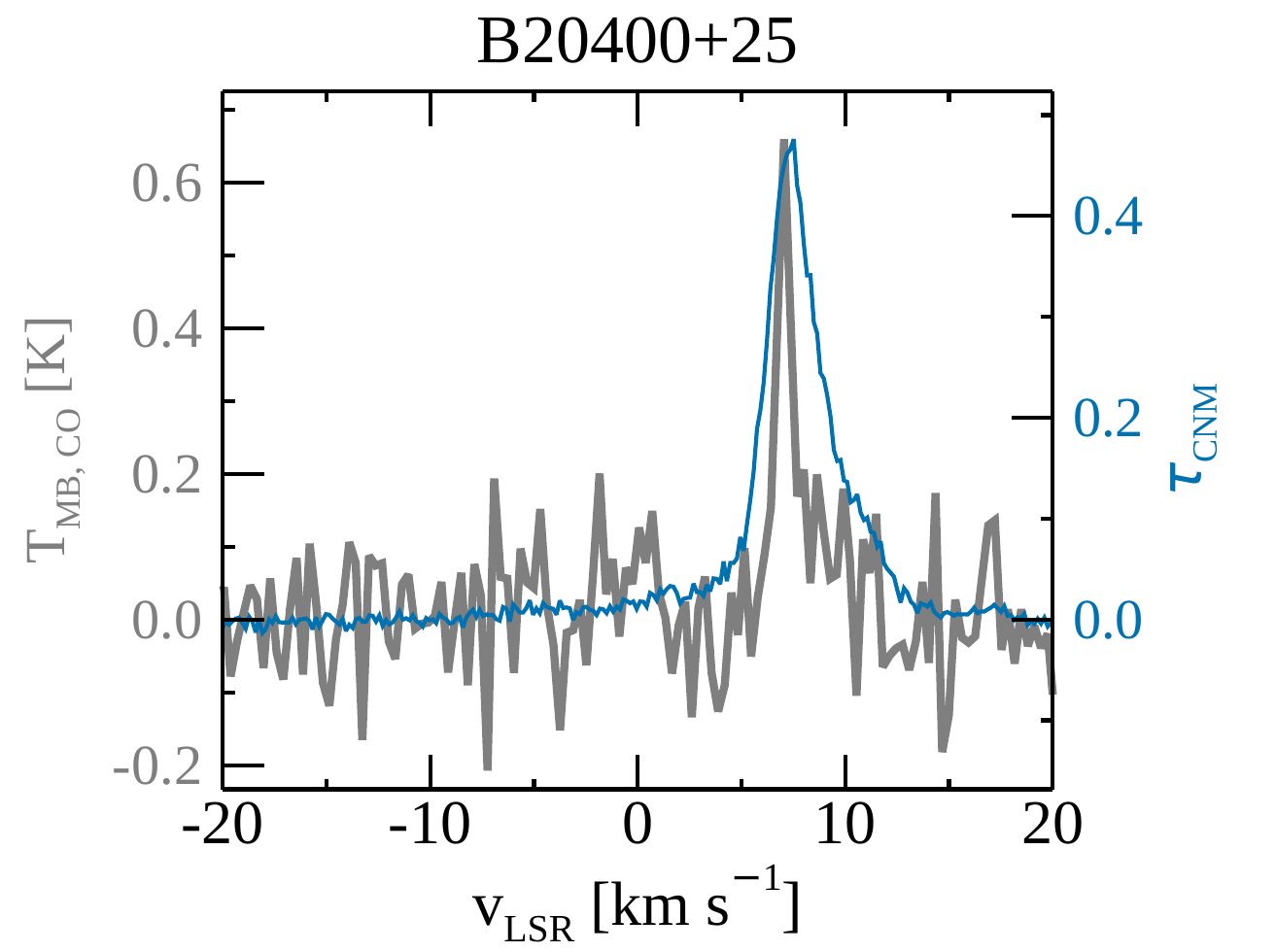}
    \includegraphics[scale=0.25]{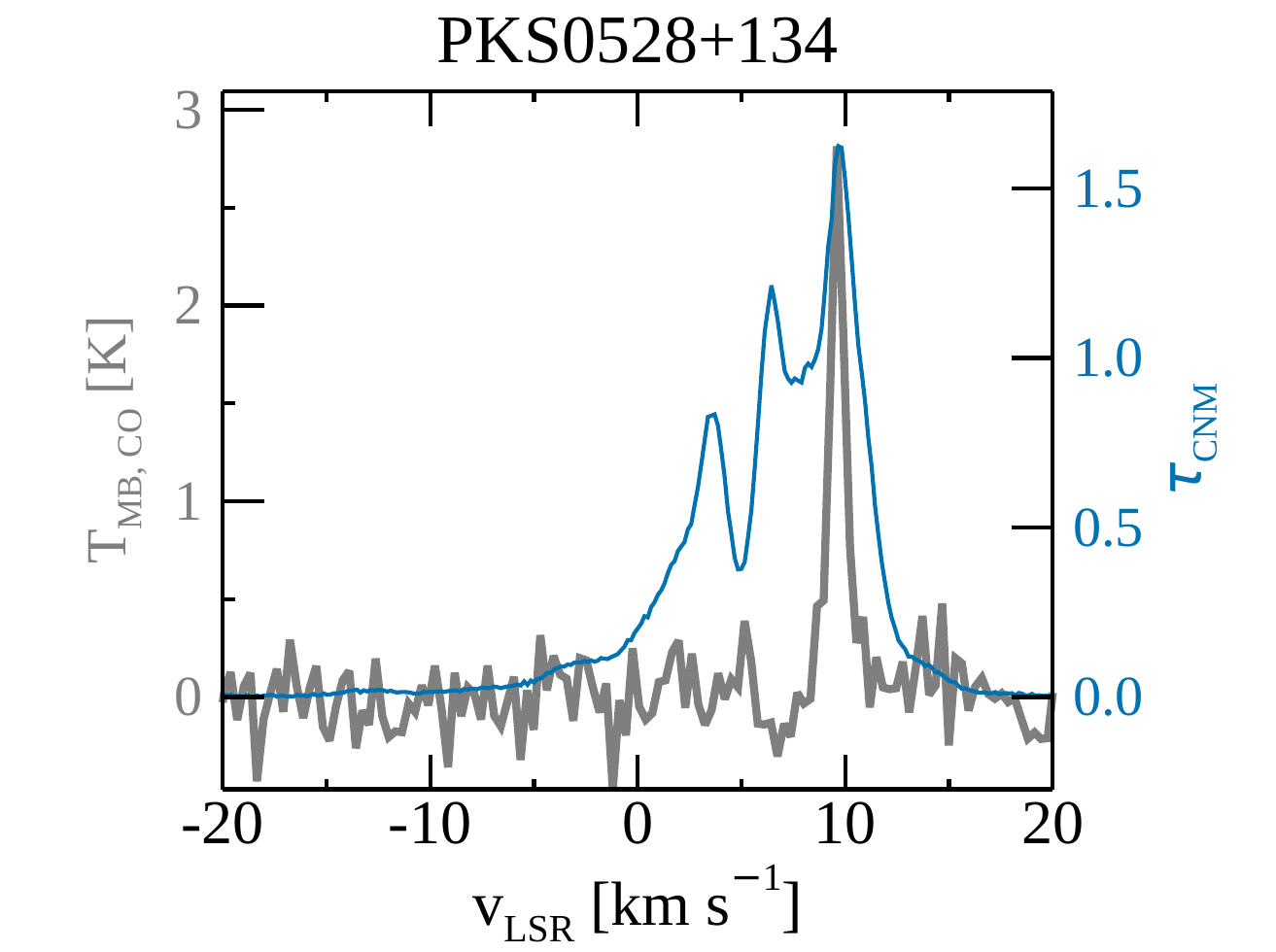}
    \caption{\label{f:GNOMES_HI_CO} Comparison between the \HI absorption (thin blue) and CO emission (thick gray) spectra toward the 19 CO-detected LOSs. The spectra are aligned at peaks for ease of comparison.} 
\end{figure*}

In Figure \ref{f:GNOMES_HI_CO}, we compare the \HI absorption and CO emission spectra toward the 19 CO-detected LOSs. The measured optical depth ($\tau_{\rm CNM}$) and main-beam brightness temperature ($T_{\rm MB, CO}$) are aligned at peaks for ease of comparison and are presented in blue and gray, respectively.   

\begin{figure*}
    \centering
    \includegraphics[scale=0.25]{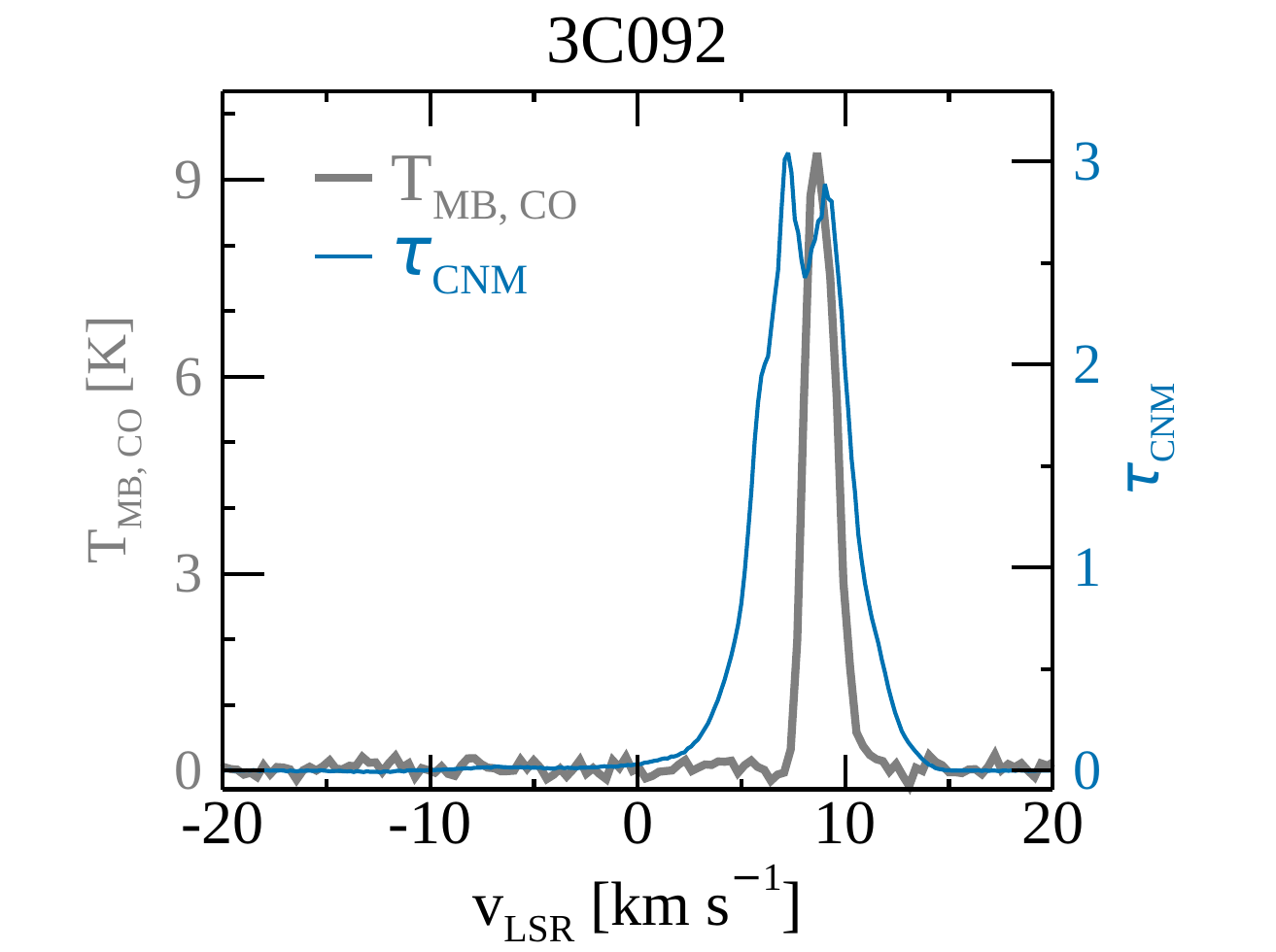}\vspace{0.1cm}\hspace{0.2cm}
    \includegraphics[scale=0.25]{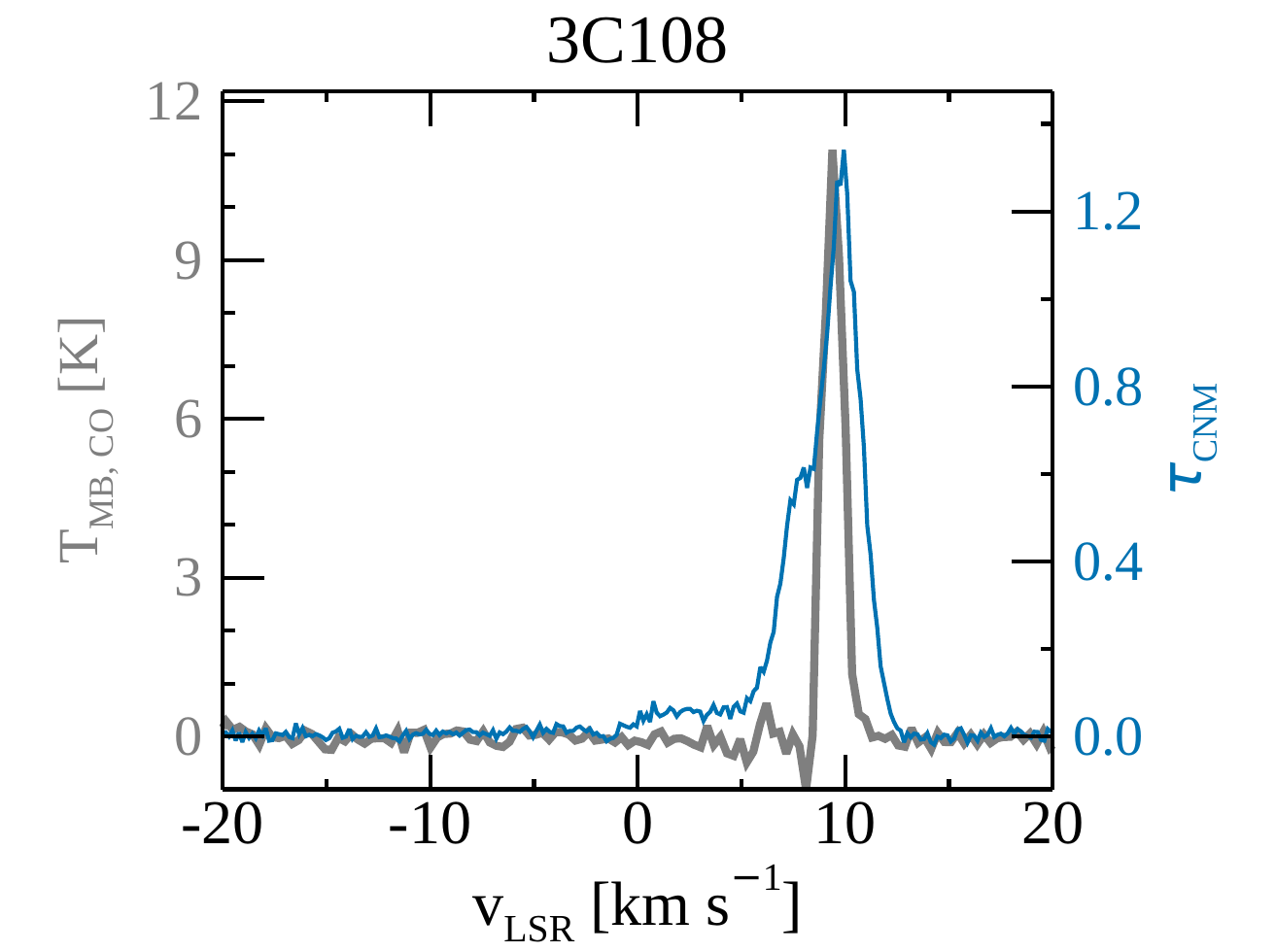}\vspace{0.1cm}\hspace{0.2cm} 
    \includegraphics[scale=0.25]{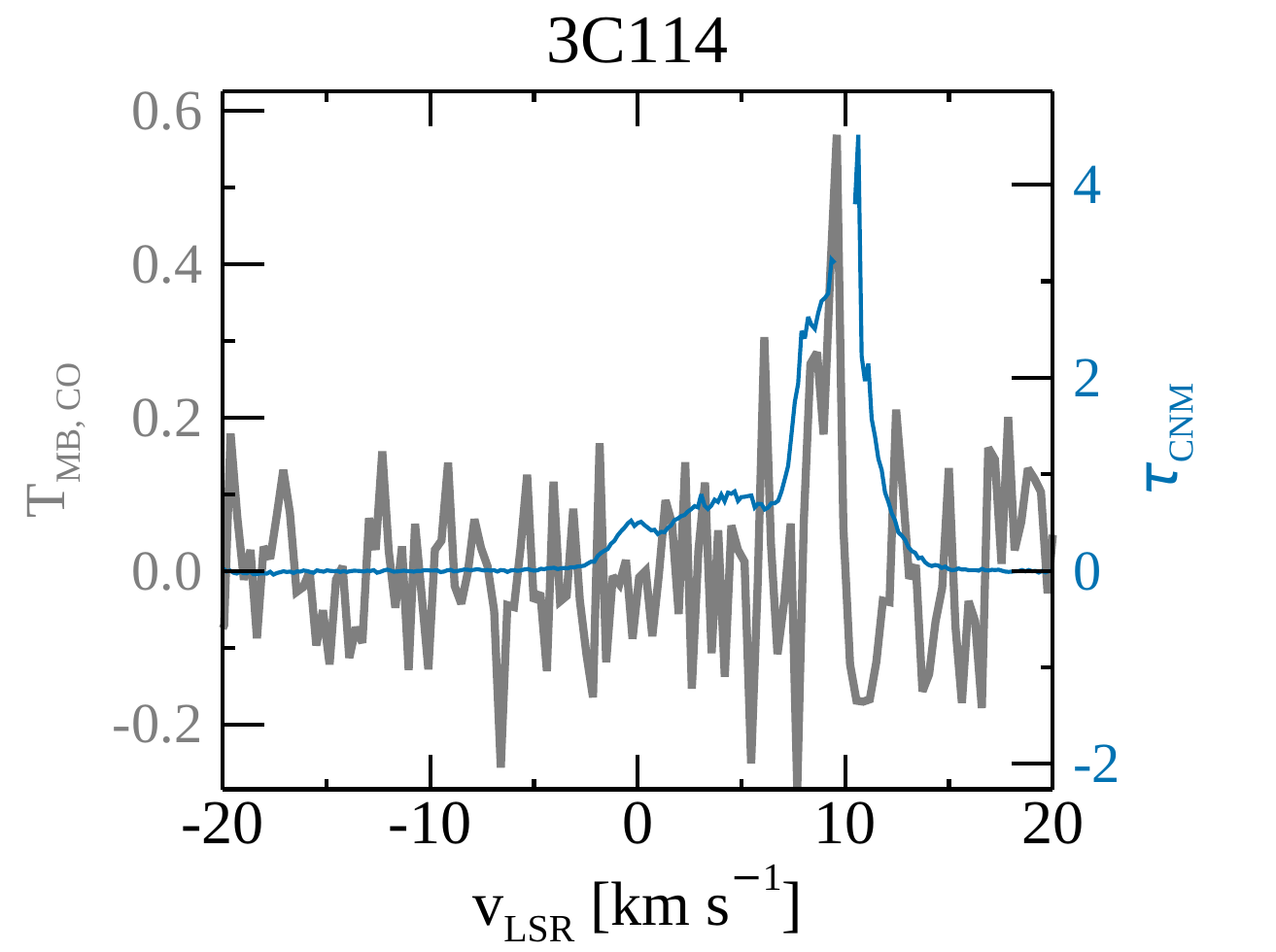}
    \includegraphics[scale=0.25]{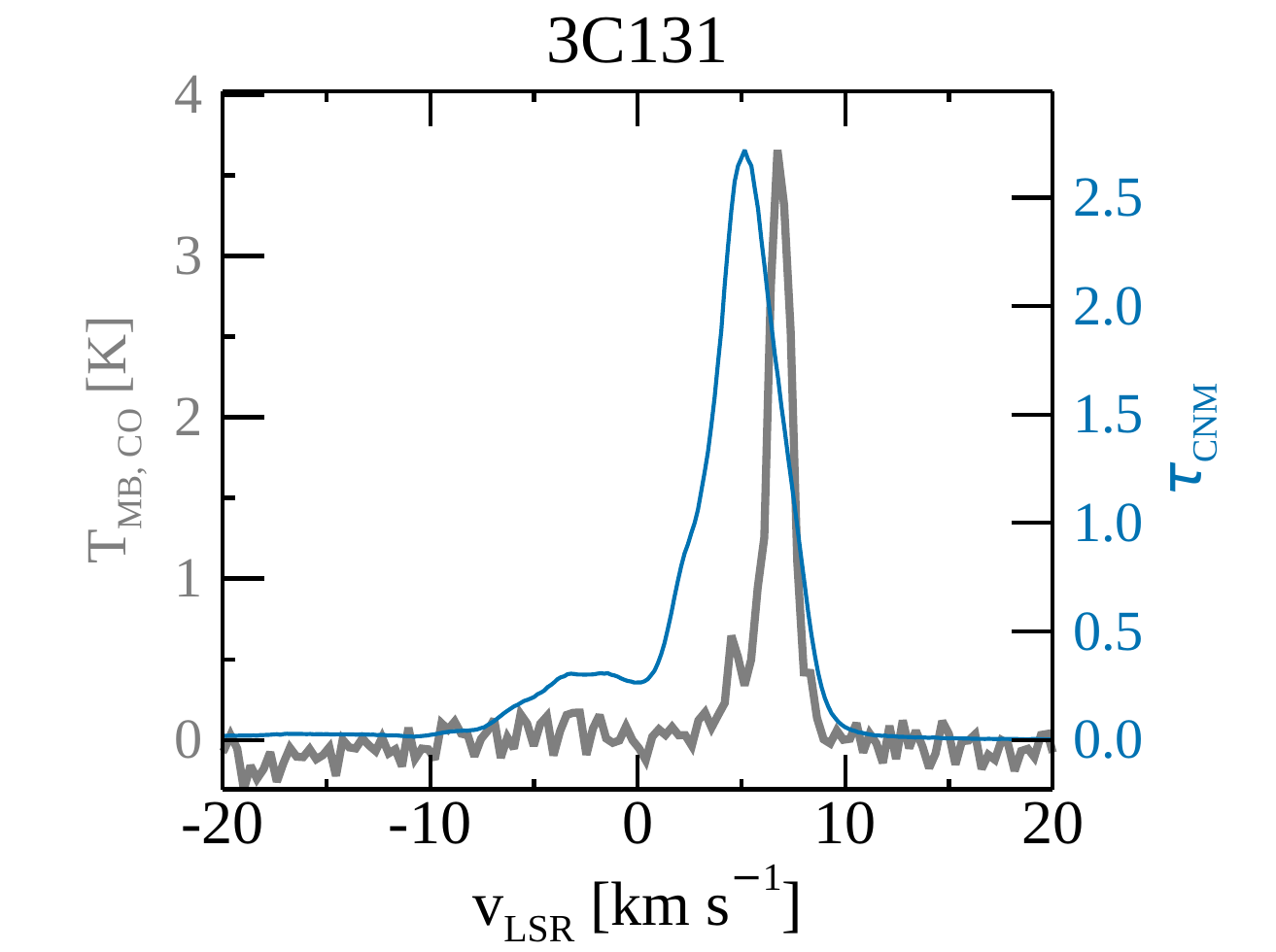}\vspace{0.1cm}\hspace{0.2cm} 
    \includegraphics[scale=0.25]{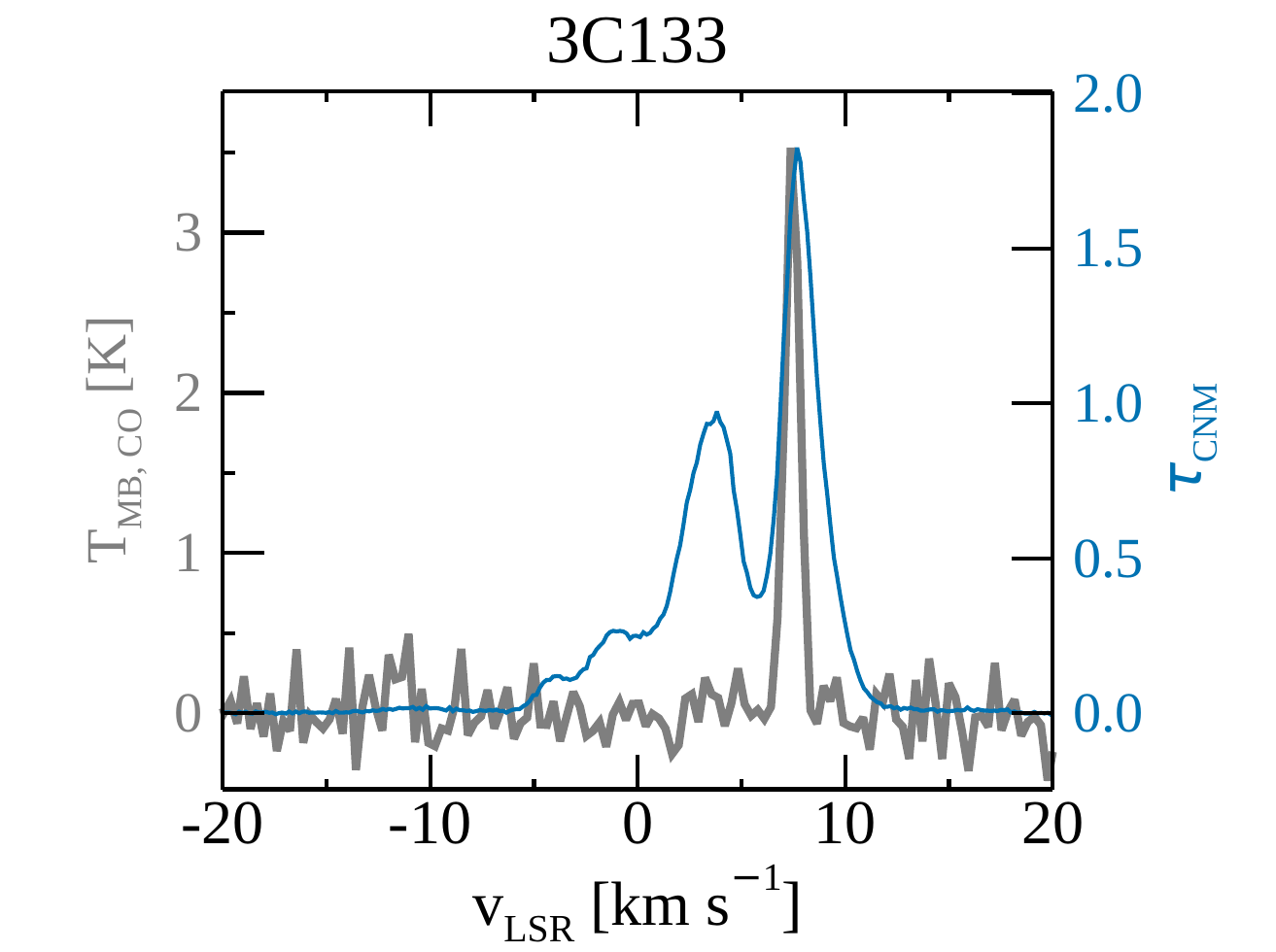}\vspace{0.1cm}\hspace{0.2cm}
    \includegraphics[scale=0.25]{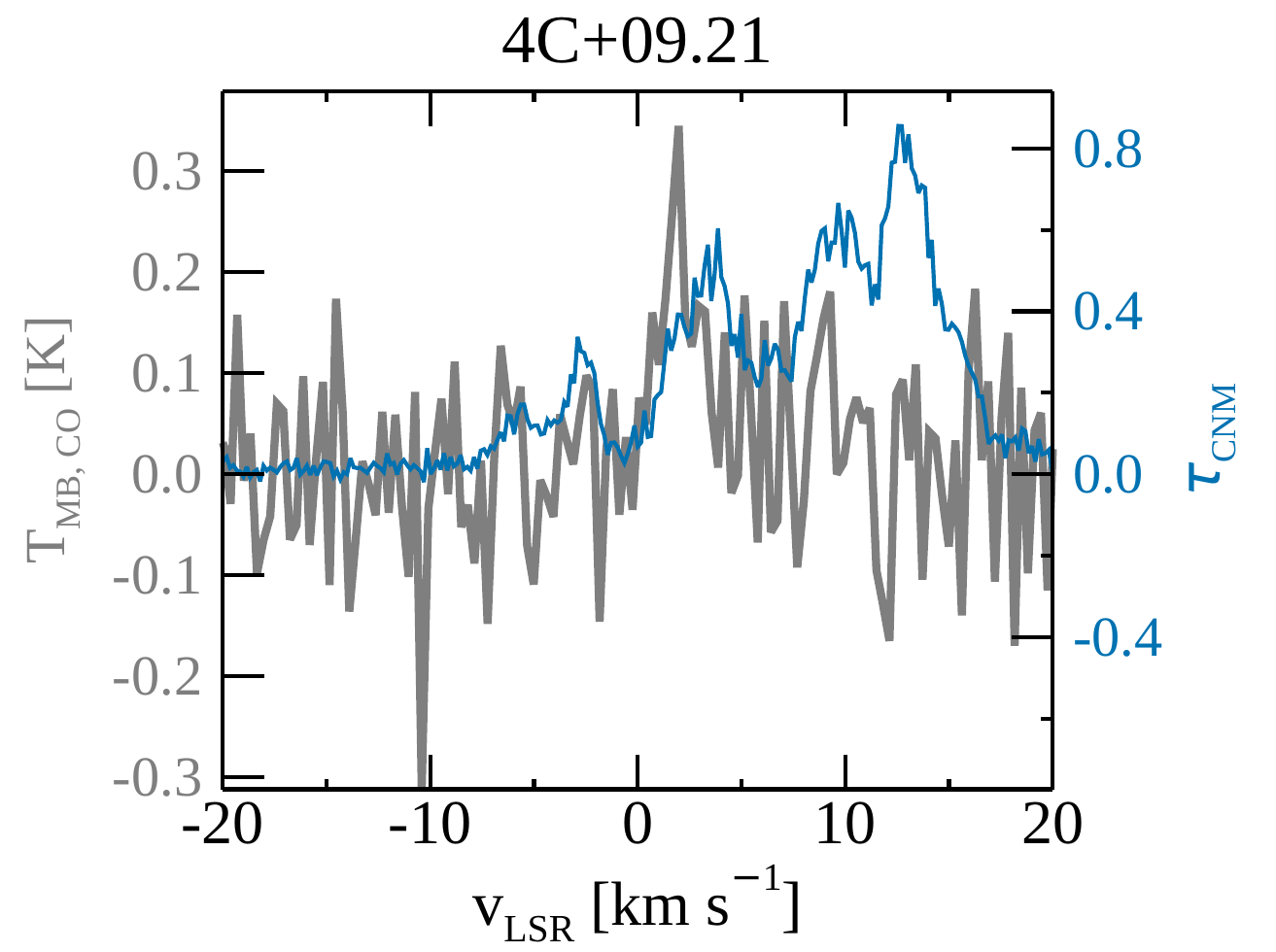}
    \includegraphics[scale=0.25]{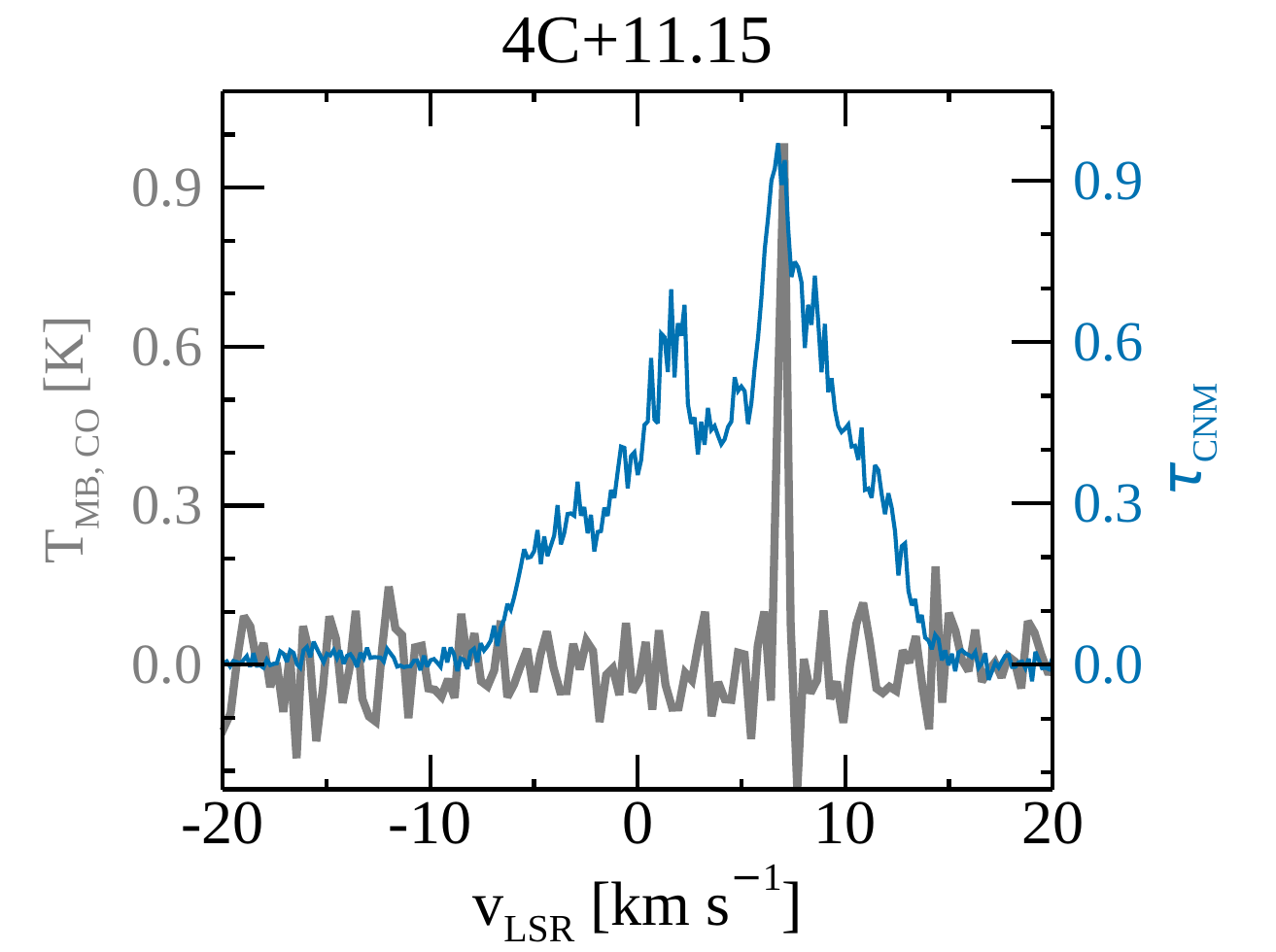}\vspace{0.1cm}\hspace{0.2cm}
    \includegraphics[scale=0.25]{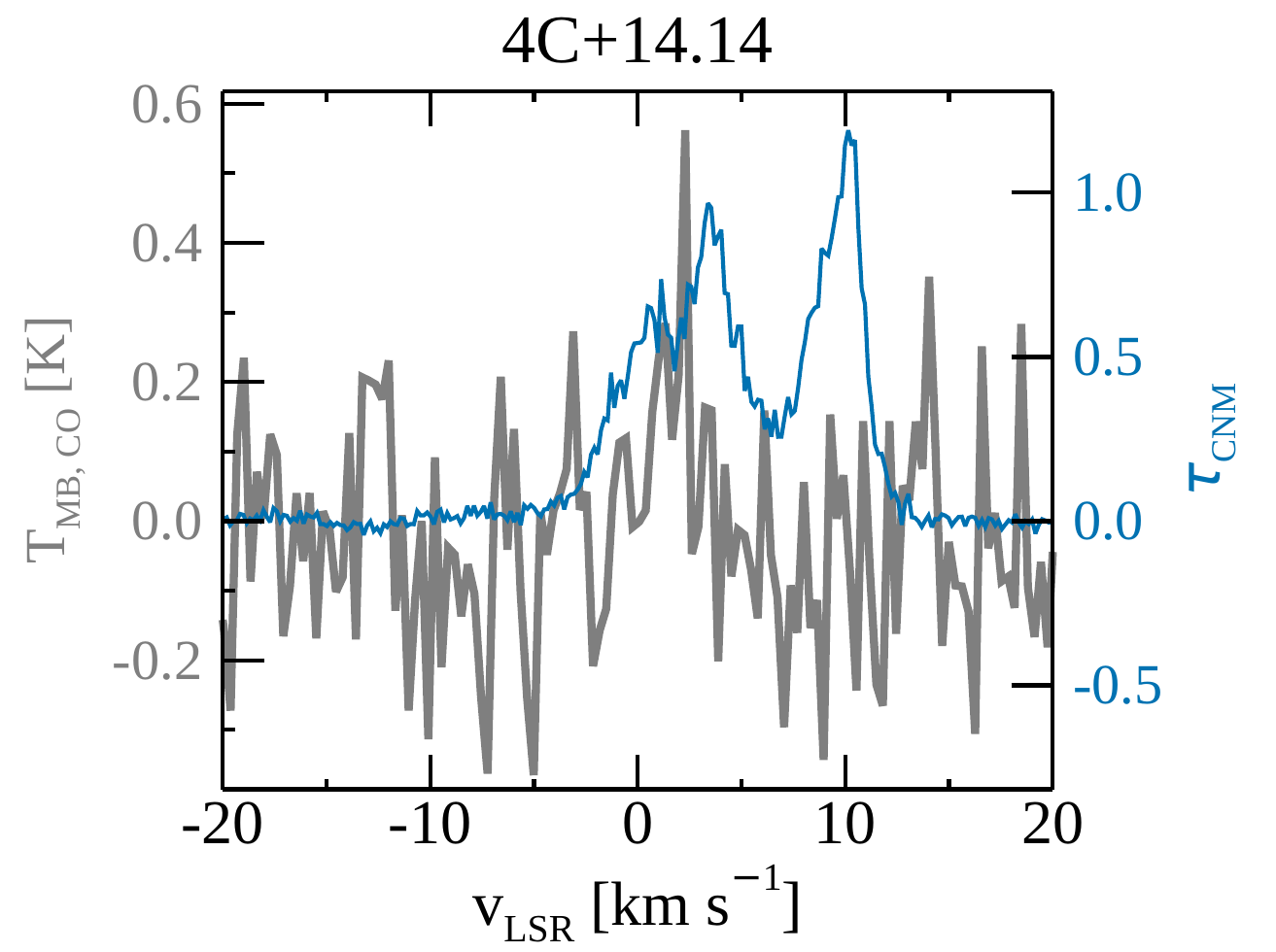}\vspace{0.1cm}\hspace{0.2cm} 
    \includegraphics[scale=0.25]{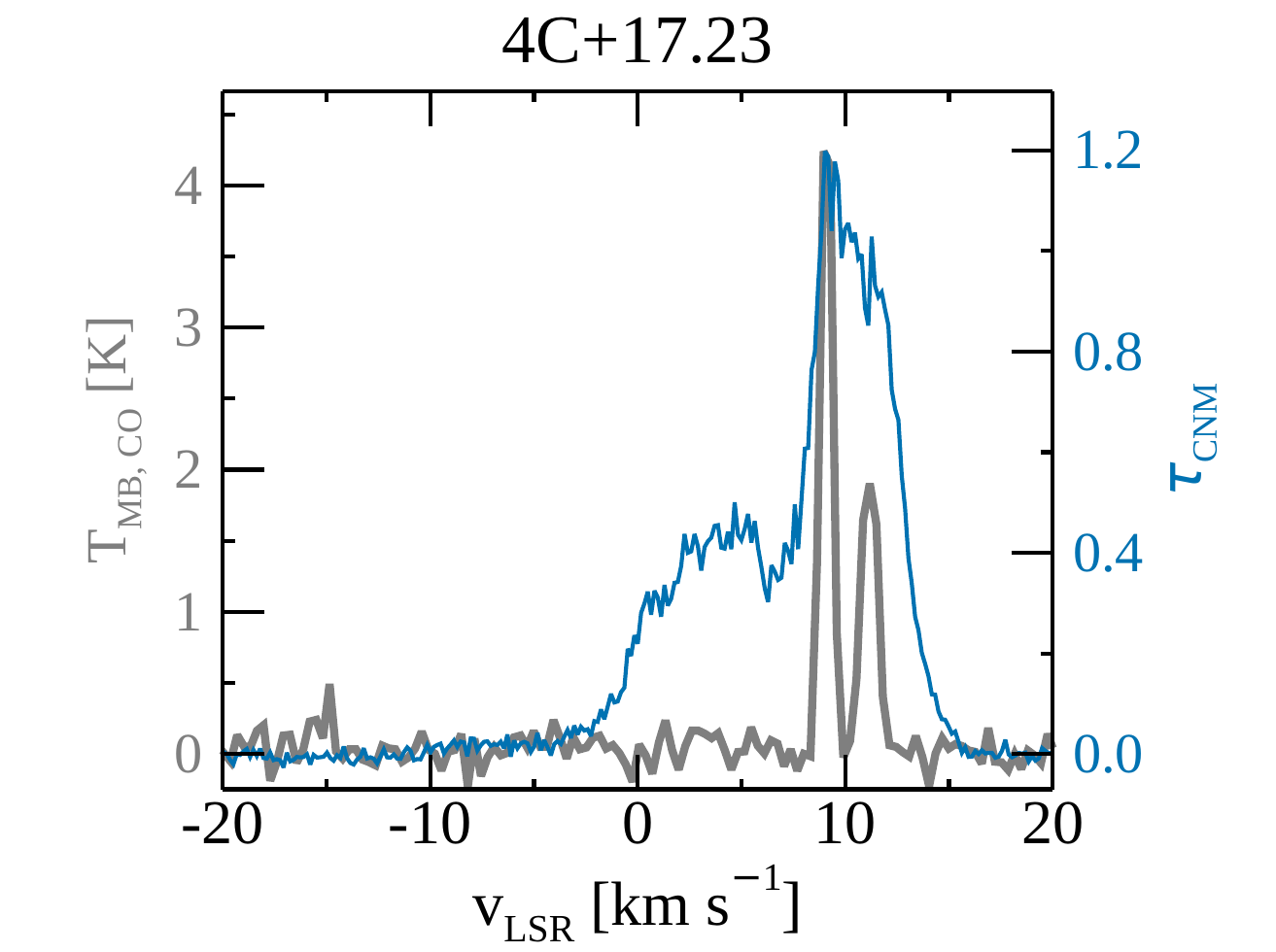}
    \includegraphics[scale=0.25]{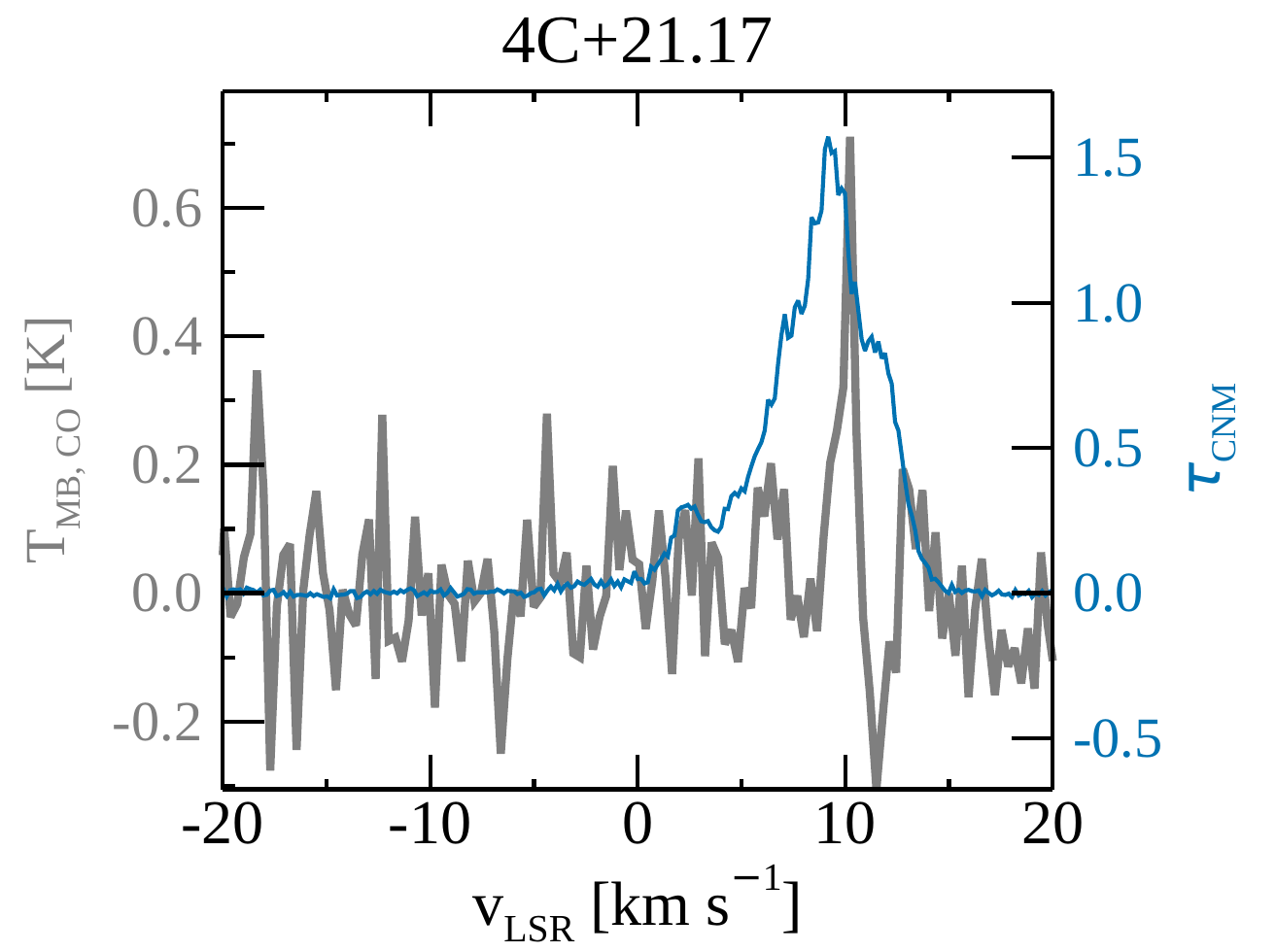}\vspace{0.1cm}\hspace{0.2cm}
    \includegraphics[scale=0.25]{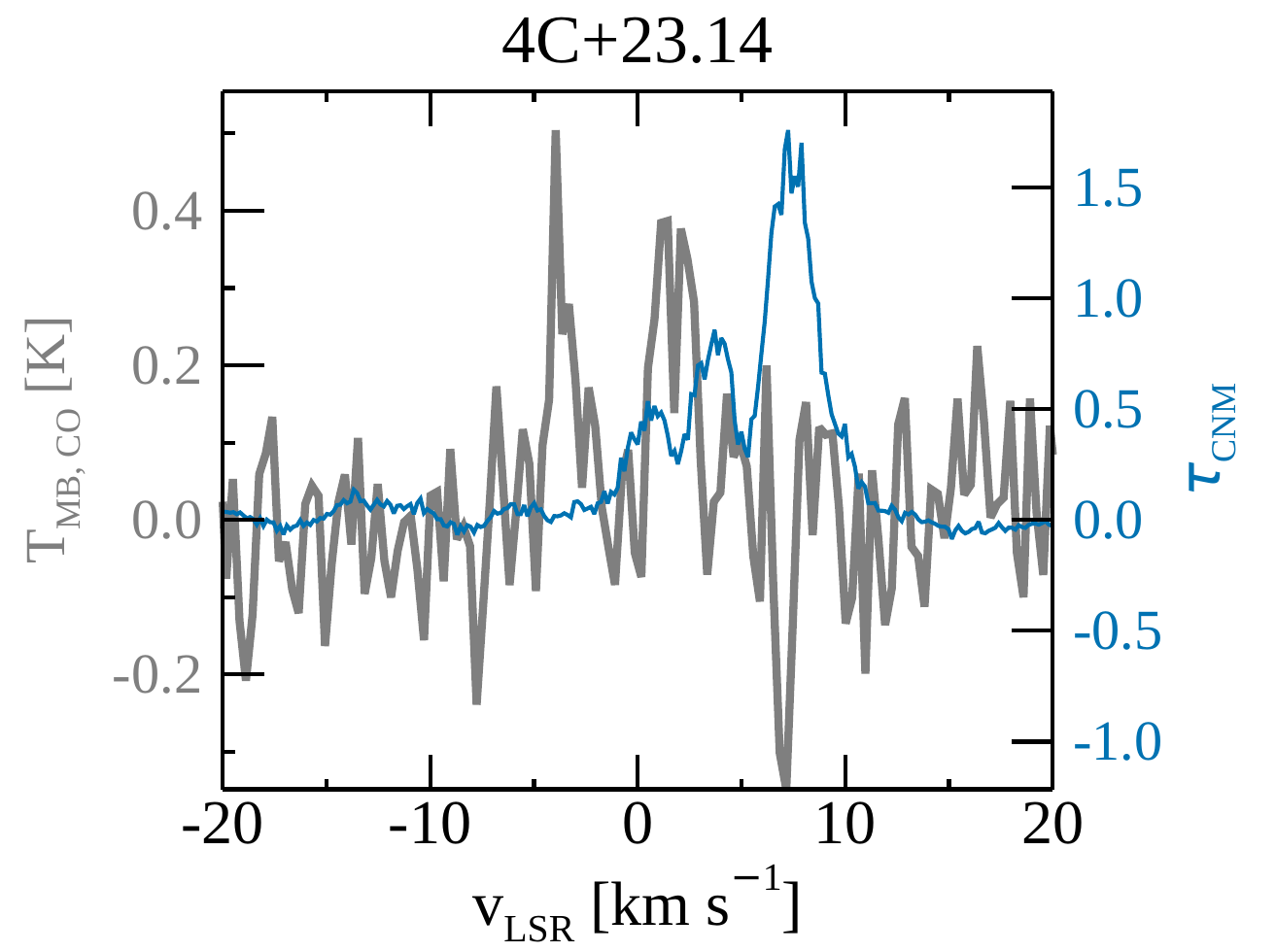}\vspace{0.1cm}\hspace{0.2cm}
    \includegraphics[scale=0.25]{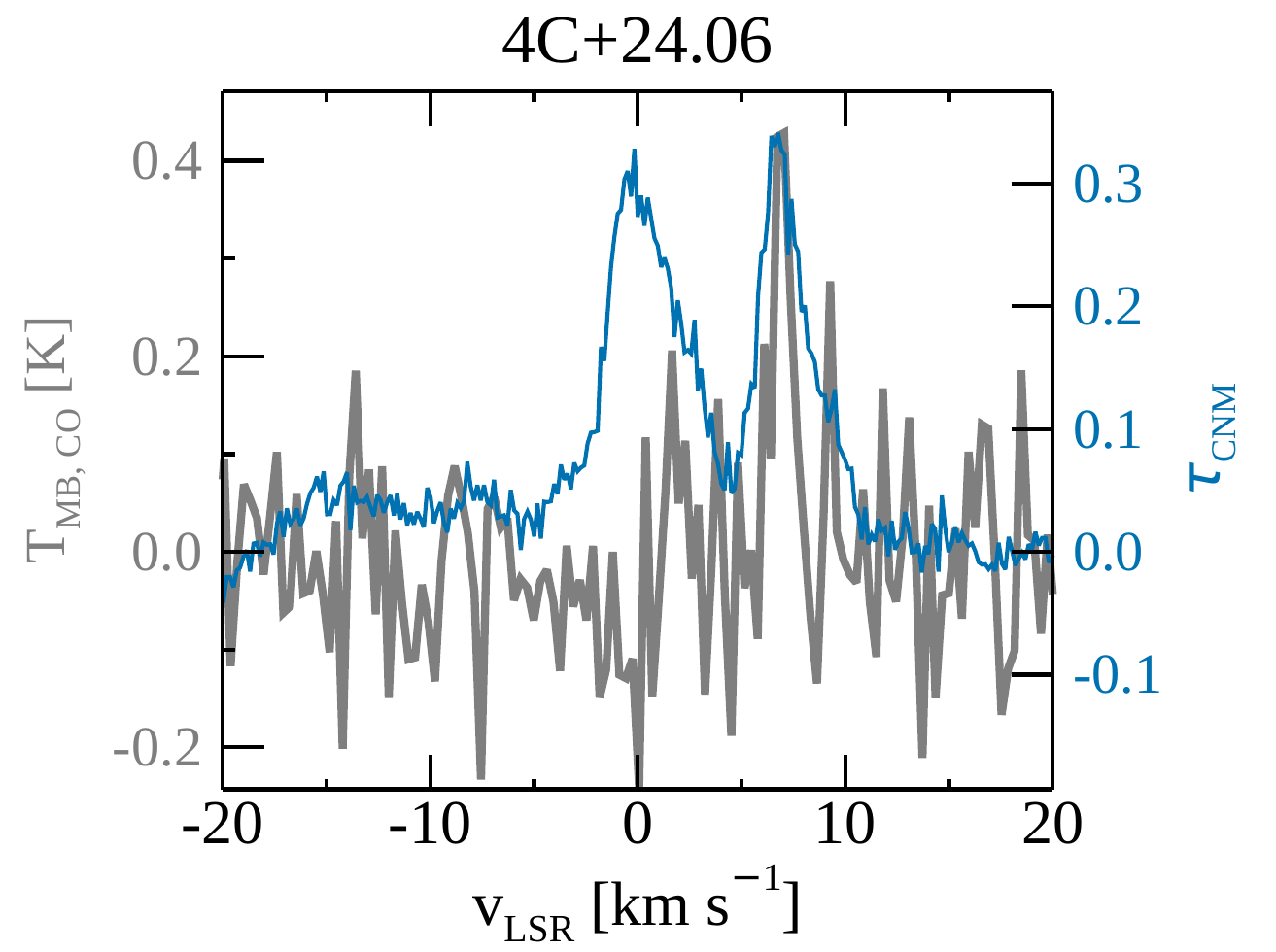}
    \includegraphics[scale=0.25]{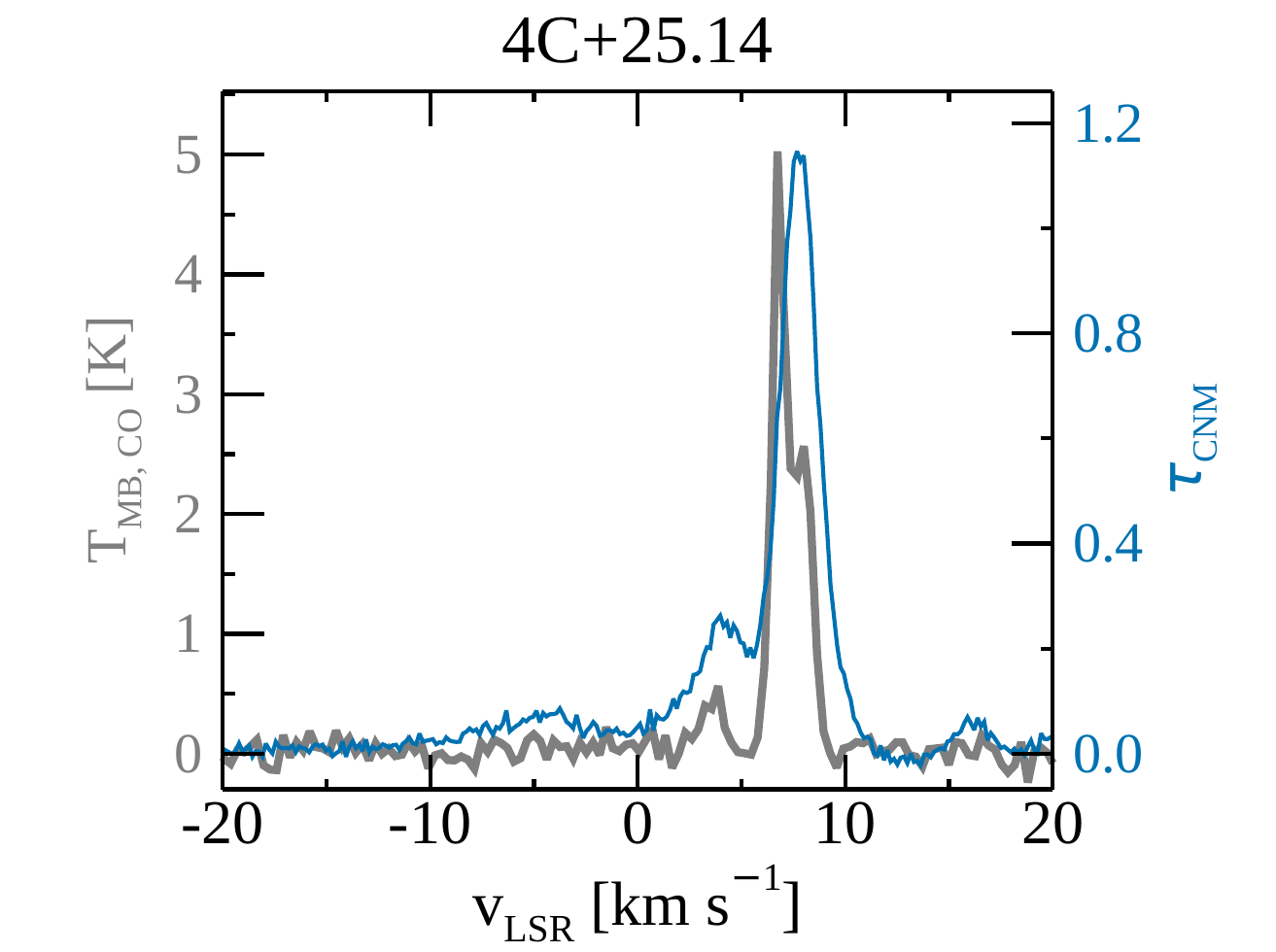}\vspace{0.1cm}\hspace{0.2cm}
    \includegraphics[scale=0.25]{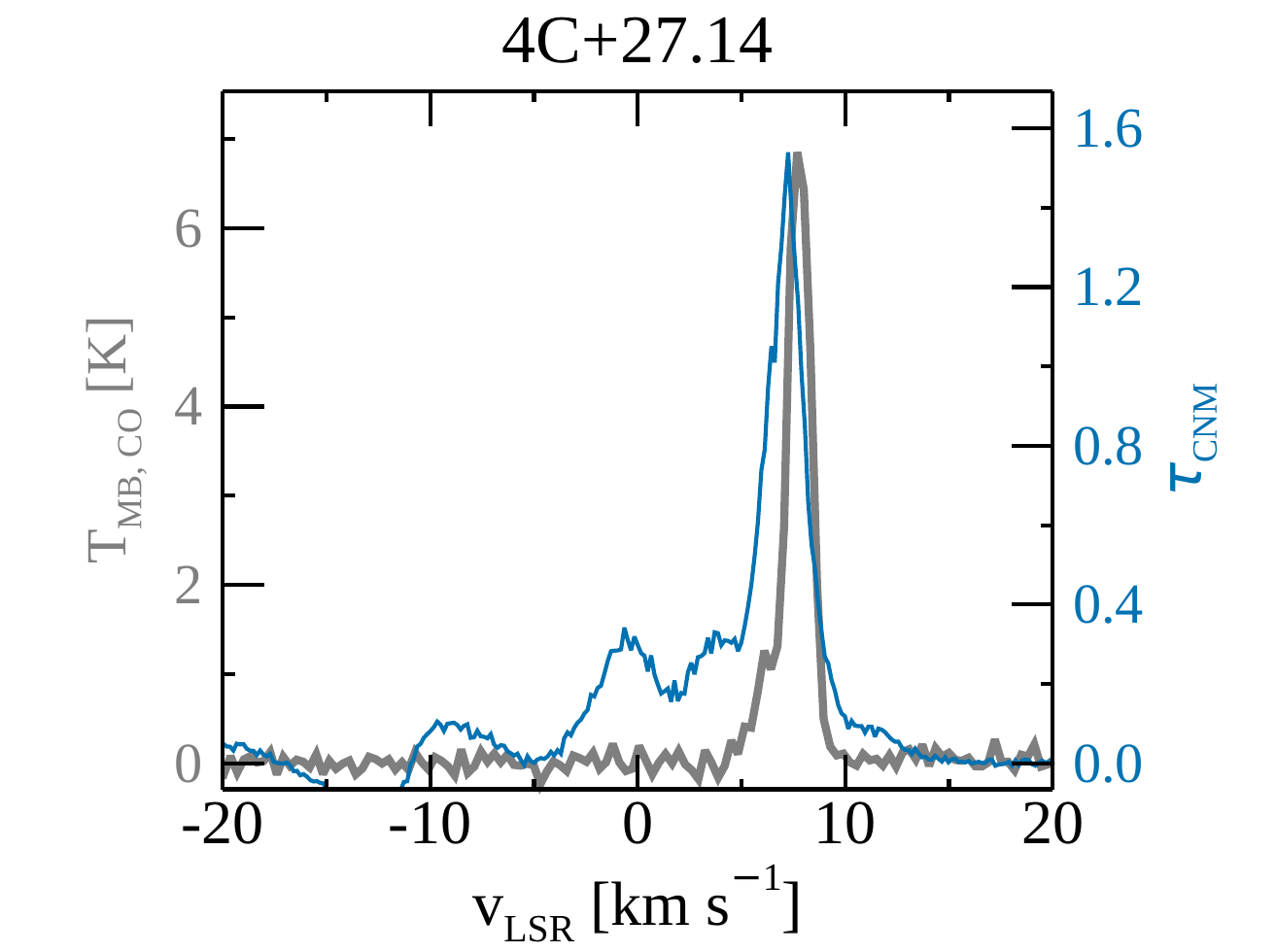}\vspace{0.1cm}\hspace{0.2cm}
    \includegraphics[scale=0.25]{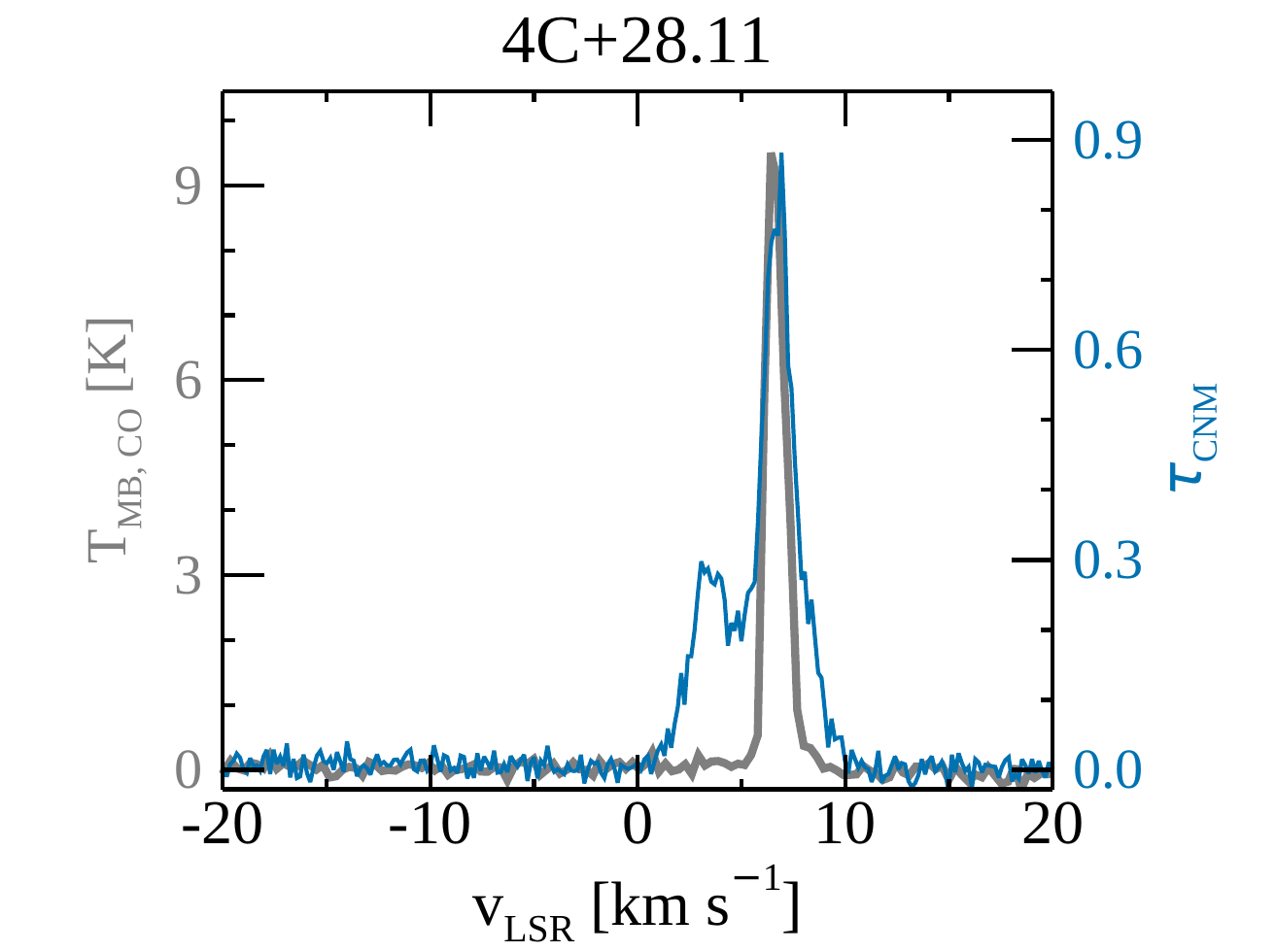}
    \caption{(\textit{continued})}\vspace{0.3Cm}
\end{figure*}

\section{Comparison to the ALMA-NOEMA/21-SPONGE Survey}
\label{s:SPONGE_comparison} 

Recently, \citet{rybarczyk2022} observed 20 LOSs from the 21-SPONGE survey 
\citep[21~cm Spectral Line Observations of Neutral Gas with the Karl G. Jansky Very Large Array;][]{murray2015,murray2018a} using the Atacama Large Millimeter/submillimeter Array (ALMA) and the Northern Extended Millimeter Array (NOEMA) and obtained HCO$^{+}$, HCN, HNC, and C$_{2}$H absorption spectra. By comparing the observed molecular species with the existing \HI and dust properties, the authors found that molecular absorption is clearly detected toward LOSs with $A_{V}$ $\gtrsim$ 0.25~mag. 
In addition, they revealed that molecular gas is preferentially associated with cold ($T_{\rm s}$ $<$ 80~K) and optically thick ($\tau_{\rm CNM}$ $>$ 0.1) CNM structures. 

\begin{figure*}
    \centering
    \includegraphics[scale=0.35]{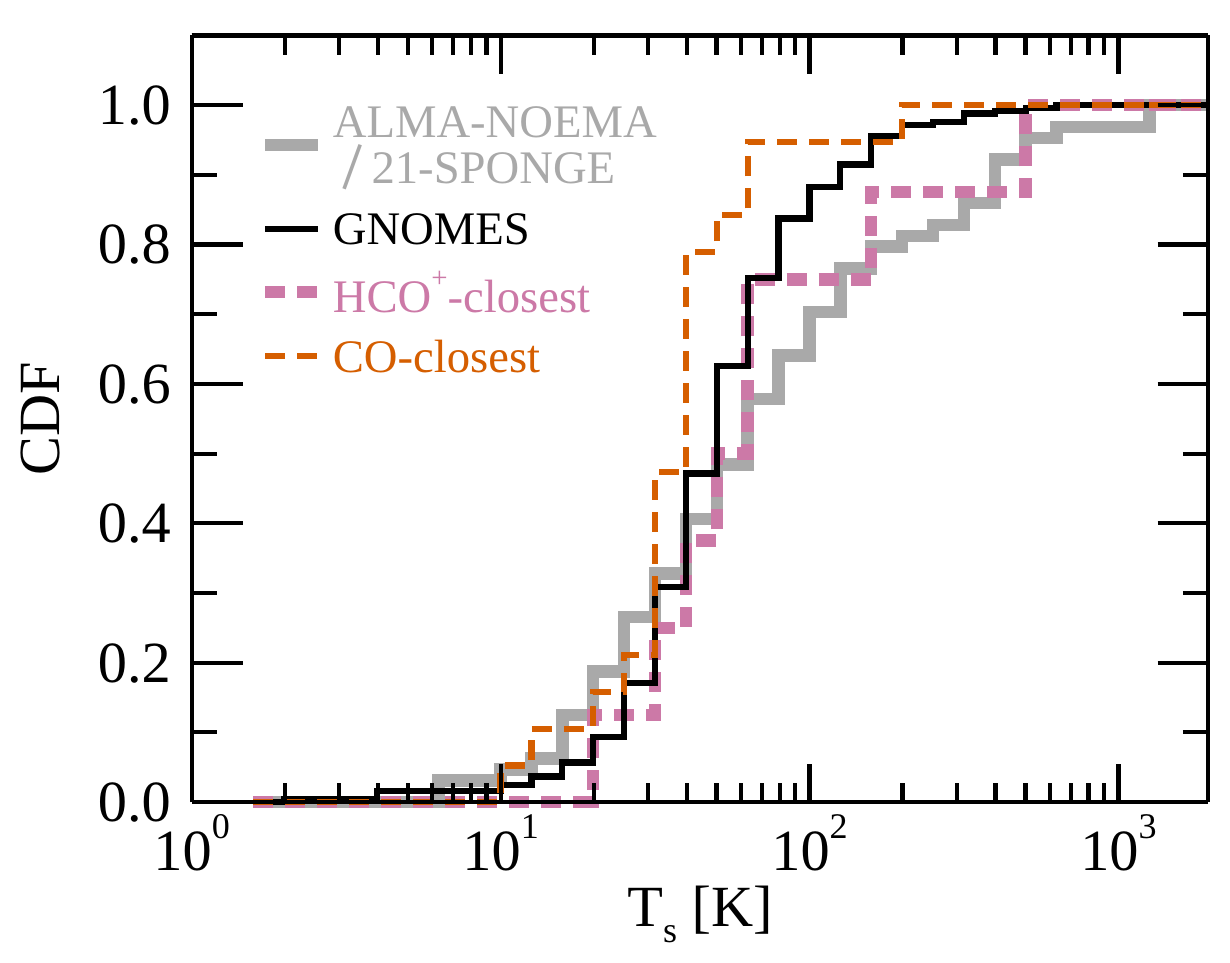}\vspace{0.3Cm}\hspace{0.3Cm}
    \includegraphics[scale=0.35]{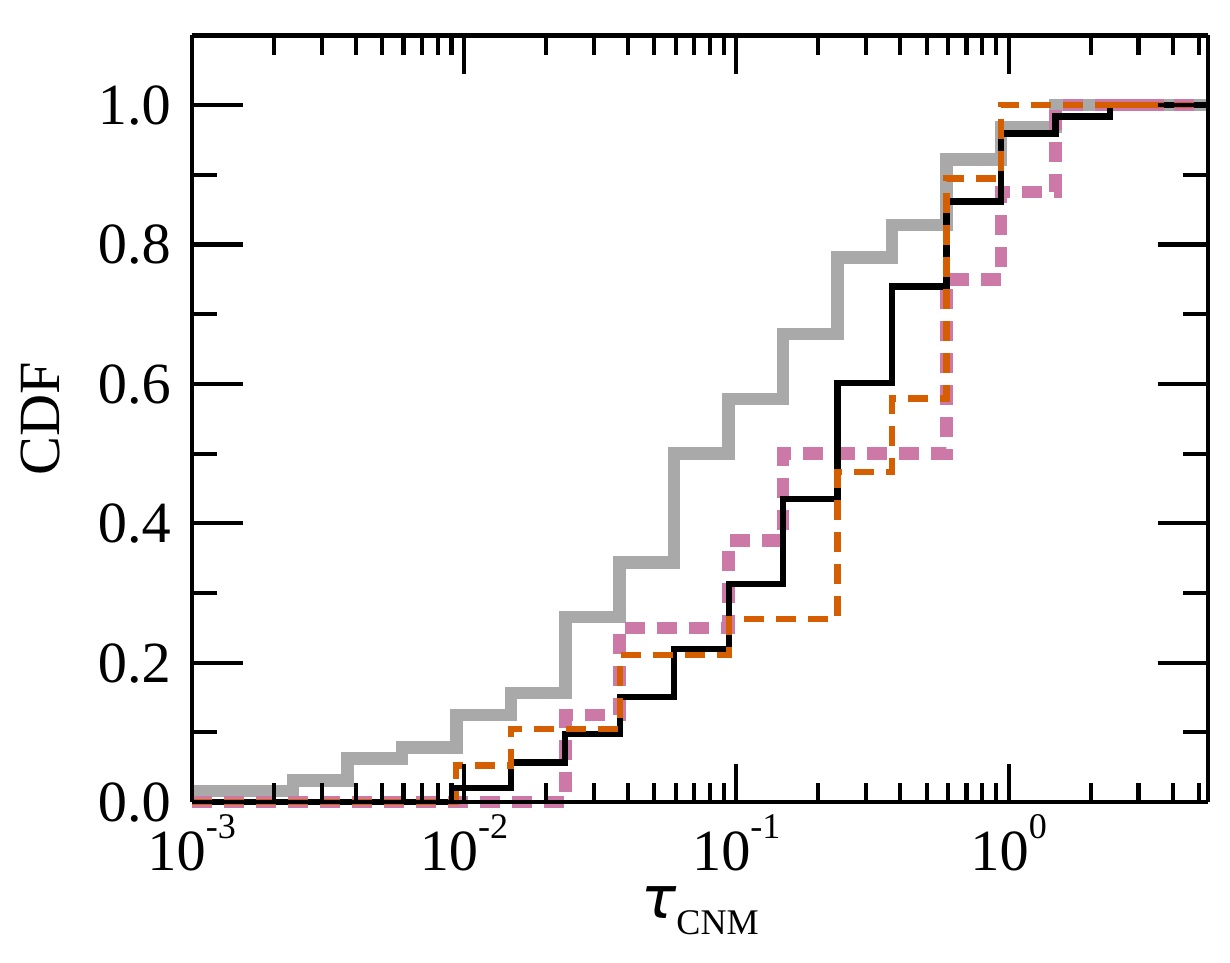}
    \includegraphics[scale=0.35]{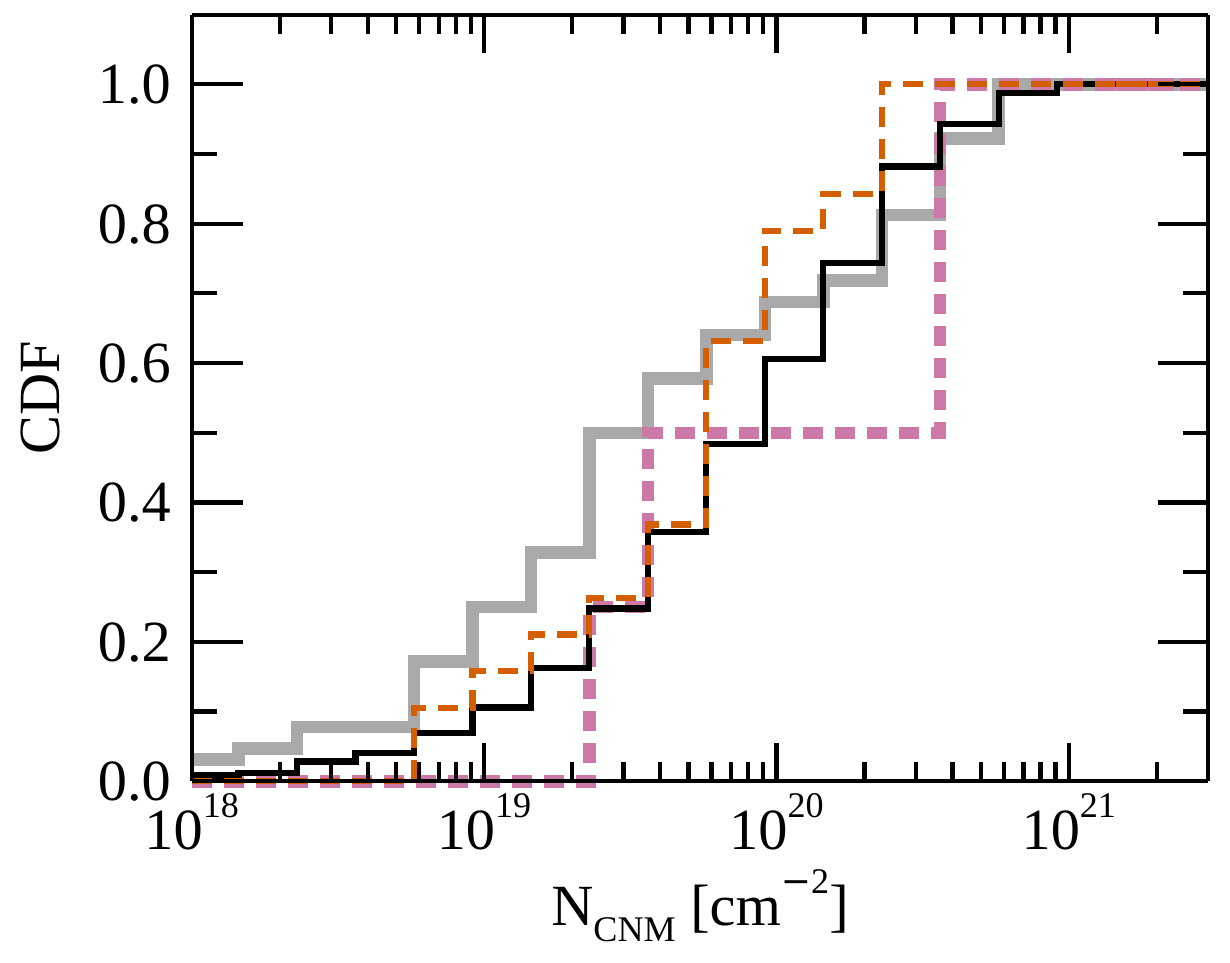}\hspace{0.3Cm}
    \includegraphics[scale=0.35]{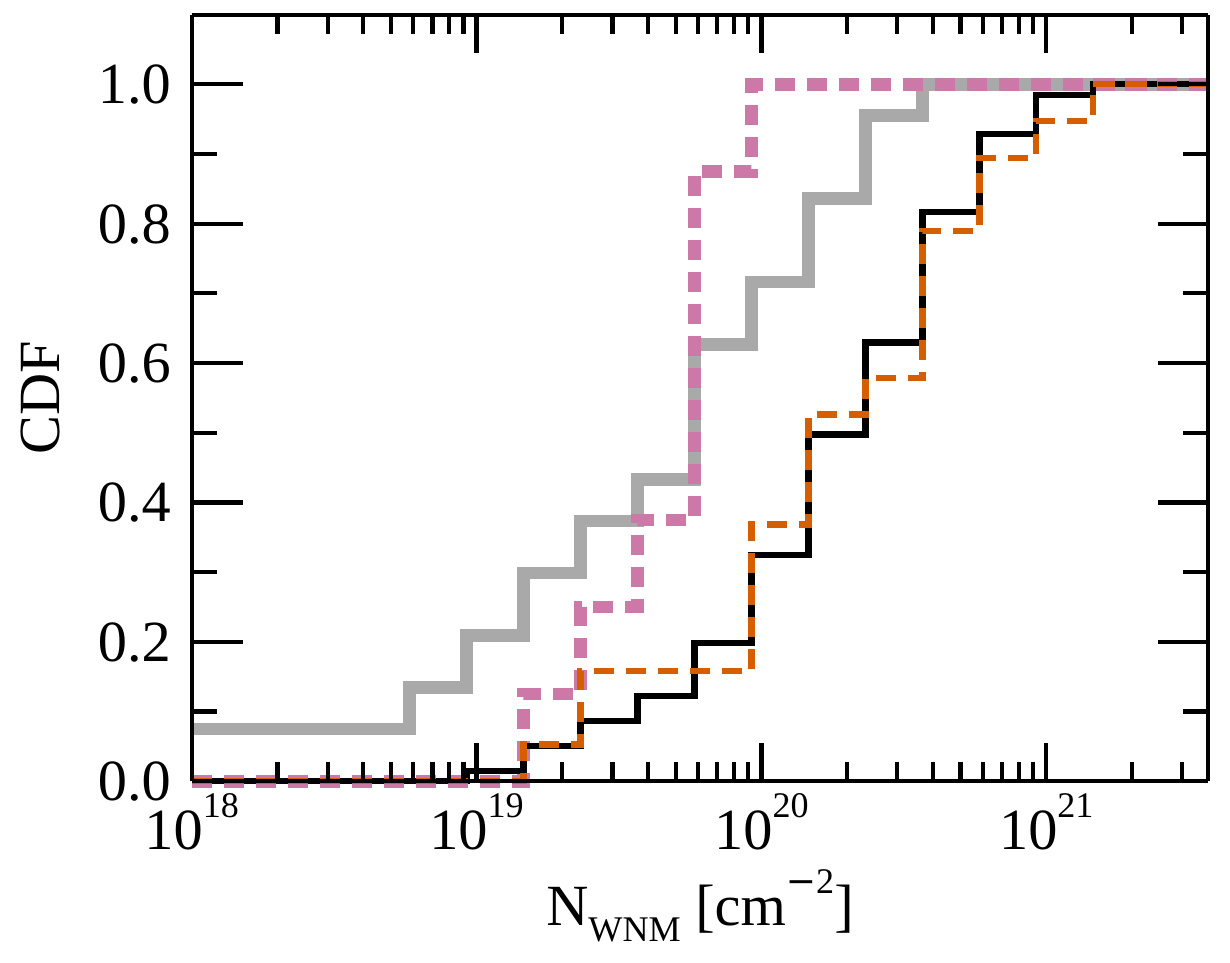}
    \caption{\label{SPONGE_individual} CDFs of the spin temperature, optical depth, CNM and WNM column density. All \HI components toward LOSs at $b$ $<$ $-5{\degree}$ from the GNOMES and ALMA-NOEMA/21-SPONGE surveys are shown in thin black and thick gray (solid lines). Among these components, those in close proximity to CO and HCO$^{+}$ in velocity are selected and presented in thin tan and thick pink (dashed lines).}
\end{figure*}

While both the GNOMES and ALMA-NOEMA/21-SPONGE surveys attempt to probe the conditions for the formation of molecular gas, our approach is uniquely different. First, unlike the ALMA-NOEMA/21-SPONGE survey targeting random LOSs throughout the Milky Way, we focus on individual molecular clouds and their surrounding environments. Second, we employ the most commonly used tracer of molecular gas, CO(1--0) emission, for our analyses. To examine the difference between the two surveys, we selected 11 LOSs at $b$~$<$~$-5{\degree}$ from the ALMA-NOEMA/21-SPONGE survey and extracted 64 CNM and 67 WNM components. In addition, we identified \HI components that are kinematically closest (minimum absolute velocity difference) to the observed CO emission (GNOMES; 19 LOSs) and HCO$^{+}$ absorption (ALMA-NOEMA/21-SPONGE; 8 LOSs) and compared their properties in Figure \ref{SPONGE_individual} and Table \ref{t:GNOMES_21SPONGE}. 

Figure \ref{SPONGE_individual} shows that the CNM and WNM at $b$~$<$~$-5{\degree}$ from the two surveys have in general systematically different properties: i.e., the GNOMES components have lower spin temperatures, higher optical depths, and higher column densities (black and gray CDFs). This difference persists for $T_{\rm s}$ and $N_{\rm WNM}$, 
when we focus on \HI components that are closely associated with molecular gas (tan and pink CDFs). One of the likely reasons for this difference is that the GNOMES survey probes higher column density environments by concentrating on molecular clouds and their surroundings. For example, the CO-detected GNOMES LOSs have a median $A_{V}$ of 2~mag, which is a factor of two higher than that for the HCO$^{+}$-detected 21-SPONGE LOSs. Another possibility is that CO and HCO$^{+}$ formation requires different conditions. In summary, bearing in mind the small number of LOSs in the ALMA-NOEMA/21-SPONGE survey, we conclude that the two surveys sample slightly different populations of the CNM and WNM, while showing consistent results regarding the evolution of \HI properties toward molecular gas. The two surveys are thus highly complementary to each other.

\begin{deluxetable*}{l c c c c c}
    \centering
    \tablecaption{\label{t:GNOMES_21SPONGE}Physical Properties of Individual \HI Components}
    \tablewidth{2pt}
    \setlength{\tabcolsep}{10pt}
    \tabletypesize{\small}
    \tablehead{
    \colhead{(1)} & \colhead{Properties} & \colhead{$T_{\rm s}$} & \colhead{$\tau_{\rm CNM}$} & \colhead{$N_{\rm CNM}$} & \colhead{$N_{\rm WNM}$}\\
    \colhead{} & \colhead{} & \colhead{({$\rm K$})} & \colhead{} & \colhead{($\rm 10^{20}~cm^{-2}$)} & \colhead{($\rm 10^{20}~cm^{-2}$)}
    }
    \startdata
    \multicolumn{6}{c}{GNOMES}\\
    \hline
    (2) & All & 1.99$-$725.42 & 0.01$-$3.41 & 0.01$-$13.70 & 0.10$-$18.57\\
    & & 52.28 & 0.27 & 1.00 & 2.32\\
    (3) & CO-closest & 10.82$-$228.91 & 0.01$-$1.38 & 0.07$-$2.94 & 0.20$-$16.11\\
    & & 40.11 & 0.45 & 0.81 & 2.13\\
    \hline
    \multicolumn{6}{c}{ALMA-NOEMA$/$21-SPONGE}\\
    \hline
    (4) & All & 7.19$-$1551.67 & 9.37E-4$-$1.68 & 3.63E-3$-$8.86 & 2.32E-18$-$5.37\\
    & & 65.47 & 0.10 & 0.37 & 0.75\\
    (5) & HCO$^{+}$-closest & 20.46$-$619.19 & 0.02$-$1.65 & 0.28$-$5.36 & 0.16$-$0.95\\
    & & 65.07 & 0.62 & 3.88 & 0.70\\
    \enddata
    \begin{tablenotes}
        \item \tablecomments{(1) Physical properties. The ranges are given, and the median values are provided below the ranges; (2) \HI components toward the 58 GNOMES LOSs; (3) \HI components that are closest to the measured 19 $T_{\rm peak, CO}$; (4) \HI components toward the 11 ALMA-NOEMA$/$21-SPONGE LOSs; (5) \HI components that are closest to the observed 8 $\tau_{\rm peak, HCO^{+}}$.}
    \end{tablenotes}    
\end{deluxetable*}

\clearpage
\bibliography{GNOMES_ref}
\bibliographystyle{aasjournal}



\end{document}